\documentclass[twocolumn,showpacs,unsortedaddress,preprintnumbers,amsmath,amssymb,pre]{revtex4}

\usepackage{graphicx}
\usepackage{dcolumn}
\usepackage{bm}

\begin{document}


\title{Correlations, spin dynamics, defects: the highly-frustrated Kagom\'e bilayer}

\author{David~Bono}
\altaffiliation[Present address: ]{ Kamerlingh Onnes Laboratorium,
Leiden University, P.O. Box 9504, 2300 RA Leiden, The
Netherlands.}
\author{Laurent~Limot}
\altaffiliation[Present address: ]{Institut f\"ur Experimentelle
und Angewandte Physik, Christian-Albrechts-Universit\"at zu Kiel,
D-24098 Kiel, Germany.}
\author{Philippe~Mendels}%
\email{mendels@lps.u-psud.fr}
\affiliation{%
Laboratoire de Physique des Solides, UMR 8502, Universit\'e
Paris-Sud, 91405 Orsay, France
}%

\author{Gaston~Collin}
\affiliation{ Laboratoire L\'eon Brillouin, CE Saclay, CEA-CNRS,
91191 Gif-sur-Yvette, France
}%

\author{Nicole~Blanchard}%
\affiliation{%
Laboratoire de Physique des Solides, UMR 8502, Universit\'e
Paris-Sud, 91405 Orsay, France
}%

\date{\today}

\begin{abstract}

The SrCr$_{9p}$Ga$_{12-9p}$O$_{19}$ and
Ba$_{2}$Sn$_{2}$ZnGa$_{10-7p}$Cr$_{7p}$O$_{22}$ compounds are two
highly-frustrated magnets possessing a quasi-two-dimensional
Kagom\'e bilayer of spin $\frac{3}{2}$ chromium ions with
antiferromagnetic interactions. Their magnetic susceptibility was
measured by local Nuclear Magnetic Resonance and non-local (SQUID)
techniques, and their low-temperature spin dynamics by Muon Spin
Resonance. Consistent with the theoretical picture drawn for
geometrically frustrated systems, the Kagom\'e bilayer is shown
here to exhibit: (i)~short range spin-spin correlations down to a
temperature much lower than the Curie-Weiss temperature, no
conventional long-range transition occurring; (ii)~a Curie
contribution to the susceptibility from paramagnetic defects
generated by spin vacancies; (iii)~low-temperature spin
fluctuations, at least down to 30~mK, which are a trademark of a
dynamical ground state. These properties point to a spin-liquid
ground state, possibly built on Resonating Valence Bonds with
unconfined spinons as the magnetic excitations.
\end{abstract}


\maketitle

\section{Introduction}

\subsection{Highly frustrated magnets}

\begin{figure}[tbp!] \center
\includegraphics[width=0.45\linewidth]{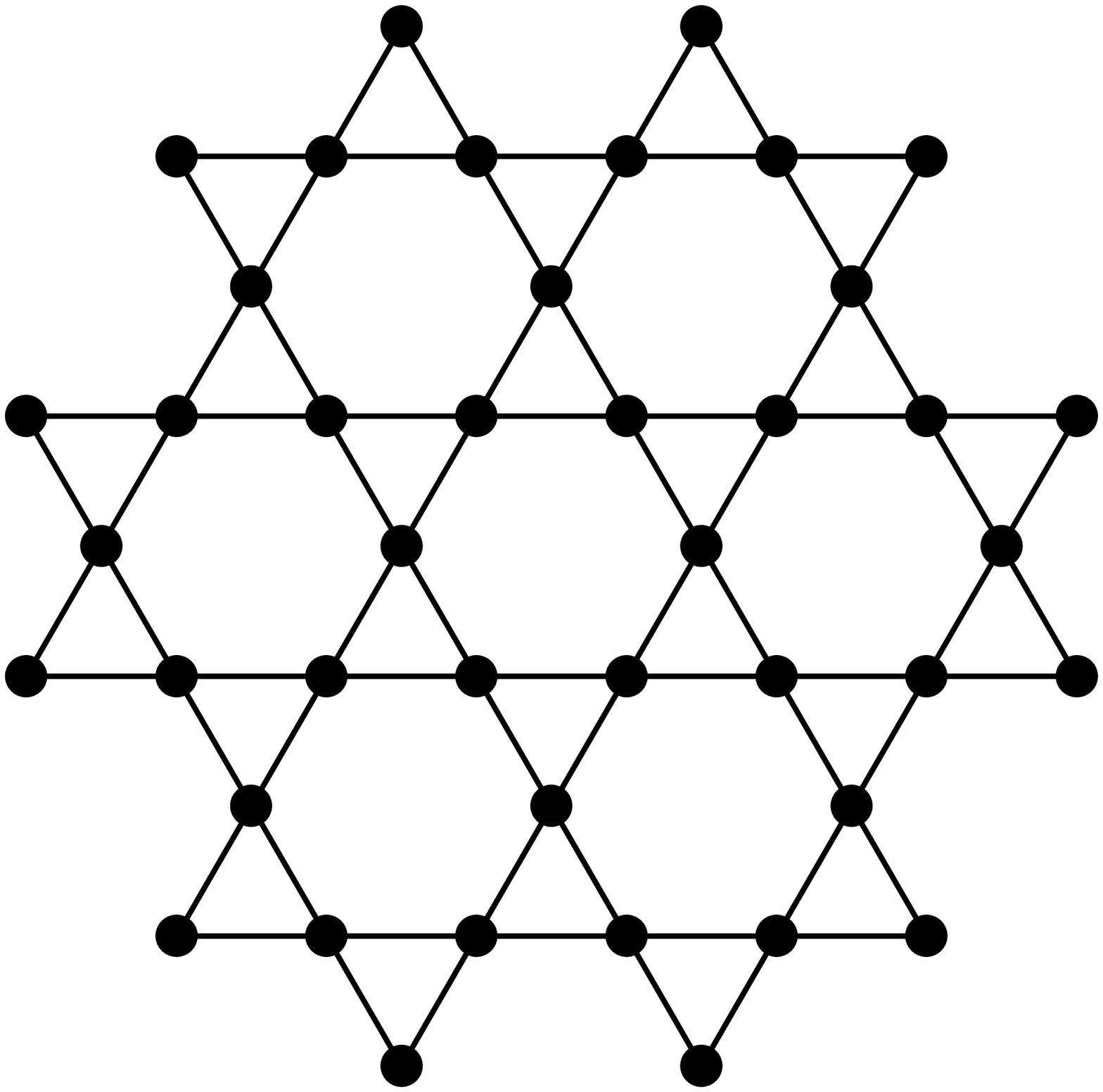} \hfill
\includegraphics[width=0.45\linewidth]{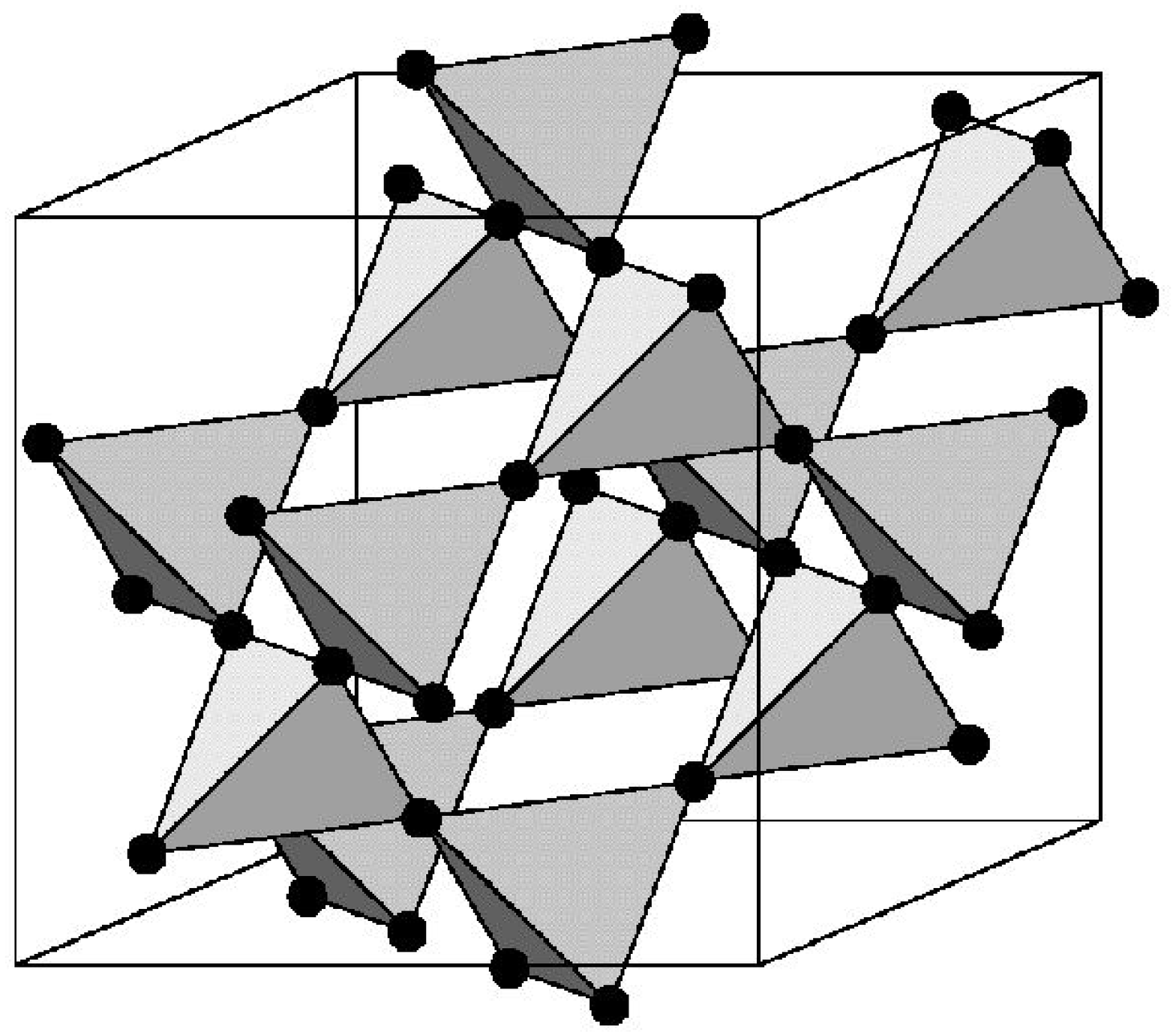}
      \caption{ \label{kagome} The Kagom\'e (left) and the pyrochlore (right \cite{MoessnerHFM01}) corner-sharing lattices.
      The coordinance of each site is $4$ and $6$, respectively.
      }
      \end{figure}

\begin{figure}[tbp!] \center
\includegraphics[width=1\linewidth]{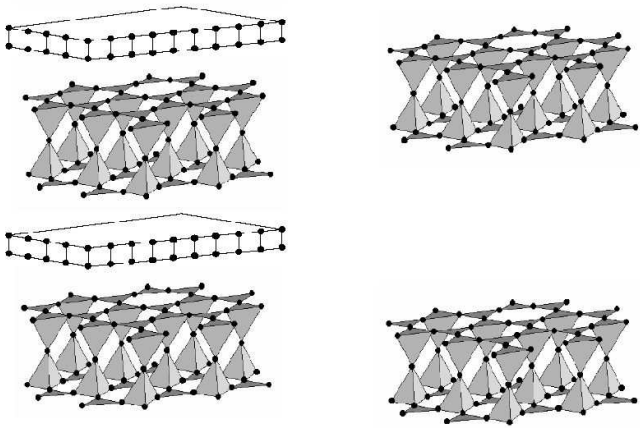}
      \caption{ \label{bikagome} Kagom\'e bilayers in SCGO (left) and
      in BSZCGO (right). Two different sites are present in the
      Kagom\'e bilayer lattice, one of coordinance $5$,
      the other of coordinance $6$. The magnetism of both compounds arises
      from the stacking of magnetically decoupled Kagom\'e bilayers, ensured in
      SCGO by the presence of the non-magnetic singlets and in BZSCGO by a large
      inter-bilayer distance.}
      \end{figure}

Anderson's initial proposal of a Resonating Valence Bond (RVB)
state was intended as a possible alternative for the N\'eel ground
state of a triangular network with Heisenberg spins coupled by an
antiferromagnetic interaction \cite{Anderson73}. The RVB state,
known also as a ``spin-liquid'' state because of the short-range
magnetic correlations and spin fluctuations down to $T=0$, was
also proposed to explain the high-$T_{\text{c}}$ behavior of
cuprates \cite{Anderson87} and, more recently, of the
superconducting compound Na$_{x}$CoO$_{2}\cdot y$H$_{2}$O
\cite{Baskaran03}. Although Anderson's conjecture was proven to be
wrong for the triangular network \cite{Bernu94}, there is a
growing consensus that the RVB state is the ground state of the
so-called highly-frustrated networks \cite{Ramirez01,HFM04}. Like
the triangular network, the magnetic frustration of these systems
is exclusively driven by the triangular geometry of their lattice
provided that spins are coupled through an antiferromagnetic (AFM)
interaction. This is different from spin glasses, where
frustration arises from the randomness of the magnetic
interactions \cite{Binder86}. However, compared to a triangular
network, these networks have a corner-sharing geometry, as in the
two-dimensional Kagom\'e (corner sharing triangles,
Fig.~\ref{kagome}), the three-dimensional pyrochlore (corner
sharing tetrahedras, Fig.~\ref{kagome}), and the
quasi-two-dimensional Kagom\'e bilayer (corner sharing triangles
and tetrahedras, Fig.~\ref{bikagome}) lattices. The corner-sharing
geometry introduces a ``magnetic flexibility'', thereby enhancing
the frustration of these networks (thus the term ``highly'') and
destabilizing the N\'eel order. A highly-frustrated network, and
the experimental counterpart known as highly-frustrated magnet
(HFM), is therefore an ideal candidate to possess a RVB ground
state. Moreover, since HFMs are insulators, the experimental
investigation of their ground state, and the related theoretical
modelling, is greatly simplified by the absence of phenomena such
as superconductivity, charge ordering or itinerant magnetism.

The original features of the ground state of the Heisenberg
two-dimensional (2D) Kagom\'e network, which is the central issue
of this paper, is already apparent through a \emph{classical}
description of their magnetism
\cite{Huse92,Chalker92,Ritchey93,MisguichDiep}, yielding a
dynamical ground state with infinite degeneracy. More precisely,
the ``order by disorder'' mechanism selects a coplanar spin
arrangement in the ground state. However, the system can swap from
one coplanar spin configuration to the other by a rotation of a
finite number of spins, with no cost in energy. These zero-energy
excitation modes, or soft modes, prevent a selection of one of the
coplanar configurations \footnote{This is different to the
situation encountered in spin glasses, where the cooling history
finishes by selecting one of the numerous local minima present in
the free energy.} and constitute a low-energy reservoir of
magnetic excitations. It results in a finite entropy per spin in
the ground state, without, still up to now, any definitive
conclusions on a possible long-range order, even at $T=0$.

In a \emph{quantum} description of the magnetism for Heisenberg
spins $\frac{1}{2}$ on the Kagom\'e network, the magnetic ground
state is disordered \cite{MisguichDiep}. It is predicted to be
RVB-like, built on a macroscopic number of singlet states
\cite{Zeng95,Mambrini00}. If any, a gap between the ground state
and the first triplet state is expected to be fairly small, of the
order of $J/20$ \cite{Waldtmann98}, where $J$ is the AFM
interaction between nearest neighbor spins \footnote{We define the
hamiltonian as $\mathcal{H}=J\sum_{\langle i,j \rangle}
\mathbf{S_{i}}\cdot\mathbf{S_{j}}$ where $i,j$ are the nearest
neighbors.}. Most strikingly and similar to the classical case,
the corner-sharing geometry and the half-integer spins lead to an
exponential density of low-lying singlet excitations
\cite{Mila98}, with energies smaller than the gap, and to a
non-zero entropy per spin at $T=0$ \cite{Misguich03}. The same
conclusions seem to emerge for the AFM pyrochlore lattice
\cite{Canals98}. The nature of the magnetic excitations, possibly
unconfined spinons, is still an open question
\cite{Lhuillier01,Misguich03}.

As a major drawback, every additional contribution to the
nearest-neighbor AFM interaction, such as next-nearest-neighbor
interactions, dipolar interactions, anisotropy,
Dzyaloshinskii-Moriya interactions, lattice distortions or
magnetic defects generated by the presence of magnetic impurities
or spin vacancies, all of which we designate by ``disorder'', may
release the frustration and stabilize an original ground state
\cite{Palmer00,Zhitomirsky00,Elhajal02,Tchernyshyov02,Dommange03,Domenge05}.
Experimentally, disorder is present in all HFMs and indeed causes
cooperative phase transitions at low temperature, reminiscent of
spin-glasses or AFM systems. This is typically observed in the
jarosites Kagom\'e family, where the disorder is governed by spin
anisotropy and by dipolar interactions \cite{WillsHFM01}
\footnote{The $S=\frac{3}{2}$ Cr jarosite is the only exception
\cite{Keren96}.}. In the spin $\frac{1}{2}$ Kagom\'e
[Cu$_{3}$(titmb)$_{2}$(OCOCH$_{3}$)$_{6}$]$\cdot$H$_{2}$O
compound, the competition between first- and second-neighbor
interactions, respectively ferromagnetic and antiferromagnetic,
yields an original ground state, with a double peak in the
specific heat and plateaus in the magnetization under particular
conditions \cite{Narumi04}. In the $S=1$ Kagom\'e staircase
Ni$_{3}$V$_{2}$O$_{8}$, the competition between first and second
neighbor interactions, combined to Dzyaloshinskii-Moriya
interactions and spin anisotropy, produces a very rich field and
temperature phase diagram \cite{Lawes04}. In three dimensions
(3D), dipolar couplings and interactions further than nearest
neighbor yield long-range magnetic order at 1~K in the pyrochlore
compound Gd$_{2}$Ti$_{2}$O$_{7}$ \cite{Raju99}. In
Y$_{2}$Mo$_{2}$O$_{7}$, lattice distortions have been proposed to
relieve the frustration in the ground state \cite{Keren01}.
However, some samples, like the pyrochlore
Tb$_{2-p}$Y$_{p}$Ti$_{2}$O$_{7}$ \cite{Hodges02,Keren04} or the
spinel ZnCr$_{2}$O$_{4}$ \cite{Lee02}, display spin dynamics down
to very low temperature. Anyway, the high spin values and/or the
3D character of the frustrated geometry of these compounds remain
far from the ideal Heisenberg Kagom\'e case we are dealing with.
Along with the spin-liquid ground state of the ideal systems, the
plethoric number of possible ground states in the presence of
disorder (including the case of ferromagnetic interactions in the
pyrochlore lattice, yielding the frustrated so-called ``spin ice''
ground state \cite{Harris99,Ramirez99}, named so as it is
equivalent to the order observed for hydrogen atoms in ice
H$_{2}$O), constitutes one of the rich aspects of the HFMs.


\subsection{SCGO and BSZCGO}

Among all HFMs, the chromium-based spin $\frac{3}{2}$ Kagom\'e
bilayer compounds SrCr$_{9p}$Ga$_{12-9p}$O$_{19}$ [SCGO$(p)$]
\cite{Obradors88} and the
Ba$_{2}$Sn$_{2}$ZnGa$_{10-7p}$Cr$_{7p}$O$_{22}$ [BSZCGO$(p)$]
\cite{Hagemann01} are the most likely candidates to possess a
spin-liquid ground state. This is not surprising in view of the
low spin value, 2D character and low level of disorder of these
compounds compared to other HFMs. They are in fact an experimental
realization of a quasi-2D Kagom\'e bilayer of spin $\frac{3}{2}$
chromium ions with antiferromagnetic interactions
(Fig.~\ref{bikagome}). Moreover, the AFM interaction between
nearest neighboring spins is dominated by a direct overlap
exchange resulting in a coupling of $J\sim40$~K
(Sec.~\ref{susceptibility}). This coupling is almost two orders of
magnitude larger than the typical disorder-related interactions,
like single-ion anisotropy, estimated to $0.08$ K
\cite{Ramirez92,Ohta96}, dipolar interactions $\sim 0.1$ K, or
next nearest neighbor interactions $< 1$ K. These compounds,
however, always have a small amount of substitutional disorder,
since a Cr-coverage higher than $p=0.95$ and $p=0.97$ cannot be
reached in SCGO$(p)$ \cite{Limot02} and in BSZCGO$(p)$
\cite{BonoRMN}, respectively. Hence there are always at least some
percent of the magnetic Cr$^{3+}$ ions which are substituted with
non-magnetic Ga$^{3+}$ ions. These low amounts of spin vacancies,
and the related magnetic defects that they produce, turn out to be
insufficient to destroy the spin-liquid behavior of the ground
state. We will show that local techniques like Nuclear Magnetic
Resonance (NMR) and Muon Spin Relaxation ($\mu$SR) are the most
suited to probe frustration-related properties in these systems,
since they can bypass the magnetic contribution of these defects.

      \begin{figure}[tbp!] \center
\includegraphics[width=0.8\linewidth]{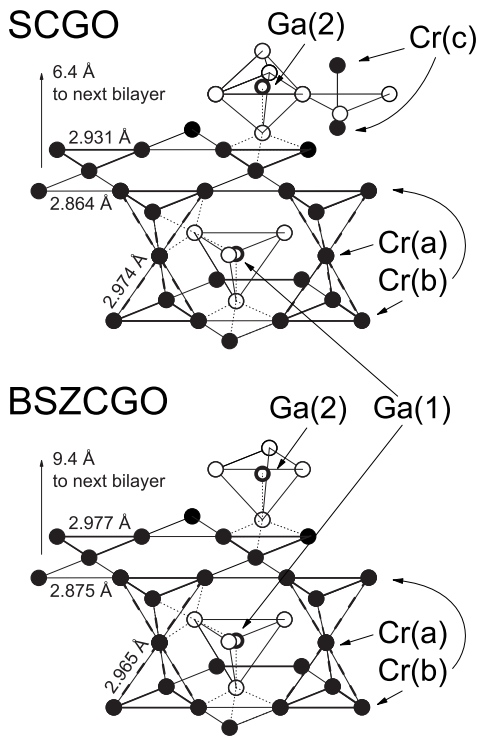}
      \caption{ \label{cell} Cr$^{3+}$ network of SCGO and BSZCGO with their
      oxygen ($\bigcirc$) environments, along with the two Ga$^{3+}$ sites.
      The dotted lines represent some of the Ga-O-Cr hyperfine
      coupling paths.}
      \end{figure}

Figure~\ref{cell} presents a simplified chemical structure for the
ideal $p=1$ unit cell of SCGO and of BSZCGO, highlighting the
chromium Kagom\'e bilayers \footnote{The official labelling are
the following for SCGO [BSZCGO]: Ga(4f) [Ga(2d)] for Ga(1), Ga(4e)
[Ga(2c)] for Ga(2), Cr(1a) [Cr(2a)] for Cr(a), Cr(12k) [Cr(6i)]
for Cr(b) and Cr(4f$_{vi}$) for Cr(c) in SCGO. We used simplified
notations for clarity in the comparison of both systems.}. As it
can be seen, BSZCGO is a pure Kagom\'e bilayer, on the contrary of
SCGO where there are two additional chromium sites, labelled
Cr(c), between the bilayer. These pairs of Cr-spins are coupled by
an AFM exchange constant of $216$~K with a weak coupling to the
Kagom\'e bilayers ($\sim 1$~K) \cite{Lee96}. They form an isolated
singlet at low temperatures, and the resulting susceptibility is
negligible at $T\lesssim 50$~K compared to the susceptibility of
the Kagom\'e bilayer in a pure compound. Moreover, the Ga/Cr
substitutions are substoechiometric on these sites (see
Sec.~\ref{DefPara}). The full crystal structure is obtained by the
stacking of the unit cells presented in Fig.~\ref{cell}. The
magnetism of both compounds arises by magnetically decoupled
Kagom\'e bilayers, ensured in SCGO by the presence of the
non-magnetic singlets at low temperature (only weakly affected by
the Ga/Cr substitution, see Sec.~\ref{DefPara}), and in BSZCGO by
the large interbilayer distance of $9.4$ {\AA} (in SCGO of $6.4$
{\AA}). These compounds can therefore be considered as ideal
two-dimensional systems.

      \begin{figure}[tbp!] \center
\includegraphics[width=1\linewidth]{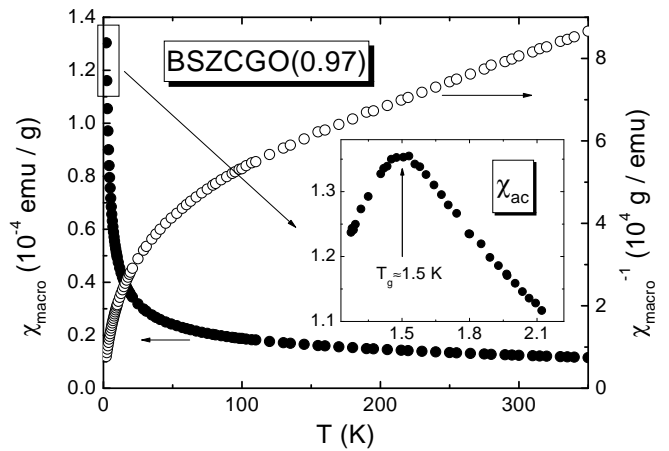}
      \caption{ \label{SQUID} Macroscopic static susceptibility
      $\chi _{\text{macro}}(T)$ of BSZCGO(0.97) measured with a SQUID
      under 100~G (left scale) and $\chi _{\text{macro}}^{-1}(T)$
      (right scale). Inset~: ac susceptibility (78~Hz) below 2~K focusing
      on the spin-glass-like transition ($T_{g}$).}
      \end{figure}

\begin{figure}[tbp!] \center
\includegraphics[width=1\linewidth]{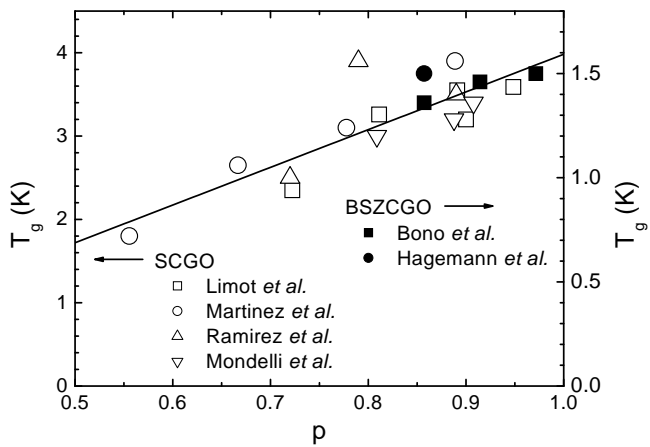}
      \caption{ \label{Tg} Collection of spin-glass-like transition temperatures $T_\text{g}$
      for SCGO and for BSZCGO versus Cr-concentration $p$
      (our data and from Ref.~\onlinecite{Ramirez92,Martinez92,Mondellipriv,Hagemann01}).
      The line is a guide to the eye.}
      \end{figure}

Figure~\ref{SQUID} shows the typical macroscopic susceptibility
$\chi _{\text{macro}}$ in these compounds. The linearity of $\chi
_{\text{macro}}^{-1}$ at high temperature and the extrapolation of
this line to $\chi _{\text{macro}}^{-1}=0$ to a negative
temperature is typical of AFM interactions and gives an order of
magnitude of the Curie-Weiss temperature $\theta_{CW}=zS(S+1)J/3$
($600$~K and $350$~K for SCGO and BSZCGO). However, contrary to
the case of ``conventional'' antiferromagnets \cite{deJongh01}, no
kink is evidenced in $\chi _{\text{macro}}$ around $\theta_{CW}$.
This is actually one of the common signatures of the frustration
in HFMs \cite{Ramirez01}: the magnetic correlation length cannot
increase because of frustration, and mean field theory remains
valid for $T\sim\theta_{CW}$. At low temperature, a
spin-glass-like transition occurs in both systems, around a
freezing temperature $T_{g}\sim3$~K in SCGO$(p)$ and 1.5~K in
BSZCGO$(p)$ (Fig.~\ref{Tg}).

However, unlike in other HFMs, the spin-glass transition is
unconventional. In fact, in \emph{conventional 3D spin glasses},
the nonlinear susceptibility diverges at $T_\text{g}$, the
specific heat is proportional to $T$ at low temperature, and
$T_\text{g}$ is proportional to the number of defects. In the
\emph{Kagom\'e bilayers}, however, (i)~the nonlinear
susceptibility diverges at $T_\text{g}$ \cite{Ramirez90}, but
$T_\text{g}$ increases when the number of spin vacancies decreases
(Fig.~\ref{Tg}), indicating that the freezing is related to an
intrinsic property of the frustrated bilayer. (ii)~The specific
heat is proportional to $T^{2}$, as in \emph{AFM 2D long-range
ordered systems}. Its unusual large value at low temperature
unveils a high density of low-lying excitations
\cite{Ramirez90,Hagemann01}, and its insensitivity to an external
magnetic field suggests a large contribution from singlet
excitations, which is one of the most striking features predicted
for the $S=\frac{1}{2}$ Kagom\'e lattice
\cite{Lecheminant97,Ramirez00,Sindzingre00}. (iii)~Magnetic
fluctuations and short-range correlations, consistent with a 2D
magnetic network, are encountered at low temperature
\cite{Broholm90,Uemura94}. The Kagom\'e bilayers therefore combine
properties of 3D spin glasses, 2D AFM ordering, 2D fluctuating
magnetic states and original properties expected in
$S=\frac{1}{2}$ Kagom\'e systems.

Given these two temperature scales of the macroscopic
susceptibility, $\theta_{CW}$ and $T_{g}$, Ramirez proposed the
definition of a ``frustration ratio'', $f=\theta_{CW}/T_{g}$, with
$f>10$ in all HFMs \cite{Ramirez94}. The Kagom\'e bilayers SCGO
and BSZCGO respectively display $f\sim 150$ and 230
\cite{BonoRMN}, the largest ratios reported so far in compounds
where the frustration is driven by corner sharing equilateral
triangles \cite{Ramirez01}. The large frustration, the Heisenberg
spins along with the low disorder, make them the archetypes of HFM
and maybe the best candidates for a RVB spin-liquid ground state.

We present a review of the magnetic properties of SCGO$(p)$ and
BSZCGO$(p)$ \emph{powder samples}, which covers a large range of
Cr-concentration $p$. The comparison between samples of different
concentration allows to identify the frustrated and
disorder-related magnetic properties of the Kagom\'e bilayer.
Through local probes it is then possible to determine the true
nature of the bilayer's static susceptibility (by NMR), as well as
the bilayer's low-temperature dynamics and ground state magnetic
excitations (by $\mu$SR). The outline of the paper is as follows.
Section~\ref{susceptibility} is dedicated to the susceptibility of
the Kagom\'e bilayer. It is shown that the susceptibility of the
bilayer of SCGO and of BSZCGO can be accessed directly through
gallium NMR experiments
\cite{Keren98,Mendels00,Limot01,Limot02,BonoHFM,BonoRMN}. Both
susceptibilities exhibit a maximum in temperature at $T\sim
40~\text{K}\sim J$ and decrease below this maximum down to at
least $J/4$. This behavior indicates that short-range magnetic
correlations persist at least down to $10$~K, consistent with the
existence of a small spin gap. The comparison of the Kagom\'e
bilayer's susceptibility with $\chi _{\text{macro}}$ shows that
the spin vacancies of the bilayer generate paramagnetic defects
responsible for the low-temperature Curie upturn observed in the
macroscopic susceptibility (both systems have also extra Curie
contributions, coming from broken Cr(c) spin-pairs in SCGO and
from bond length distribution in BSZCGO). Section~\ref{MuSR} is
dedicated to the $\mu$SR study of the low-temperature spin
dynamics of these HFMs \cite{Uemura94,Keren00,BonoHFM,BonoMuSR}.
It is shown, that magnetic excitations persist down to at least
$30$~mK (the lowest temperature that could be accessed). This
temperature sets an upper limit for the value of a spin gap. A
qualitative and quantitative analysis of the data shows that these
excitations are possibly coherent unconfined spinons of a RVB
ground state. The energy scale $T_{g}$ would correspond then, in
this phenomenological approach, to the signature of this coherent
resonating state. A summary and concluding remarks can be found in
Sec.~\ref{conclusion}.


\section{Frustrated vs. disorder-related susceptibility}
\label{susceptibility}

\subsection{Gallium NMR in SCGO and BSZCGO}
\label{Ga NMR}

The NMR experiments were performed on $^{69}$Ga
($^{69}\gamma/2\pi=10.219$~MHz/T, $^{69}Q=0.178\times
10^{-24}$~cm$^{2}$) and $^{71}$Ga
($^{71}\gamma/2\pi=12.983$~MHz/T, $^{71}Q=0.112\times
10^{-24}$~cm$^{2}$) nuclei in SCGO and BSZCGO using a
$\frac{\pi}{2}$-$\tau$-$\pi$ spin echo sequence, where $\gamma$
and $Q$ are, respectively, the gyromagnetic ratio and the
quadrupolar moment of the nuclei. The gallium ions are located on
two distinct crystallographic sites, labelled Ga(1) and Ga(2) in
Fig.~\ref{cell}. As pointed out in
Ref.~\onlinecite{Keren98,Limot02,BonoHFM,BonoRMN}, the interest of
Ga NMR lies in the coupling between gallium nuclei with their
neighboring magnetic Cr$^{3+}$ ions through a Ga-O-Cr hyperfine
bridge (Fig.~\ref{cell}). In particular, the $^{69,71}$Ga(1)
nuclei are exclusively coupled to the Cr$^{3+}$ ions of the
Kagom\'e bilayer: to nine chromium ions of the upper and lower
Kagom\'e layers (labelled Cr(b) in Fig.~\ref{cell}), and to three
chromium ions in the linking site in between the two Kagom\'e
layers (labelled Cr(a) in Fig.~\ref{cell}). Through the NMR of
Ga(1), it is possible to probe locally the magnetic properties of
the Kagom\'e bilayers of both HFMs. This is the central topic of
the NMR study presented in this paper.

Before going any further, we briefly recall some relationship
between the contribution of each gallium nucleus to the Ga(1) NMR
line and its local magnetic environment in order to underline what
can be exactly probed through Ga(1) NMR. We suppose first that the
susceptibility in the Kagom\'e bilayer varies from chromium site
to chromium site and label it, in a generic manner, by $\chi$. A
Ga(1) at site $i$ will contribute to the NMR spectrum at a
position depending upon the number of the nearest neighbors (NN)
occupied chromium sites and their susceptibility $\chi$. This
corresponds to the shift $K^{(i)}$ in the NMR spectrum for a
gallium at site $i$
\begin{equation}
\label{eqshift} K^{i}=\sum_{\text{occupied NN Cr}(a,b)}A\chi \ ,
\end{equation}
where $A$ is the hyperfine constant (the chemical shift is
neglected here). The average shift, $K$, of the NMR line, is
simply related to the average $\overline{K^{i}}$ over all the
gallium sites. Hence, it is practically proportional to the
average susceptibility $\chi  _{\text{Kag}}$ of the Kagom\'e
bilayer. On the other hand, the distribution of $K^{i}$ around $K$
defines the magnetic width of the Ga NMR line and reflects the
existence of a distribution of $\chi$. As we show in the following
sections, the NMR width probes a susceptibility related to
disorder in BSZCGO and in SCGO.

Along with this hyperfine interaction, the $^{69,71}$Ga
Hamiltonian in SCGO and in BSZCGO has a quadrupolar contribution.
The quadrupole interaction of gallium nuclei in both SCGO and
BSZCGO is a consequence of the coupling of the nucleus to the
electric field gradient (EFG) produced by the surrounding
electronic charges. Following the usual notations, the nuclear
Hamiltonian may be expressed as
\begin{equation}
\label{NMRHamiltonian}
    \mathcal{H}=-\gamma \hbar \mathbf{I}\left(\bar{1}+ \bar{K} \right)\mathbf{H_{0}} +
               \frac{h \nu_{Q}}{6}\left[3I_{z}^{2}-I^{2}+
                        \eta\left(I_{x}^{2}-I_{y}^{2}\right)\right]  \ ,
\end{equation}
where $\mathbf{H_{0}}$ is the applied magnetic field, $\nu_{Q}$
the quadrupolar frequency and $0\leq\eta\leq1$ the asymmetry
parameter. The principal axes of the shift tensor $\bar{K}$ are
collinear with the direction of the nuclear spin operators
$I_{x}$, $I_{y}$ and $I_{z}$.

\begin{table}[tbp!]
\caption{ \label{ParamQuad} Quadrupole parameters for the
$^{71}$Ga sites in SCGO \cite{Keren98,Limot02} and in BSZCGO
\cite{BonoRMN}. Notice that $^{69}\nu_{Q}=(^{69}Q/^{71}Q)
\nu_{Q}\approx1.59^{71}\nu_{Q}$. Quadrupole effects are therefore
more pronounced for the $^{69}$Ga isotope than for the $^{71}$Ga.}
\begin{ruledtabular}
\begin{tabular}{l l l l}
& Ga site & $^{71}\nu_{Q}$ (MHz) & $\eta$ \\
\hline
SCGO    &Ga(1) & \ \,2.9(2)     & 0.005(6)  \\
        &Ga(2) & 20.5(3)    & 0.050(35) \\
BSZCGO  &Ga(1) & \ \,3.5(1.0)   & 0.60(15) \footnote{Determined using point charge simulation.}  \\
        &Ga(2) & 12.0(5)    & 0.04(3)   \\
\end{tabular}
\end{ruledtabular}
\end{table}

\begin{figure}[tb!] \centerline{
\includegraphics[width=1\linewidth]{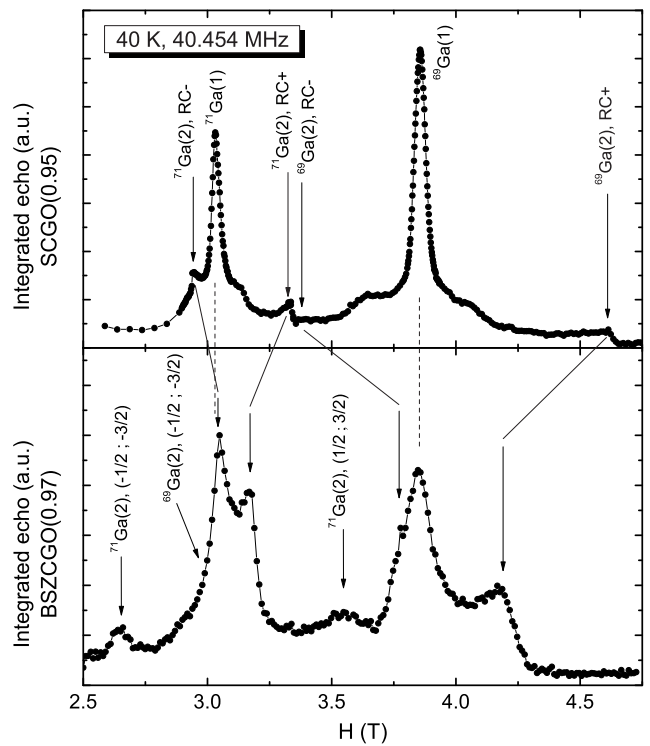}}
\caption{ \label{NMRcomparison} Comparison of the $^{69,71}$Ga NMR
spectra in SCGO and BSZCGO obtained by sweeping the field with a
constant frequency $\nu_{l}\approx40~$MHz, at $T=40$~K. The arrows
point the quadrupolar singularities for the Ga(1) and Ga(2),
respectively located inside and outside the bilayers
(Fig.~\ref{cell}). ``CL'' points the central line
($\frac{1}{2}\leftrightarrow-\frac{1}{2}$ transition)
singularities. The continuous and dashed lines show the shift of
the in-range singularities of the Ga sites, from one system to the
other. We cannot identify the first order satellites of the Ga(1)
in BSZCGO on this spectrum.}
\end{figure}

As shown in Fig.~\ref{cell} the gallium sites of both HFMs have
different crystallographic environments. Some simple arguments can
be used to anticipate differences concerning their spectra (Tab.~\ref{ParamQuad}):

(i) The Ga(1) sites of both HFMs have a nearly-tetrahedral oxygen
environment. Given this nearly-cubic symmetry, the EFG is small,
i.e. $\nu_{Q}$ is small. However, whereas the Ga(1) site of SCGO
is only occupied with Ga$^{3+}$ ions, the Ga(1) site of BSZCGO is
randomly occupied either by Zn$^{2+}$ or by Ga$^{3+}$ ions. As a
consequence, point charge simulations of the EFG on the Ga(1) site
of BSZCGO show that there is a large distribution of $\eta$ and
$\nu_{Q}$, contrary to the Ga(1) of SCGO.

(ii) The Ga(2) site of SCGO is surrounded by a bi-pyramid of 5
oxygen ions, whereas the Ga(2) of BSZCGO lies in an
oxygen-elongated tetrahedra. Both environments have no cubic
symmetry which yields a large quadrupolar frequency. For this
site, point charge simulations yield a $\nu_{Q}$ twice larger in
SCGO than in BSZCGO.

(iii) The ratio Ga(1)$:$Ga(2) is 2$:$1 in SCGO, whereas it is
1$:$2 in BSZCGO. Consequently, the relative spectral weight of the
Ga(1) and Ga(2) sites will be opposite in the two compounds.

\begin{figure}[tbp!]
\center
\includegraphics[width=1\linewidth]{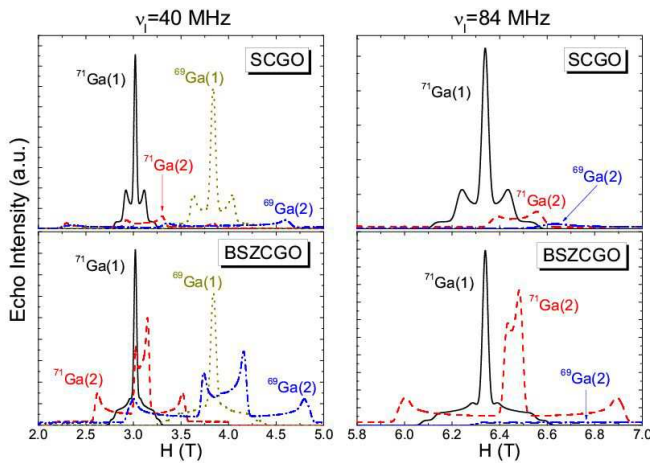}
\caption{Quadrupolar powder pattern simulations for both sites and
isotopes with the parameters of Tab.~\ref{ParamQuad}, for
frequencies $\nu_{l}=40~$MHz (left) and 84~MHz (right) in SCGO and
BSZCGO. The area of each contribution is obtained considering the
stoichiometry of each site in the samples and the natural
abundance of each isotope. The full NMR spectra
are obtained by adding all the contributions (not shown).}%
\label{NMRsimul}%
\end{figure}

\begin{figure}[tb]
\center
\includegraphics[width=0.9\linewidth]{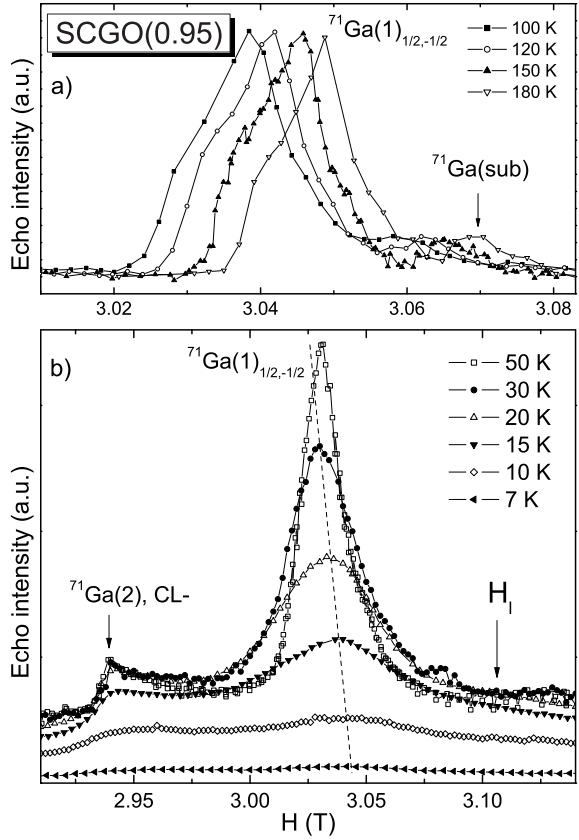}
\caption{$^{71}$Ga spectra in SCGO
($H_{l}=2\pi\nu_{l}/^{71}\gamma$). a)~High temperature. The
$^{71}$Ga(2) contribution appears as a flat background in this
field range. b)~Low temperature, lines are broadened. The
$^{71}$Ga(2) contribution remains flat at the position of the
$^{71}$Ga(1) line,
which shifts to higher fields in this temperature region when the temperature decreases. }%
\label{spectresSC}%
\end{figure}

An extensive study of the $^{69,71}$Ga NMR spectra in SCGO was
presented in Ref.~\onlinecite{Keren98,Limot02}. It allowed to
identify the NMR lines and to extract the parameters of the
nuclear Hamiltonian, among them the quadrupolar parameters given
in Tab.~\ref{ParamQuad}. It was shown that the Ga NMR spectrum of
both isotopes is the sum of the Ga(1) and Ga(2) sites, plus a
small extra contribution related to the presence of the
non-stoichiometric gallium substituted on the chromium sites
(labelled Ga(sub) in Fig.~\ref{spectresSC}). A similar analysis
allowed to identify the NMR lines in BSZCGO, and to determine the
related parameters of the nuclear Hamiltonian \cite{BonoRMN}, also
listed in Tab.~\ref{ParamQuad}.

Neglecting the Ga-substituted sites, the Ga NMR spectrum displays
four sets of lines corresponding to the two isotopes distributed
on both Ga(1) and Ga(2) sites. In a powder sample, as used here,
for a given site and a given isotope, the line shape results from
the distribution of the angles between the magnetic field and the
EFG principal axes. This yields singularities rather than well
defined peaks (the so-called powder line shape), as in the Ga NMR
spectra of SCGO(0.95) and of BSZCGO(0.97) presented in
Fig.~\ref{NMRcomparison}.

Powder-pattern simulations for two different r.f. frequencies
($\nu_{l}=40~$MHz and 84~MHz) are presented in
Fig.~\ref{NMRsimul}, using the parameters of Tab.~\ref{ParamQuad}.
They perfectly agree with the experimental NMR spectra and clearly
show that the BSZCGO spectrum is more intricate than the SCGO one,
because of the more pronounced quadrupolar contribution. This is
probably the reason why the Ga(sub) line cannot be resolved in
BSZCGO, contrary to SCGO.

Most of the SCGO data were acquired in a field-sweep spectrometer
with a r.f. frequency $\nu_{l}\approx40~$MHz, because then the
$^{69,71}$Ga(1) NMR lines can be isolated, as shown in
Fig.~\ref{NMRsimul} \cite{Limot02}. At high temperatures
(Fig.~\ref{spectresSC}a), the NMR line shape remains unaltered and
shifts with decreasing temperature. At low temperature ($T<50$~K),
the line broadens with decreasing temperature
(Fig.~\ref{spectresSC}b). The $^{69,71}$Ga(1) shift is still
qualitatively visible down to 5 K, where the Ga NMR signal is lost
(see below). Fits of the low-temperature spectra (using a
convolution of a high-temperature unbroadened spectrum with a
Gaussian) allow to extract quantitatively the shift and the width
of the NMR lines for all the samples studied ($0.72\leq p\leq
0.95$). All display similar shifts with temperature, but different
broadening, the line shape increasing with Ga/Cr substitution at a
given temperature \cite{Limot02}.

\begin{figure}[tb]
\center
\includegraphics[width=0.9\linewidth]{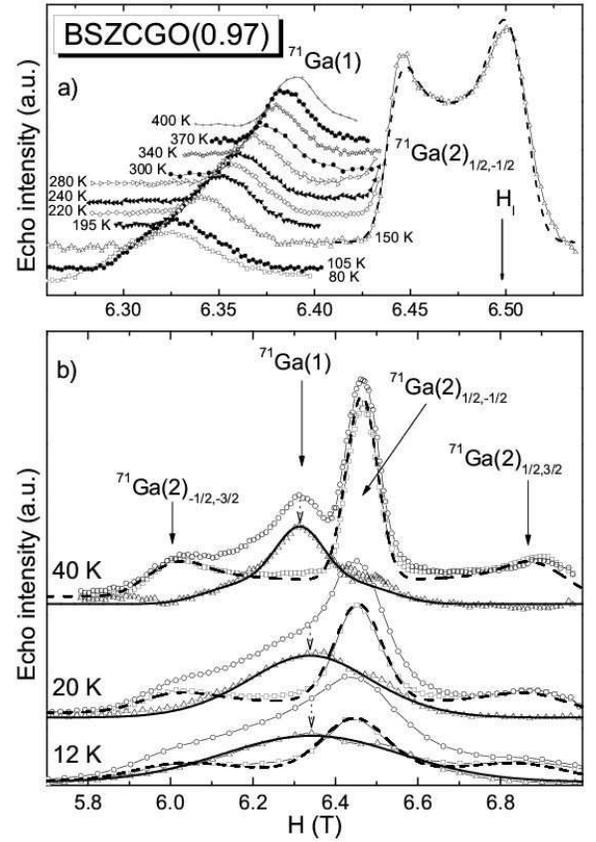}
\caption{$^{71}$Ga spectra in BSZCGO. a)~High temperature. The
$^{71}$Ga(2) first order quadrupolar contribution appears as a
flat background in this field range at the $^{71}$Ga(1) line
position. b)~Low temperature, lines are broadened. $\circ$ and
$\square$ are the short $\tau$ and the rescaled long $\tau$
spectra respectively. $\triangle$ is for the Ga(1) contribution,
given by their subtraction. The dotted arrows point at the center
of the Ga(1) line, which shifts to higher fields when $T$
decreases. Continuous (dashed) lines are gaussian broadened
$^{71}$Ga(1) [$^{71}$Ga(2)] quadrupolar powder pattern simulation,
with $\nu_{Q}=3.5$~MHz
(12~MHz) and $\eta=0.6$ ($\eta=0.04$). }%
\label{spectres}%
\end{figure}

In the case of BSZCGO, it can be seen from Fig.~\ref{NMRsimul}
that it is advantageous to work at a frequency higher than 40 MHz,
in order to isolate the Ga(1) contribution \footnote{Since the
Ga(2) line of BSZCGO is governed by quadrupolar effects, its width
is $\propto \nu_{Q}^{2}/\nu_{l}$ \cite{Cohen57}. Therefore this
line is broader at low field. On the contrary, the quadrupolar
frequency of the Ga(1) is smaller and the linewidth is
proportional to the applied external field.}. Field-sweep
$^{71}$Ga NMR experiments were performed at a r.f. frequency
$\nu_{l}\approx84$~MHz. At high temperature
(Fig.~\ref{spectres}a), a $^{71}$Ga(1) shift is evidenced similar
to the one in SCGO \footnote{Notice that the high temperature
analysis of both $^{71}$Ga(1) and $^{71}$Ga(2) lines in BSZCGO
allows us to compare their shifts, which yields an estimate of the
chemical shift of 0.15(3)\%.}. At low temperature, the
$^{71}$Ga(1) and $^{71}$Ga(2) lines broaden and start to overlap.
However, the two lines can be resolved by exploiting the different
transverse relaxation times $T_{2}$ of the two gallium sites
($T_{2,Ga(1)}\sim 20~\mu$s, $T_{2,Ga(2)}\sim 200~\mu$s). The first
step consists in using a large time separation between the two
r.f. pulses ($\tau=200~\mu$s) in order to eliminate the Ga(1)
contribution. The isolated Ga(2) line is extracted using a
gaussian convolution of the quadrupolar powder pattern. In the
second step, a short time separation is employed between the
pulses ($\tau\approx10~\mu$s) so that the Ga(1) contribution is
recovered. The Ga(1) line is finally isolated by subtracting the
long $\tau$ spectrum corrected in intensity to account for the
shorter time separation employed \footnote{We checked that the
transverse relaxation time $T_{2}$ is homogeneous over the whole
$^{71}$Ga(2) line. Its shape is hence used for the subtraction
whereas its intensity is multiplied by
$\sim\exp[2(\tau_{l}-\tau_{s})/T^{71}_{2,2}]$.}. Such a procedure
can be employed down to 10~K. The width and the shift of the Ga(1)
line can then be extracted from a fit using a
$^{71}\nu_{Q}=3.5$~MHz quadrupolar powder pattern convoluted by a
gaussian line shape, and are little affected by the uncertainty on
$^{71}\nu_{Q}$. The ratio between the intensities of the Ga(2) and
Ga(1) lines (integrated area below the lines) was found to stay
close to 2 down to 10~K, in agreement with the 2:1 stoichiometric
ratio for the Ga(2) and Ga(1) sites of BSZCGO.

\begin{figure}[tb]
\center
\includegraphics[width=1\linewidth]{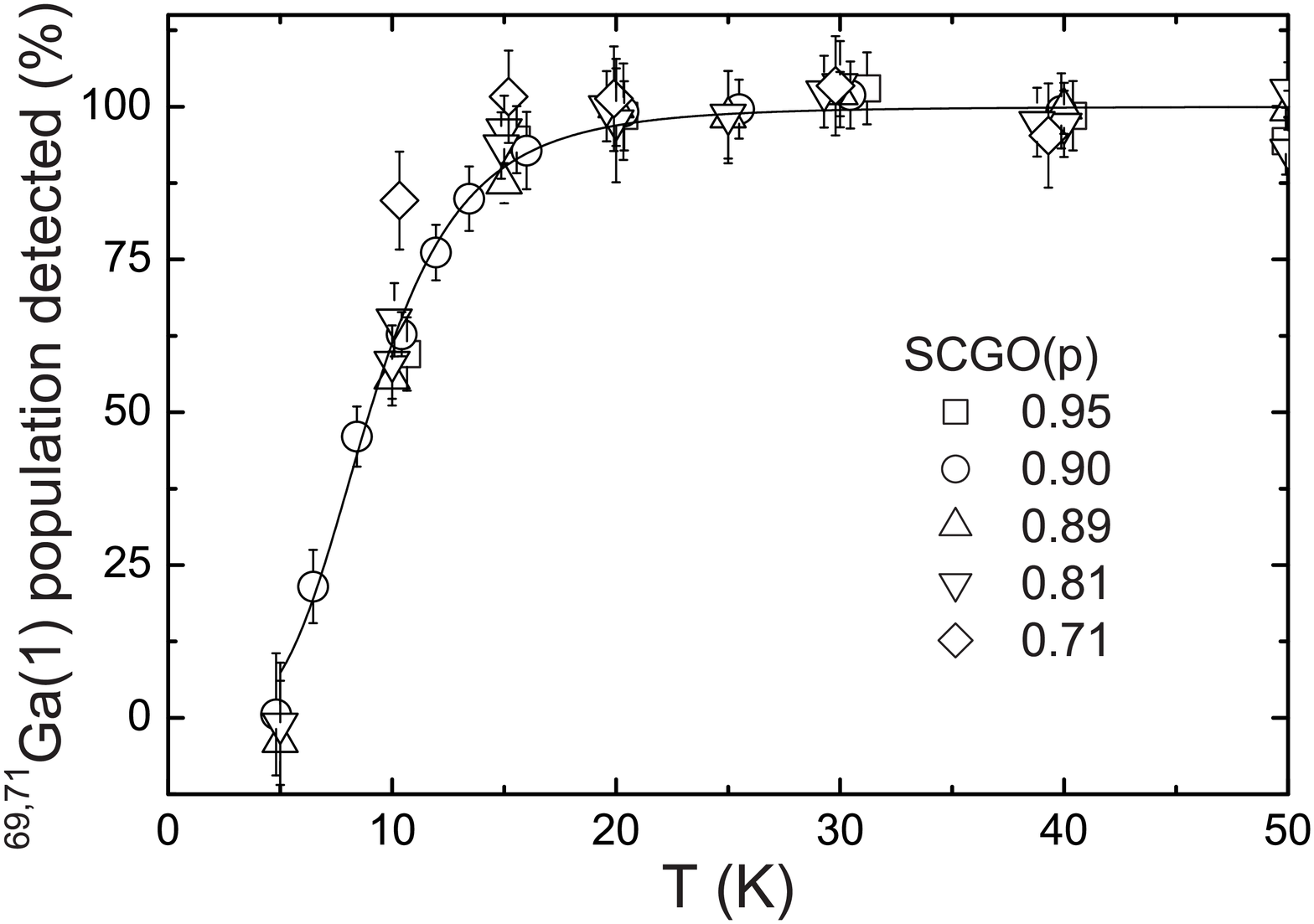}
\caption{NMR signal intensity in several SCGO$(p)$ samples. A loss
of $^{69,71}$Ga sites is observed for $T\lesssim5T_{g}$
($\sim16$~K). The line is a guide to the eye. }%
\label{NMRintensity}%
\end{figure}

The temperature dependence of the intensity of the $^{71}$Ga(1)
NMR line, which corresponds to the number of detected $^{71}$Ga(1)
nuclei, is presented in Fig.~\ref{NMRintensity} for SCGO. A
similar trend is observed in BSZCGO, with larger error bars due to
the preponderant Ga(2) contribution. As temperature decreases the
spin dynamics slows down, and results in a decreasing longitudinal
relaxation time $T_{1}$ (see Sec.~\ref{MuSR}), hence in a
progressive loss of the NMR signal below 15 K. Such a wipeout is
reminiscent of the one observed in spin-glasses at temperatures
near $T_{g}$ \cite{Alloul76}. However, contrary to spin-glasses
where ultimately the signal is recovered below $T_{g}$, there is
no evidence for such a recovery in SCGO, nor in BSZCGO.

\subsection{Susceptibility of the Kagom\'e bilayer}
\label{SusKagBil}

\begin{figure}[t]
\center
\includegraphics[width=0.7\linewidth]{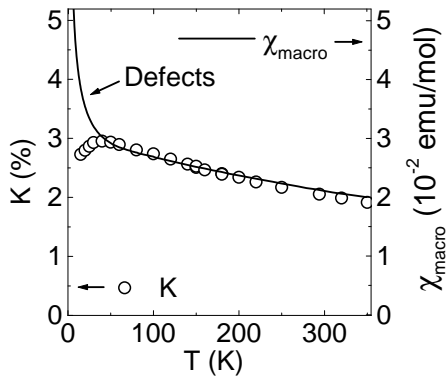}
\caption{$K$ (left) and $\chi _{\text{macro}}$ (right) for
SCGO(0.95). The NMR shift and the SQUID susceptibility do not
follow exactly the same law at high temperature, since $\chi
_{\text{macro}}$ also probes the susceptibility of the Cr(c)-Cr(c)
spin pairs of SCGO (a detailed analysis can
be found in Fig.~13 of Ref.~\onlinecite{Limot02}).}%
\label{KvsChi}%
\end{figure}

\begin{figure}[tbp!] \center
\includegraphics[width=1\linewidth]{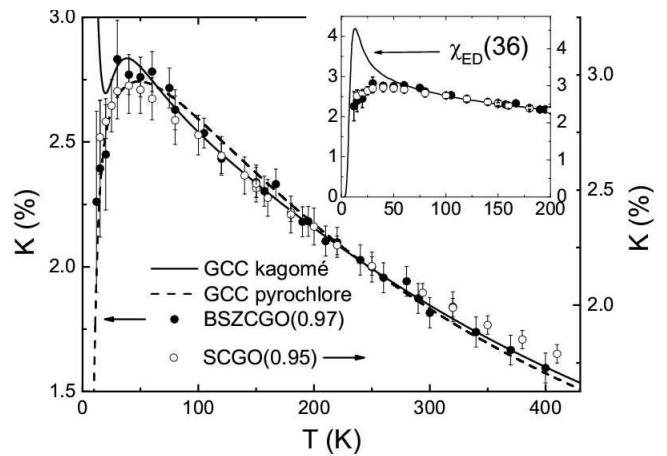}
      \caption{ \label{shift} Shift of the $^{71}$Ga(1) line in SCGO and
      BSZCGO. Lines are GCC calculations (see text).
      Inset: Full scale down to 0. The line is the 36 spin $S=\frac{1}{2}$
      cluster calculation mentioned in the text ($\chi_\text{ED}$).}
\end{figure}

The Kagom\'e bilayer susceptibility $\chi _{\text{Kag}}$ is probed
through the shift $K$ of the $^{69,71}$Ga(1) NMR lines
\cite{Mendels00,Limot02,BonoRMN}, presented in Fig.~\ref{shift}
for the purest samples of both systems in a temperature range
displaying no significant loss of intensity ($T\gtrsim 10$ K). The
shift (and $\chi _{\text{Kag}}$) increases when the temperature
decreases up to $45$~K, where a maximum is reached, and then
decreases again as the temperature is lowered. This behavior is
common to both systems and is field- and dilution-independent
\cite{Mendels00}. Most importantly, it is quite clear from
Fig.~\ref{KvsChi} that there is a discrepancy between the
low-temperature behavior of $\chi _{\text{Kag}}$ and  of $\chi
_{\text{macro}}$. Instead of a maximum, the susceptibility probed
by SQUID measurements follows a Curie-like law at low temperature.
This establishes that $\chi _{\text{macro}}$ is a two-component
susceptibility \footnote{In the case of SCGO, $\chi
_{\text{macro}}$ is a three-component susceptibility, the third
contribution coming from the Cr(c)-Cr(c) spin pairs, which we drop
here for clarity.}, resulting from the sum of the Kagom\'e
bilayer's susceptibility and, as evidenced in Sec.~\ref{Defects},
of a susceptibility related to magnetic defects ($\chi
_{\text{def}}$):
\begin{equation}
\label{Chimacro}
\chi _{\text{macro}}=\chi _{\text{Kag}}+\chi _{\text{def}} \ .
\end{equation}

Concerning the high-temperature ($T\geq80$~K) behavior of $K$, a
phenomenological Curie-Weiss law, $K=C _{\text{NMR}}/(T+\theta
_{\text{NMR}})$, is an accurate fit, yielding $\theta
_{\text{NMR}}=440\pm5$~K and $380\pm10$~K, respectively for
SCGO(0.95) and BSZCGO(0.97). The ``Curie-Weiss'' constant $C
_{\text{NMR}}$ found from the fit is 20\% larger in SCGO than in
BSZCGO, which indicates that hyperfine couplings are stronger in
SCGO, likely because of the shorter Ga-O-Cr bonds
(Fig.~\ref{cell}). Although very commonly employed to extract
information from the HFMs susceptibility, this phenomenological
law is a rather crude approximation, since the linear behavior of
the inverse of the susceptibility is expected to hold, within a
mean field approach, only when $T\gtrsim 2\theta _{\text{NMR}}$,
which is not the case here. From $\theta _{\text{NMR}}$, one can
therefore only grossly estimate the coupling constants $J$ of the
Kagom\'e bilayers. In the absence of any prediction for
$S=\frac{3}{2}$, we used the high temperature series expansion of
$\chi _{\text{Kag}}$, derived for a $S=\frac{1}{2}$ Kagom\'e
lattice, to correct $\theta _{\text{NMR}}$ by a factor 1.5
\cite{Harris92}. The coupling constant is finally extracted from
the mean-field relation $\theta _{\text{NMR}}=1.5zS(S+1)J/3$. The
coordinance is $z=5.14$ and corresponds to the average number of
nearest neighbors for a chromium ion of the bilayer. The couplings
then read $J=45$~K and 40~K for SCGO and BSZCGO, respectively.
These values are consistent with the couplings observed in other
chromium-based compounds \cite{Motida70,Samuelsen70}, and they
indicate that the Cr-Cr AFM interaction stems from the direct
overlap between orbitals of neighboring chromium ions, rather than
from an oxygen-mediated superexchange coupling. A direct
interaction results in a coupling $J$ very sensitive to the Cr-Cr
distance ($d$), and follows a phenomenological law of
\cite{Limot02}
\begin{equation}
\label{eqcoupling}
 \delta J /\delta d=450~\text{K/\AA} \ .
\end{equation}
The slightly stronger coupling found in SCGO compared to BSZCGO
results then, following this viewpoint, from the shorter Cr-Cr
bond lengths of its Kagom\'e bilayer (Fig.~\ref{cell}).

We now turn to the low-temperature behavior of $\chi
_{\text{Kag}}$, and in particular we discuss the maximum around
45~K. The first possible interpretation of this maximum is the
existence of a spin gap. Such a gap $\Delta$ is an important issue
for the Kagom\'e bilayer, since it would be the signature of the
existence of a singlet ground state, found in the theoretical
quantum description of a $S=\frac{1}{2}$ Kagom\'e layer
\cite{Waldtmann98}. Its value is predicted to be $\Delta=J/20$,
and should be of the same order for higher spins
\footnote{C.~Lhuillier, Private communication.}. Since Ga NMR
cannot access lower temperatures than 10~K, which corresponds for
SCGO and BSZCGO to $J/4$, we cannot conclude whether the maximum
in temperature is related to a gap. From an experimental point of
view, its observation would require temperatures $T<\Delta$ in
order to observe the exponential decrease predicted for the
susceptibility \cite{Sindzingre00}. In the inset of
Fig.~\ref{shift}, we compare our experimental data to the
susceptibility computed with exact diagonalization on a 36 spin
$S=\frac{1}{2}$ Kagom\'e cluster \cite{Lhuillier01}. We see that
the position of the predicted maximum does not match the
experimental one. Also, the sharp peak of the susceptibility of
the calculation is not seen in $\chi _{\text{Kag}}$. Clearly,
larger cluster sizes and the Kagom\'e bilayer geometry are needed
for a better comparison to the NMR data. For the time being, it
may be safely concluded, that the behavior of the NMR shift, hence
of $\chi _{\text{Kag}}$, is only consistent with a spin gap
smaller than $J/10$.

Rather than the signature of a gap, the maximum in $\chi
_{\text{Kag}}$ was assigned in Ref.~\onlinecite{Mendels00} to the
signature of a moderate increase of the spin-spin correlations of
the Kagom\'e bilayer which, however, must remain short ranged
given the absence of any phase transition. This conclusion was
confirmed by neutron measurements on SCGO for temperatures ranging
from 200 K down to 1.5 K \cite{Mondelli99B}, and by susceptibility
calculations performed on the Kagom\'e and the pyrochlore lattices
with Heisenberg spins, in which the spin-spin correlation length
is kept of the order of the lattice parameter \cite{Garcia01}.
These calculations, using the so-called ``generalized constant
coupling'' method (GCC) \cite{Garcia01}, consists in computing the
susceptibility of isolated spin-clusters with zero-magnetic
moments (triangles for the Kagom\'e and tetrahedras for the
pyrochlore) and in coupling these clusters following a mean field
approach. In the case of the pyrochlore lattice, an exponential
decrease of the susceptibility is obtained at low temperature. In
this case, the ground state of a cluster is non magnetic and the
spin gap is between the $S = 0$ ground state and the magnetic $S =
1$ excited states. In the case of the Kagom\'e lattice, the ground
state of the triangle is magnetic ($S_{\text{total}}=\frac{1}{2}$)
which gives rise to the non-physical divergence in the
susceptibility as $T \rightarrow 0$, due to the choice of a
particular cluster \cite{Garcia01}. The GCC simulations for the
$S=\frac{3}{2}$ Kagom\'e and pyrochlore lattices, in between which
the Kagom\'e bilayer's susceptibility is expected to lie, are
presented in Fig.~\ref{shift}. As shown, they agree with the data
down to $T=40~\text{K}\ll \theta _{\text{NMR}}$, yielding a
maximum and a behavior around the maximum in agreement with the
experimental findings. From the simulations, the values of
$\theta_{CW}$ are $260$~K for SCGO(0.95) and $235$~K for
BSZCGO(0.97), which correspond to $J=40$~K and $J=37$~K, close to
the previous values obtained through the high-temperature
Curie-Weiss law.

In conclusion, the GCC computation fits quite well the data down
to $T\sim J\ll \theta_{CW}$, where $J\approx40$~K. It only shows
that the strongest underlying assumption of this cluster
mean-field approach is relevant, i.e., the spin-spin correlation
length is of the order of the lattice parameter, which prevents
any magnetic transition to a long range ordered state.

\subsection{Defect-related susceptibility}
\label{Defects}

In this section, we focus on the defects contribution to the
susceptibility. The NMR linewidth and the SQUID data at low
temperature show that the paramagnetic susceptibility $\chi
_{\text{def}}$ observed in $\chi
 _{\text{macro}}$ comes from the dilution of the Kagom\'{e}
bilayer and from other kinds of defects, namely the Cr(c) pairs in
SCGO and bond defects in BSZCGO. The Ga(1) NMR width not only
enables to evidence that the vacancy of a spin on the network,
i.e. the dilution, generates a paramagnetic defect, but also
allows to shed light on the more fundamental question concerning
the extended nature of the defects. We first present the
experimental facts concerning the magnetic defects, as measured
quantitatively through the NMR linewidth and the SQUID
susceptibility $\chi _{\text{macro}}$ and then elaborate on their
nature in both compounds.

\begin{figure}[tbp!] \center
    \includegraphics[width=0.65\linewidth]{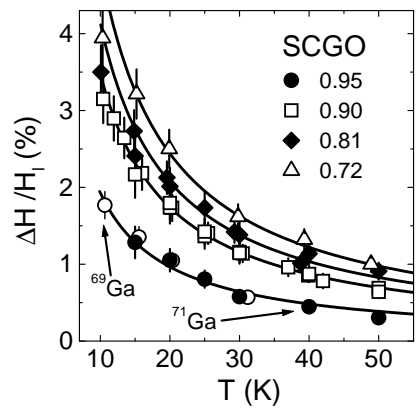}
    \caption{Ga(1) NMR width versus temperature for SCGO samples of
    different Cr-concentration. The width $\Delta H$ is normalized by
    the reference field $H_{l}=\nu _{l}/^{69,71}\gamma$ to superimpose
    the results from the two gallium isotopes (as explicitly shown for
    $p=0.95$). The solid lines are $\propto 1/T$ Curie-like fits.}
\label{largSCGOdil}
\end{figure}

\begin{figure}[tbp!] \center
\includegraphics[width=1\linewidth]{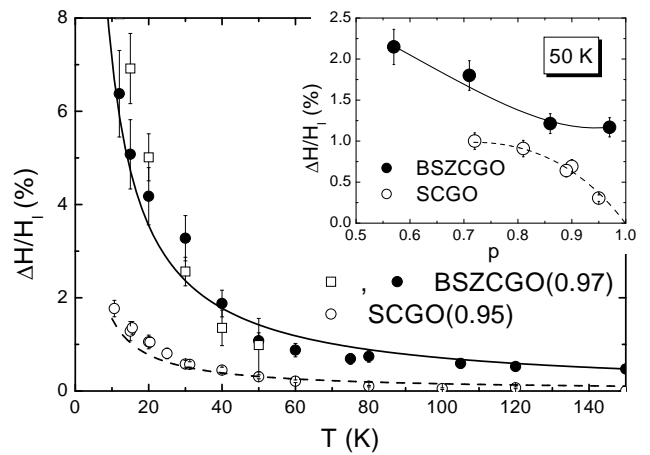}
      \caption{ \label{larg} Magnetic contribution to the NMR
      linewidth $\Delta H/H_{l}$. For BSZCGO(0.97), the data
      for the $^{71}$Ga(2) ($\square$) are rescaled by a
      factor 6, corresponding to the ratio of the coupling
      constants, deduced from the high temperature shifts.
      $\bullet$ are for $\Delta H/H_{l}[^{71}$Ga(1)].
      The lines are $\propto1/T$ Curie-like fits.
      Inset: $p$-dependence of $\Delta H^{71}[^{71}$Ga(1)] at 50~K for SCGO$(p)$ and BSZCGO$(p)$.
      The lines are guides to the eye.}
\end{figure}

\subsubsection{NMR linewidth}
\label{DefData}

\begin{figure}[tbp!] \center
    \includegraphics[width=0.65\linewidth]{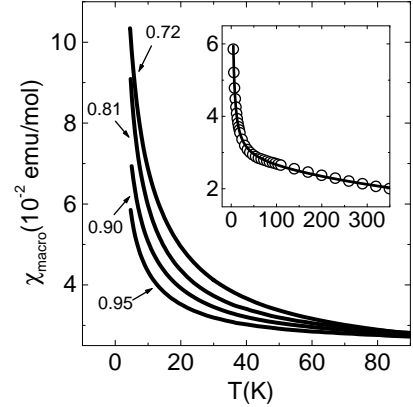}
    \caption{Low-temperature macroscopic susceptibility (SQUID)
for some SCGO Cr-concentrations. Inset: SCGO($0.95$) data (open circles) fitted with Eq.~\ref{fitSQUID}.}
\label{SQUIDSCGOdil}
\end{figure}

The Ga(1) NMR linewidth of SCGO and of BSZCGO was measured for
various Cr-concentrations $p$. Figure~\ref{largSCGOdil} presents
the typical $p$-dependence of the linewidth in both samples ---~in
the figure, only for SCGO. In agreement with the qualitative
presentation of the raw spectra in Sec.~\ref{Ga NMR}, the width
increases as temperature drops and is well described by $1/T$
Curie law (lines in Fig.~\ref{largSCGOdil}). It is very sensitive
to the lattice dilution, similarly to the low-temperature behavior
of $\chi  _{\text{macro}}$ (Fig.~\ref{SQUIDSCGOdil}), but differs
from the NMR shift. The perfect scaling of the widths of the two
$^{69,71}$Ga(1) isotopes (e.g. SCGO(0.95) in
Fig.~\ref{largSCGOdil}) underlines the magnetic origin of the
low-temperature broadening.

Figure~\ref{larg} presents the linewidth obtained for SCGO(0.95)
and for a comparable dilution in BSZCGO(0.97). Surprisingly, the
Curie upturn in BSZCGO(0.97) is four times larger than in
SCGO(0.95). This ratio cannot be explained by a higher hyperfine
coupling constant for BSZCGO, since, as mentioned in
Sec.~\ref{SusKagBil}, the constants $C _{\text{NMR}}$ obtained
from the high-temperature Curie-Weiss analysis of the shift in
SCGO and BSZCGO point to the opposite variation. Further insight
concerning this mismatch between the two HFMs can be gained by
plotting the widths as a function of dilution at a given
temperature (inset of Fig.~\ref{larg}). In SCGO the width
extrapolates to 0 when $p=1$, which shows that the width is only
related to dilution. On the contrary, in BSZCGO the width reaches
an asymptotic non-zero value for $0.86\leq p<1 $. As a first
assumption, one could wonder whether the dilution of the lattice
is larger than the nominal one. This can be ruled out since
(i)~the expected evolution of the line shape with $p$ is found at
300~K (see the following) (ii)~muon spin relaxation measurements
indicate a regular evolution of the dynamical properties with $p$
(Sec.~\ref{MuSR}). Therefore, the increased low-temperature upturn
in BSZCGO comes from dilution-independent paramagnetic defects,
which are not present in SCGO.

\begin{figure}[tbp!] \center
\includegraphics[width=1\linewidth]{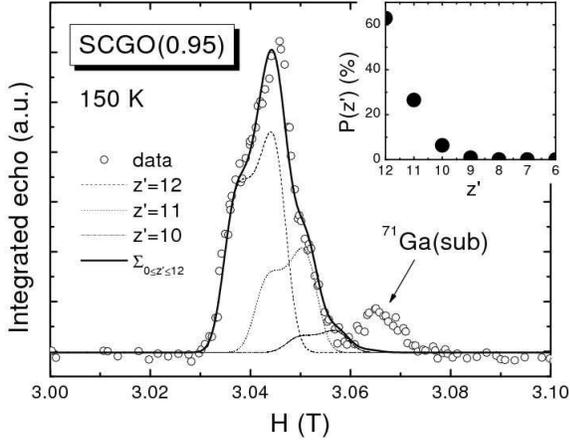}
      \caption{ \label{proba} $^{71}$Ga(1) spectrum at 150~K in
      SCGO(0.95) ($\nu_{l}\approx40$~MHz).
      The lines are quadrupolar powder pattern simulations
      with the parameters of Tab.~\ref{ParamQuad},
      shifted with a value proportional to $z^{\prime}$,
      the number of NN Cr$^{3+}$ for a site.
      Their area is weighted by the probability $P(z^{\prime})$
      when $p=0.95$, presented in the inset.}
\end{figure}

We may already rule out possible scenarios which could be
responsible for the low-temperature broadening. For instance, the
broadening generated by the suppression of chromium ions in the
nuclear environment. Such a distribution is present in the Ga(1)
NMR spectra of SCGO and of BSZCGO, but is only responsible for a
minor broadening of the line as shown in Fig.~\ref{proba} for
SCGO(0.95). This broadening mechanism results from the fact that
each Ga(1) nucleus has a probability $P(z^{\prime})$ (inset of
Fig.~\ref{proba}) to have $z^{\prime}$ chromium neighbors ($0\leq
z^{\prime}\leq 12$). Assuming that the Ga/Cr substitution does not
affect the Ga-O-Cr hyperfine couplings and that all the chromium
ions have the same susceptibility, the shift of a Ga(1) nucleus
surrounded by $z^{\prime}$ neighboring chromium ions is then
proportional to $z^{\prime}$, according to Eq.~\ref{eqshift}. To
construct the spectrum, each gallium nucleus is associated to a
quadrupole line simulated with the parameters of
Tab.~\ref{ParamQuad}, with a shift reflecting its Cr-environment
and an intensity weighted by the total number of gallium nuclei
having the same environment, i.e. weighted by $P(z^{\prime})$. The
simulated line perfectly matches the experimental Ga(1) NMR line
at 150~K (Fig.~\ref{proba}). However, the broadening resulting
from this distribution scales with the susceptibility, hence with
the shift $K$. Since $K$ decreases at low temperature, this
broadening mechanism cannot explain the low-temperature upturn of
the width. For the same reason, a spatial distribution of the
hyperfine constant, which yields also a width $\propto K$, cannot
justify the broadening observed and also has to be ruled out
\footnote{A dipolar broadening due to the diluted paramagnetic
defects \cite{Walstedt74} would be to small to justify the
broadening of the $^{69,71}$Ga spectra.}. Even if a such refined
analysis is not possible in BSZCGO because of a stronger
broadening than SCGO, similar conclusions are derived for this
compound \cite{BonoHFM}. Hence, the explanation for the
low-temperature broadening must be searched elsewhere (see
Sec.~\ref{DefPara}).

\subsubsection{SQUID}

The SQUID measurements were carried out over a wide range of
Cr-concentrations ($0.29\leq p \leq0.97$) for temperatures down to
$1.8$~K (no difference was observed between the Field Cooled and
Zero Field Cooled susceptibility above the freezing temperature
$T_{g}$ in a field of 100~G). Figure~\ref{SQUIDSCGOdil} presents
the typical low-temperature macroscopic susceptibility of these
HFMs. Like the NMR linewidth, the susceptibility exhibits a
low-temperature upturn, increasing with growing dilution. In
Sec.~\ref{SusKagBil}, by a comparison between the NMR shift and
the macroscopic susceptibility, we established that $\chi
_{\text{macro}}$ must be a two-component susceptibility, and, in
particular, that it should possess, compared to the shift, a
paramagnetic susceptibility $\chi _{\text{def}}$ to explain its
low-temperature upturn. The dilution-dependent upturn of $\chi
_{\text{macro}}$ in Fig.~\ref{SQUIDSCGOdil} therefore establishes
that $\chi _{\text{def}}$ is related to the Ga/Cr substitution in
SCGO$(p)$. However, we also know from NMR that there also extra
paramagnetic contributions, related to intrinsic defects in BSZCGO
for example.

\begin{figure}[tbp!] \center
\includegraphics[width=1\linewidth]{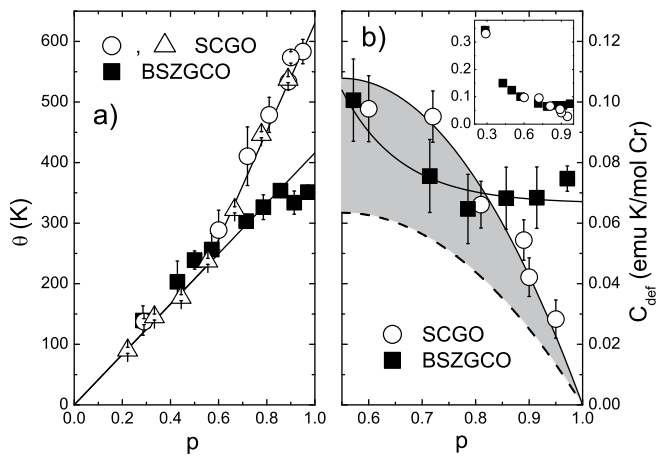}
      \caption{ \label{SQUIDfits} Fits of the SQUID data (Eq.~\ref{fitSQUID}).
      a)~$\triangle$ come from Ref.~\onlinecite{Martinez92}.
      The lines are guide to the eye.
      b)~The dashed line represents the Kagom\'e bilayers defects
      contribution to $C_{\text{def}}(p)$ in SCGO$(p)$, the other lines are guide to the eye.
      The grey area represent the Cr(c) contribution in $C_{\text{def}}(p)$ for SCGO$(p)$ (see text).
      Inset: same data as (a) for a broader range of $p$. }
\end{figure}

In order to quantify the low-temperature dependence of the
susceptibility, we follow Ref.~\onlinecite{Schiffer97} and fit
$\chi _{\text{macro}}$ by a two component expression:
\begin{equation}
\label{fitSQUID}
\chi _{\text{macro}}=\frac{C}{T+\theta}+\frac{C _{\text{def}}}{T+\theta _{\text{def}}} \ .
\end{equation}
The first Curie-Weiss term roughly takes into account the
contribution from the Kagom\'e bilayer and governs the behavior of
$\chi _{\text{macro}}$ at high temperature. The second term, the
more relevant for this section, also \emph{a priori} Curie-Weiss,
quantifies the contribution of $\chi _{\text{def}}$. A more
accurate fit of $\chi _{\text{macro}}$ would consist in using a
low-temperature susceptibility mimicking the NMR shift. Such a
fit, however, yields the same qualitative results as presented
below.

From these fits (e.g. inset of Fig.~\ref{SQUIDSCGOdil}), an
effective moment per Cr$^{3+}$ is extracted,
$p_{eff}=4.1\pm0.2/\mu_{B}$ in both SCGO \cite{Limot02} and
BSZCGO, close to the 3.87$/\mu_{B}$ expected for $S=\frac{3}{2}$
spins. In BSZCGO, the Curie-Weiss temperature is found to increase
with $p$ (Fig.~\ref{SQUIDfits}a), in agreement with mean-field
theory. From the linear variation, following the procedure
described in Sec.~\ref{SusKagBil}, we extract $J=40$~K
\cite{BonoRMN}, consistent with our NMR results. However, two
linear regimes of $\theta(p)$ are observed for SCGO$(p)$, below
and above $p\sim0.55$. This was first reported in
Ref.~\onlinecite{Martinez92} and cannot be understood in terms of
simple mean field theory. Taking into account a third term in
Eq.~\ref{fitSQUID}, corresponding to the susceptibility of the
Cr(c) pairs in SCGO, derived in Ref.~\cite{Limot02} and
vanishingly small below 50~K, does not affect this result.

The second term in Eq.~\ref{fitSQUID} is introduced to fit the
low-temperature behavior of the defects. For all samples studied,
it is close to a pure Curie law, within error bars. The values of
the Curie constants $C _{\text{def}}$ extracted are presented in
Fig.~\ref{SQUIDfits}b. A $p$-dependence qualitatively similar to
the NMR width is found and $C _{\text{def}}$ extrapolates again to
0 for $p=1$ in SCGO, whereas it displays a finite value in BSZCGO.
When the Cr-concentration is $p\le0.8$, the dependency of $C
_{\text{def}}$ is qualitatively the same in both systems (inset of
Fig.~\ref{SQUIDfits}b). To give an order of magnitude, this term
represents, for the purest samples, a number of $S=\frac{1}{2}$
paramagnetic spins equal to the number of spin vacancies in
SCGO$(p)$ \cite{Limot02} and to 15-20\% of the number of Cr$^{3+}$
spins in BSZCGO$(p)$ \cite{BonoRMN}.

\subsubsection{Paramagnetic defects: discussion}
\label{DefPara}

The NMR and the macroscopic susceptibility measurements indicate
that there are paramagnetic defects in SCGO and BSZCGO. Some of
them are related to the spin vacancies in the Kagom\'e bilayers
and affect the neighboring chromium ions in such a way that their
overall behavior is paramagnetic. These defects therefore yield
information on how the correlated spin network of the Kagom\'e
bilayer responds to the presence of the vacancy. In the following,
we discuss this point as well as the presence of other types of
defects in both compounds.

The second type of defects result from lattice imperfections in
BSZCGO, and from the dilution of the spin-pair sites in SCGO.
Unfortunately, in the case of BSZCGO the second type of defect
dominate the paramagnetic response, and in this compound, on the
contrary of SCGO, a fine analysis of the spin-vacancy defects
cannot be carried out, even by Ga(1) NMR.

\paragraph{Spin vacancies.}

In AFM systems such as the 1D spin chains
\cite{Takigawa97,Tedoldi99} the quasi-2D spin ladders \cite
{Azuma94} and the 2D cuprates \cite{Ouazi04} it is now well
established that a vacancy (or a magnetic impurity) generates a
long range oscillating magnetic perturbation and creates a
paramagnetic-like, i.e. not necessarily $\propto 1/T$, component
in the macroscopic susceptibility. The symmetric broadening of the
NMR line observed in these systems is related to the oscillating
character of the perturbation. The dilution effects of the
Kagom\'{e} bilayer would then be described by the same physics of
these correlated systems.

In SCGO, the symmetric feature of the Ga(1) NMR line indicates
that spin vacancies induce a perturbation extended in space. This
rules out descriptions of the defect in terms of uncorrelated
paramagnetic centers \cite{Schiffer97,Moessner99}, since the
broadening of the line would then be asymmetric. Hence, given the
AFM interactions of the Kagom\'e bilayer, the perturbation was
assigned to a staggered polarization of the network
\cite{Limot02}. The exact diagonalization of 36 spin $\frac{1}{2}$
clusters on the Kagom\'e lattice with spin vacancies by Dommange
\emph{et al.} confirms this interpretation \cite{Dommange03}. They
show that within an RVB ground state, the spin-spin correlations
around a vacancy are enhanced, and that the magnetization is
staggered around it, a picture that should also hold for higher
spin values. However, it seems that the image of a simple envelope
of the staggered magnetization, like the Lorentzian now granted in
High $T_{\text{c}}$ cuprates \cite{Ouazi04}, is not correct here
since this localization of singlets around the vacancies shifts
the staggered cloud from the non-magnetic impurity. Although the
agreement is qualitative at the moment, further work is required
within this framework for a quantitative modelling of the NMR
lineshape.

\paragraph{Bond disorder in BSZCGO.}

We previously saw that in BSZCGO$(p)$, low temperature macroscopic
susceptibility and NMR width data can be both satisfactorily
accounted for, only if a novel $p$-independent defect-like
contribution is taken into consideration, a result somehow
surprising in view of the close similarity between the Kagom\'e
bilayers of SCGO and BZSCCGO. The only major change lies in the
1:1 random occupancy of the Ga(1) site by Ga$^{3+}$ or Zn$^{2+}$
ions which induces different electrostatic interactions with the
neighboring ions. As an example, distances to O$^{2-}$ in a
tetrahedral environment vary from $r_{Ga^{3+}-O^{2-}}=0.47$~\AA\
to $r_{Zn^{2+}-O^{2-}}=0.60$~\AA\ which is at the origin of a
change of the average Ga(1)-O bonds, from 1.871~\AA\ for
SCGO~\cite{Lee96} to 1.925~\AA\ in BSZCGO~\cite{Hagemann01}.
Similarly, one expects that the Cr$^{3+}$ will be less repelled by
Zn$^{2+}$ than by Ga$^{3+}$, which certainly induces magnetic
bond-disorder, i.e. a modulation of exchange interactions $J$
between neighboring Cr$^{3+}$. From Eq.~\ref{eqcoupling}, one can
estimate that a modulation as low as $0.01$~\AA\ in the
Cr$^{3+}$-Cr$^{3+}$ distances would yield a $10$\% modulation of
$J$.

The presence of non-perfect equilateral triangles classically
induces a paramagnetic component in the susceptibility of the
frustrated units \cite{Moessner99}. The fact that we do not
observe any extra Curie variation as compared to SCGO$(p)$ in the
average susceptibility, probed through the NMR shift $K$
(Fig.~\ref{shift}), indicates that such unbalanced exchange
interactions only induce a staggered response in the same manner
as spin vacancies. Whether this could be connected with the
localization of singlets in the vicinity of defects and a
staggered cloud around them, like for spin vacancies, should be
more deeply explored. This might highlight the relevance of a RVB
approach to the physics of the Kagom\'e network. Further insight
into the exact nature of the defects likely requires a better
determination of the bond disorder through structural studies at
low $T$, to avoid the usual thermal fluctuations at room $T$.

\paragraph{Broken spin-pairs in SCGO.}

The Ga/Cr substitution on a Cr(c) site of SCGO could in principle
break a Cr(c)-Cr(c) spin pair and free a paramagnetic spin, which
then could contribute to the paramagnetic upturn of
$\chi_{\text{macro}}$. However, this contribution turns out to be
small. That the broken spin-pair susceptibility is negligible, can
be deduced by comparing NMR and SQUID measurements for SCGO and
BZSCGO. The starting point is to notice that the crossing between
the variation of the curie constants $C_\text{def}$ of both
samples determined by the fits of Eq.~\ref{fitSQUID} to the
macroscopic susceptibility (Fig.~\ref{SQUIDfits}b), is not
observed in the NMR width (inset of Fig.~\ref{larg}). Considering
from our NMR analysis that there is a 20\% difference between the
hyperfine coupling, we can evaluate that the scaling factor
between $\Delta H/H_{l}(p)$ and $C_{\text{def}}(p)$ should be 20\%
larger in BSZCGO$(p)$ than in SCGO$(p)$. An error bar of 10\%
comes from the possible difference between the bilayer
susceptibility measured with NMR and SQUID, since the Ga(1) nuclei
do not probe stoichiometrically the magnetic sites Cr(b) and Cr(a)
\footnote{Ga(1) nuclei are coupled to 9 Cr(b) and 3 Cr(a). The
measured susceptibility is hence $\chi_\text{{Kag,NMR}}\propto 9
\chi_{b} + 3 \chi_{a}$, where $\chi_{b}$ and $\chi_{a}$ are the
Cr(b) and Cr(a) susceptibilities. On the other hand, the
macroscopic susceptibility probes these susceptibilities with
their stoichiometric ratio, i.e., $\chi_\text{{Kag,macro}}\propto
6 \chi_{b} + \chi_{a}$ \cite{Limot01}. A small difference between
the ratios $\chi_{b}/\chi_{a}$ would yield a slightly different
ratio $[\Delta H/H_{l}(p)]/C_{\text{def}}(p)$.}. We first
determine the scaling factor $[\Delta
H/H_{l}(p)]/C_{\text{def}}(p)$ in BSZCGO$(p)$, where 100\% of the
magnetic ions belong to the Kagom\'e bilayers. With the same
ratio, the $\Delta H/H_{l}(p)$ data for SCGO$(p)$ yield therefore
a lower $C_{\text{def}}(p)$ than the measured one, out of the
error bars [dashed line in Fig.~\ref{SQUIDfits}b]. This difference
is likely due to the Cr(c) pairs dilution in SCGO$(p)$. Their
contribution which, is not probed with the Ga(1) NMR [grey area in
Fig.~\ref{SQUIDfits}b], corresponds to a $0.04p$ Cr$^{3+}$
$S=\frac{3}{2}$ paramagnetic term, whereas a $2p/9$ proportion
would be expected in the case of a stoichiometrically substituted
sample. This substoichiometric contribution of the Cr(c) site is
consistent with neutron diffraction measurements \cite{Limot02}.


\section{Spin Dynamics and Magnetic Excitations}
\label{MuSR}

\subsection{Muon: an appropriate probe of dynamics}

We mentioned that an abrupt loss of the Ga(1)-NMR intensity occurs
below 10 K (Fig~\ref{NMRintensity}). Hence, Ga(1)-NMR cannot be
employed to probe the low-temperature magnetic properties of SCGO
and of BSZCGO. This is unfortunate, since the low-temperature spin
dynamics reseal, in particular, clue information concerning the
ground state of these systems. On the contrary, due to the smaller
coupling of the muon to the electronic moments and a shorter time
window than the NMR one, the Muon Spin Relaxation ($\mu$SR)
technique has proven to be a front tool for the study of quantum
dynamical states in a vast family of fluctuating systems.

The electronic dynamics is commonly probed, either for NMR or for
$\mu$SR, through the measurements of the longitudinal relaxation
time, necessary to recover the thermodynamic equilibrium after the
excitation of the nuclear spin or the muon spin systems. However,
while the out-of-equilibrium state is reached using \emph{rf}
pulses in NMR, the muon spin system is already in an excited
state. Indeed, muons ($\mu^{+}$, $S=\frac{1}{2}$) are implanted in
the sample, 100\% spin polarized along the $z$ axis. Therefore,
the muon spins always depolarize to reach the Boltzmann
distribution and $\mu$SR can be performed in zero external field,
contrary to conventional NMR \cite{Abragam}. The time dependence
of their \emph{polarization} $P_{z}(t)$ along the $z$ axis is
studied through their decay into a positron \cite{Muons} and is
directly linked to both the spin fluctuations (the time
correlation function of the local field $H_{\mu}$ is usually
exponential, i.e., $\langle
\mathbf{H_{\mu}}(0)\cdot\mathbf{H_{\mu}}(t)\rangle/\langle
H_{\mu}(0)^{2}\rangle=\exp (-\nu t)$) and the (random) local field
distribution, characterized by a width $\Delta/\gamma_{\mu}$
($\gamma_{\mu}$ is the muon gyromagnetic ratio). The longitudinal
relaxation times is related to the spin-correlation function
through $1/T_{1}\sim \int_{0}^{\infty}\langle
\mathbf{S}(0)\cdot\mathbf{S}(t)\rangle \cos(\gamma_{\mu} H_{LF} t)
dt$. Zero field and longitudinal field (LF) $\mu$SR experiments,
where the external field $H_{LF}$ is applied along the $z$ axis,
allow for instance to distinguish magnetic (randomly or ordered)
frozen states from dynamical ones \cite{Hayano79,Uemura85}.
Moreover, whereas specific heat is sensitive to all kinds of
excitations, including low-lying singlets at low temperature in
the Kagom\'e bilayer samples \cite{Ramirez00}, the muons probe
\emph{magnetic} excitations \emph{only}, with a specific range of
frequencies ($\sim10^{9}$~Hz) sitting in a typical time window
(10~ns-10~$\mu$s) in between NMR and neutron experiments.

We present here our $\mu$SR study of the spin dynamics in
SCGO$(p)$ and BSZCGO$(p)$ \cite{BonoHFM,BonoMuSR}. We first show
that while conventional $\mu$SR polarization functions can be used
in the whole temperature range for the very diluted samples and in
the high temperature regime ($T\gtrsim 3T_{g}$) for the purest
ones, they cannot be used satisfactorily in the low temperature
regime in the latter case ($p\gtrsim 0.7$). Moreover, in this
first step, we qualitatively show that the magnetic state of all
the samples is dynamical down to the experimental limit of 30~mK
(Sec.~\ref{Conventional}). We further use a model independent
analysis of the data, simply taking the time necessary for the
muon spin polarization to decrease down to the value $1/e$. The
comparison between SCGO and BSZCGO clearly shows a correlation
between the muon spin relaxation rate and $T_{g}$, suggesting that
this freezing temperature is not an impurity phase. On the
contrary, it seems to be closely linked to the slowing down of the
spin dynamics in the bulk sample (Sec.~\ref{Lambda}). Finally, a
phenomenological model for the muon relaxation, based on sporadic
dynamics due to spin excitations in a singlet sea, proposed by
Uemura \emph{et al.} \cite{Uemura94}, is extended to all fields-
and temperature- range. Its connection to the RVB picture is
discussed, and we argue that such coherent states might mediate
the interactions between ''impurities'', which induce the spin
glass freezing (Sec.~\ref{Spinons}).

\subsection{Conventional approaches}
\label{Conventional}

\subsubsection{High-temperature behavior}

\begin{figure}[tbp!] \center
\includegraphics[width=0.8\linewidth]{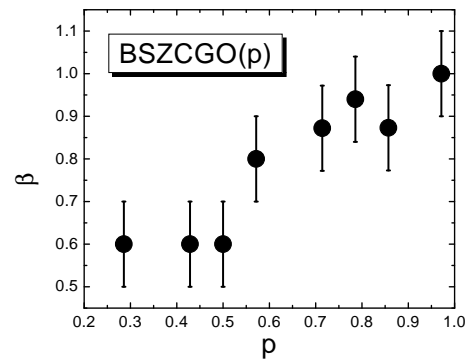}
      \caption{ \label{MuSRbeta} Exponent $\beta(p)$ in
      BSZCGO$(p)$ using the relaxation function $P_{z}(t)=exp[-(\lambda t)^{\beta}]$
      appropriate for the fast fluctuations limit at high temperature ($T\geq 10~$K.). }
      \end{figure}

At high temperature, a conventional paramagnetic behavior is found
for all the samples, with a stretched exponential variation of
$P_{z}(t)=\exp[-(\lambda t)^{\beta}]$. Figure~\ref{MuSRbeta} shows
that as expected, $\beta\rightarrow1$ in the dense magnetic cases
\cite{Hayano79} (in our case with a high coverage of the Kagom\'e
bilayer lattice with Cr$^{3+}$) and $\beta\rightarrow0.5$ in the
dilute cases \cite{Uemura85} (i.e. with a low coverage).

We can give here an estimate of the fluctuation frequency in this
temperature range and of the NMR time window which would be
required to measure this spin dynamics. At $\sim50$~K, the muon
relaxation rate is $\lambda\sim0.03~\mu$s$^{-1}$ and
$0.01~\mu$s$^{-1}$ in SCGO(0.95) and BSZCGO(0.97). In this
appropriate fast fluctuating paramagnetic limit, the fluctuation
rate can be estimated from $\nu\sim\sqrt{z}JS /k_{B}\hbar
\sim2\times 10^{13}$~s$^{-1}$ \cite{Uemura94} using the coupling
$J\approx40$~K determined previously by NMR, and the average
number $z=5.14$ of Cr nearest neighbors. We extract
$\Delta=\sqrt{\lambda\nu/2}\sim600~\mu$s$^{-1}$ and
$300~\mu$s$^{-1}$, which corresponds to an average field at the
muon sites of 7000~G and 3500~G for SCGO and BSZSCGO respectively.

Since the relationship between the longitudinal relaxation time
$T_{1}$ measured in NMR and $\mu$SR is $T_{1}^{NMR}/ T_{1}^{\mu
SR}\sim [\Delta/(^{71}\gamma ^{71}A)]^{2} $, where
$^{71}A\sim70~$kOe/$\mu_{B}$ is the hyperfine coupling constant of
the $^{71}$Ga nuclei obtained from NMR experiments, we can
estimate that $T_{1}^{NMR}/ T_{1}^{\mu SR}\sim 5$. This is
consistent with our data [$T_{1}^{NMR}\sim 100~\mu$s at high
temperature and establishes that below 5~K, the relaxation must be
$T_{1}^{\text{NMR}}\sim 1~\mu$s, which is non-measurable as it
lies out of the NMR time window. Such a small relaxation time is
responsible for the wipeout of the Ga-NMR intensity
(Fig.~\ref{NMRintensity}).

      \begin{figure}[tbp!] \center
\includegraphics[width=0.8\linewidth]{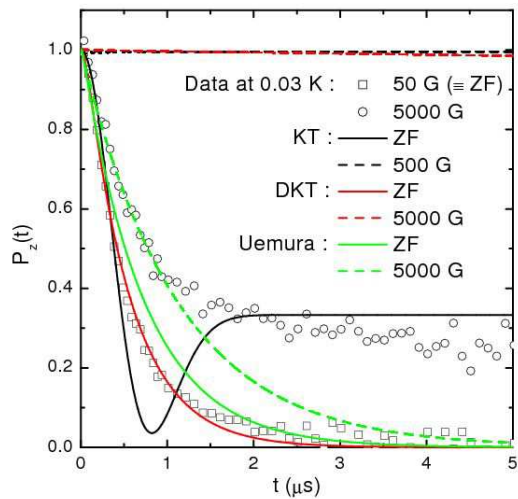}
      \caption{ \label{MuSRfits} BSZCGO(0.97) data at 0.03~K.
      The lines are fits with different models (see text).
      We used $\Delta\approx2~\mu$s$^{-1}$, $4~\mu$s$^{-1}$
      ($\nu\approx17~\mu$s$^{-1}$) and $65~\mu$s$^{-1}$ ($\nu\approx600~\mu$s$^{-1}$)for the static,
      dynamic and ``Uemura'' cases respectively. }
      \end{figure}

\subsubsection{Failure of conventional approaches at low temperature for the purest samples}

The derivation of zero field and longitudinal field $\mu$SR
relaxation functions is presented in
Ref.~\onlinecite{Hayano79,Uemura85} for textbook cases.  We first
rule out the possibility of some static freezing as the source of
the muon relaxation. Two options can be considered. (i)~In
\emph{randomly} frozen (static) magnetic states, the major results
are the following: first, the muon spin polarization displays a
``$\frac{1}{3}$ tail'' in zero external field, i.e,
$P_{z}(t\geq5/\Delta) \rightarrow \frac{1}{3}$. Indeed,
$\frac{1}{3}$ of the frozen magnetic internal fields are
statistically parallel to $z$ and do not contribute to the muon
spin depolarization. Second, $P_{z}(t)$ is ``decoupled'', i.e.,
does not relax, for longitudinal fields $H_{LF}\gtrsim10
\Delta/\gamma_{\mu}$. (ii)~In the case of \emph{ordered} magnetic
states and powder samples, the ``$\frac{1}{3}$ tail'' is still
observed in zero field experiments for the same reasons, but
oscillations appear in $P_{z}(t)$, which correspond to well
defined local fields.

One should notice that nuclear moments which appear static on the
$\mu$SR time window always contribute to the time-evolution of the
polarization. A small longitudinal field of the order of a few
10~G ($\Delta_{ND}\sim0.1~\mu$s$^{-1}$) is commonly applied to
``decouple'' this contribution. In other words, since the
electronic and dipolar contributions are multiplied, such a
contribution, if any, ``disappears'' and only the electronic
contribution, not perturbed by such low fields, is measured.
Hence, experiments performed in $H_{LF}\sim 100$~G are equivalent
to a zero field measurement without nuclear dipoles.

Figure~\ref{MuSRfits} shows $P_{z}(t)$ measured in BSZCGO(0.97) at
0.03~K for $H_{LF}=50$~G and for $H_{LF}=5000$~G. Neither
oscillations nor a $\frac{1}{3}$ tail are observed when
$H_{LF}=50$~G. Although not emphasized in the figure, one can note
that the initial polarization is Gaussian, which is even clearer
in SCGO$(p)$ \cite{Uemura94}. In a \emph{static} case, such a
shape is observed in dense magnetic randomly frozen systems, where
the polarization is given by the Gaussian Kubo-Toyabe (KT)
function \cite{Hayano79}. We fitted our data at short times with
this function in zero field, and plot the expected polarization
when $H_{LF}=500$~G in Fig.~\ref{MuSRfits}, which is found to be
completely decoupled. This contrasts with our experimental
finding, i.e., the polarization at 5000~G is hardly decoupled.
\emph{The lack of the $\frac{1}{3}$ tail, the absence of
oscillations and of decoupling effect show that the magnetic state
of BSZCGO(0.97) is fluctuating at 0.03~K}. All these
considerations remain valid in both SCGO$(p)$ and BSZCGO$(p)$
compounds when $p\gtrsim0.6$.

The computation of the \emph{dynamical} relaxation functions is
based on a strong collision approximation \cite{Hayano79}. Using
the corresponding Dynamical Kubo-Toyabe (DKT) function,
analytically derived by Keren when $\nu\geq\Delta$ \cite{Keren94},
allows us to fit the zero field data on the whole time range
(Fig.~\ref{MuSRfits}). Again, the experimental data is much less
sensitive to the applied field $H_{LF}=5000$~G than the expected
polarization.

This so called ``undecouplable Gaussian line shape'' was first
reported in Ref.~\onlinecite{Uemura94} for SCGO(0.89) and further
in other Kagom\'e compounds \cite{Keren96, Keren00,Fukaya03} or
spin singlet compounds like the doped Haldane chain
Y$_{2-x}$Ca$_{x}$BaNiO$_{5}$ \cite{Kojima95}. Uemura \emph{et al.}
proposed a relaxation model \cite{Uemura94}, presented further in
this paper, which catches some of the facets of this relaxation.
For now, we just notice that this model cannot justify our data
for all fields (Fig.~\ref{MuSRfits}), even if it shows a lower
decoupling effect, consistent with our (and previous)
observations. Before going further with more complex models
(Sec.~\ref {Spinons}), we will first, in the following, give
qualitative arguments to characterize the low temperature
magnetism of our samples.

\begin{figure}[tbp!] \center
\includegraphics[width=1\linewidth]{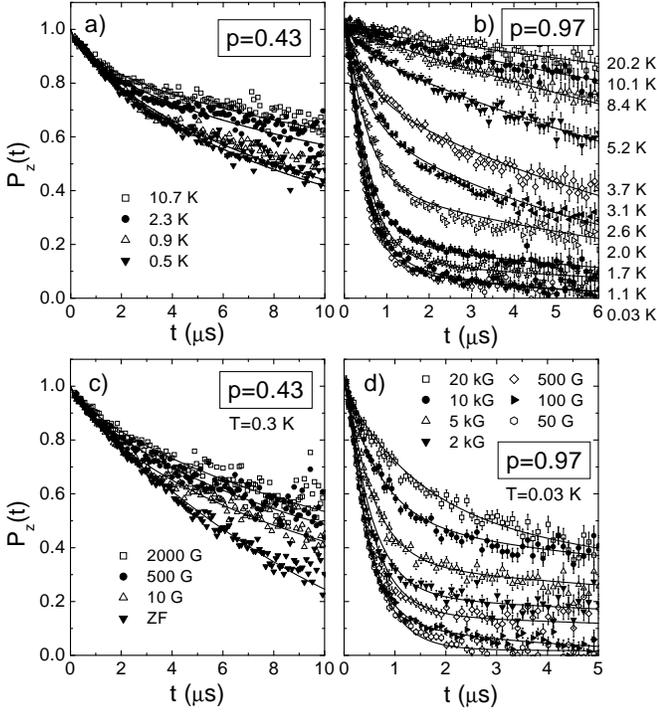}
      \caption{ \label{fig_Asym_T} a,b)~temperature-dependence of the muon
      polarization $P_{z}(t)$ in BSZCGO under weak $H_{LF}$ (10~G for
      $p=0.43$ and 100~G for $p=0.97$).
      Notice the different time scales for both samples.
      c,d)~$H_{LF}$-dependence at base-temperature.
      The line for $p=0.43$, ZF, is a square-root exponential
      times the Kubo-Toyabe function in zero field
      ($\Delta_{ND}\sim0.08~\mu$s$^{-1}$), accounting
      for the nuclear dipole contribution.
      The other lines are fits described in the text.}
      \end{figure}

In order to illustrate the evolution of the properties with $p$,
we present two typical low and high occupations $p$ of the
Kagom\'e bilayers in Fig.~\ref{fig_Asym_T} which emphasize the
qualitative differences for various frustrated network coverage
rates $p$. In addition to the high temperature behavior already
commented, the evolution of $P_{z}(t)$ at low temperature is
markedly different. For $p=0.97$, the relaxation rate increases by
more than two orders of magnitude to reach a
temperature-independent value for $T\lesssim T_{g}\approx1.5$~K
[Fig.~\ref{fig_Asym_T}(b)], with the undecouplable character
presented above. For a low coverage rate, $p=0.43$, we also find a
dynamical state but, at variance with the previous case, only a
weak temperature-dependence is observed. Also, the polarization
displays a square root exponential decay for \emph{all}
temperatures [Fig.~\ref{fig_Asym_T}(a)] and for any longitudinal
field $H_{LF}$ [Fig.~\ref{fig_Asym_T}(c)]. A weak plateau of the
relaxation rate is still observed below $0.5$~K but is no more
present for $p=0.3$, where $P_{z}(t)$ is found weakly temperature-
and $H_{LF}$-dependent. This is typical of the pure paramagnetic
fast fluctuations limit for dilute magnetic systems, which
contrasts with the results of SQUID measurements showing that
strongly frustrated antiferromagnetic interactions are still
present. This is the so called ``cooperative paramagnetism''.

\subsection{Model independent basic analysis}
\label{Lambda}

As a first step to a quantitative analysis, we estimate the muon
spin relaxation rate $\lambda$ using the $\frac{1}{e}$ point of
the polarization $P_{z}(t)$, as discussed in
Ref.~\onlinecite{Keren00}. This allows to single out information
about two important issues using a simple model-independent
analysis: the role of the spin glass-like transition and the
impact of the spin vacancies on the dynamics.

\begin{figure}[bp!] \center
\includegraphics[width=1\linewidth]{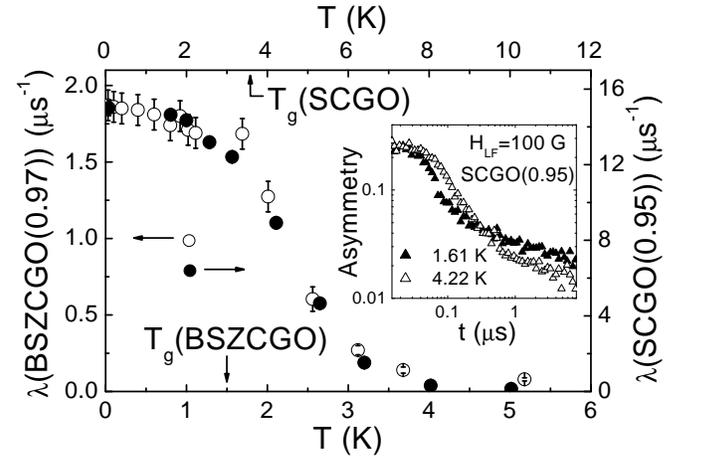}
      \caption{ \label{fig_lambda_scale}  Comparison of the evolution of
      the muon spin relaxation rate $\lambda$ vs. $T$ in BSZCGO(0.97)
      and SCGO(0.95), from a $1/e$ analysis.
      Inset: recovery of a part of the asymmetry at low
      temperature and long times in SCGO(0.95) below $T_{g}$
      (log-log scale).
       }
      \end{figure}

From high temperature, $\lambda$ increases by more than two orders
of magnitude down to $T_{g}$ for both kagome bilayers at their
highest level of purity, to reach a relaxation rate plateau
$\lambda_{T\rightarrow0}$ for $T\lesssim T_{g}$
\cite{Uemura94,Keren00}. Using a temperature-scale twice larger
for SCGO(0.95) than in BSZCGO(0.97), the temperature dependence of
$\lambda$ of both samples perfectly scales on the temperature axis
as shown in Fig.~\ref{fig_lambda_scale} \footnote{We attribute the
factor 8 between $\lambda$ in both systems to a different coupling
of the muon spin to the Cr spins. Transverse field experiments
show for instance a larger linewidth in SCGO(0.95) than in
BSZCGO(0.91) despite a lower defect term \cite{BonoRMN}. On the
other hand, $\Delta$, which is mainly created by dipolar
interactions in $\mu$SR, is $\Delta\propto\sum1/r^{3}$, where the
sum is done over all the spins in the lattice and $r$ is the
distance between the muon site and these spins \cite{Hayano79}.
Usually, the muon is located near an O$^{2-}$ ion \cite{Brewer90}.
In a very simplified approach, we computed $\Delta$, considering
that the muon would be located \emph{on} the O$^{2-}$ sites, as an
average, and find $\Delta\approx1100~\mu$s$^{-1}$ and
$800~\mu$s$^{-1}$ in SCGO(1) and BSZCGO(1). This is in perfect
agreement with the high temperature evaluation of $\Delta$. Since
$\lambda\propto\Delta^{2}$ in the fast fluctuation limit and
$\lambda\propto\Delta$ in the static limit, we expect, with this
simple approach, $\lambda$ to be 1.5-2 times larger in SCGO$(p)$.
The missing factor 4 may either come from a more complex muon
sites distribution, due to the O$^{2-}$ ions around the Cr(c)
sites, or to a different dynamics range as suggested by Neutron
Spin Echo \cite{Mutkaprep}.}. This ratio is very close to the
2.3(2) ratio between their freezing temperatures $T_\text{g}$,
which points to a link between the formation of the spin
glass-like state and the presence of fluctuations. This seems to
be a quite common feature of various systems with a singlet ground
state \cite{Uemura94,Kojima95,Keren96,Keren00,Fukaya03}.

\begin{figure}[tbp!] \center
\includegraphics[width=1\linewidth]{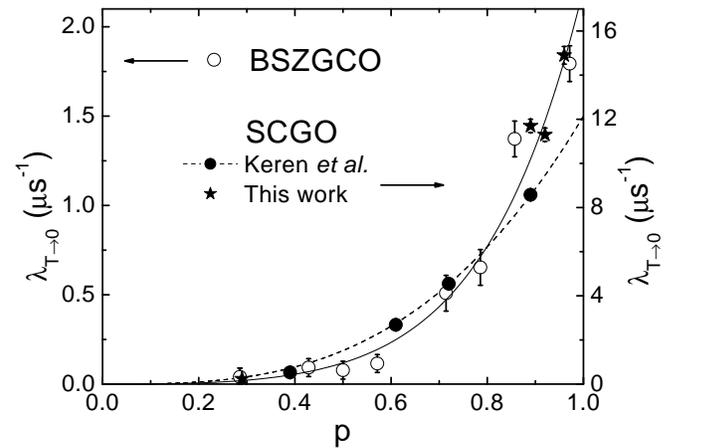}
      \caption{ \label{fig_lambda_p}  $p$-dependence of $\lambda_{T\rightarrow0}$
      for BSZCGO$(p)$ (open symbols) and SCGO$(p)$ (full symbols).
      Continuous (dashed) lines are
      $\sim p^{3}+1.9(3)p^{6}$ ($\propto p^{3}$, \cite{Keren00}) fits.
       }
      \end{figure}

In Fig.~\ref{fig_lambda_p} we report the variation of
$\lambda_{T\rightarrow0}(T)$ for various BSZCGO$(p)$ samples and
compare them to SCGO$(p)$ \cite{Uemura94,Keren00}. Although
additional $p$-independent defects are dominant in BSZCGO$(p)$, as
evidenced by SQUID and NMR measurements, it is very surprising to
find, for the two systems, a very similar quantitative
low-temperature relaxation rate $\lambda_{T\rightarrow0}$ with
increasing Cr-concentration over the \emph{entire} $p$-range
studied. Therefore we conclude that \emph{only} defects related to
\emph{dilution} of the frustrated magnetic network influence the
relaxation rate.

In a classical framework, a coplanar arrangement with zero-energy
excitation modes, involving spins on hexagons, was proposed in the
literature to be selected at low temperature \cite{Chubukov92}.
Inelastic neutron scattering experiments on SCGO and the spinel
compound ZnCr$_{2}$O$_{4}$ are indications in favor of such
excitations \cite{Broholm90,Lee02}. Therefore, one expects the
muon spin relaxation to scale with the number of fully occupied
hexagons, $\propto p^{6}$. In Fig.~\ref{fig_lambda_p}, the
BSZCGO$(p)$ data altogether with our SCGO$(p\geq0.89)$ samples
indicate that $\lambda_{T\rightarrow0}(p)$ is well accounted for
by adding a dominant $p^{6}$ term to the $p^{3}$ term proposed for
SCGO($p\leq0.89$) in Ref.~\onlinecite{Keren00}. In a case of
purely dipolar couplings of the muon with the electronic magnetic
moments, the average of the local dipolar field is given by
$\Delta\propto \langle\sum_{i} 1/r_{i}^{3}\rangle\propto p$, where
the sum is made over the Cr$^{3+}$ ions, $r_{i}$ is the distance
between the muon and the $i$th position and $\langle\cdot\rangle$
is the average over the muon positions. One therefore obtains
$\lambda(p)\sim \Delta(p)\propto p$ in the case of slow
fluctuations and $\lambda(p)\sim
2\Delta(p)^{2}\nu/(\nu^{2}+\gamma_{\mu}^{2}H_{LF}^{2}) \propto
p^{2}$ in the case of fast fluctuations, i.e., $\lambda(p)\propto
p^{\eta} $, with $\eta\leq2$, which is not consistent with our
data. This indicates that the relaxation is not induced by single
spin excitations but rather \emph{collective} excitation processes
extending at least on triangles and/or hexagons. However these
excitations still involve only a finite number of spins since the
increase of spin vacancies only affect smoothly the spin dynamics.
On the contrary, a more pronounced effect is observed in the
$S=\frac{1}{2}$ Kagom\'e-like compound volborthite
\cite{Fukaya03}.

\subsection{RVB ground state with coherent unconfined spinons}
\label{Spinons}

\begin{figure}[tbp!] \center
\includegraphics[width=1\linewidth]{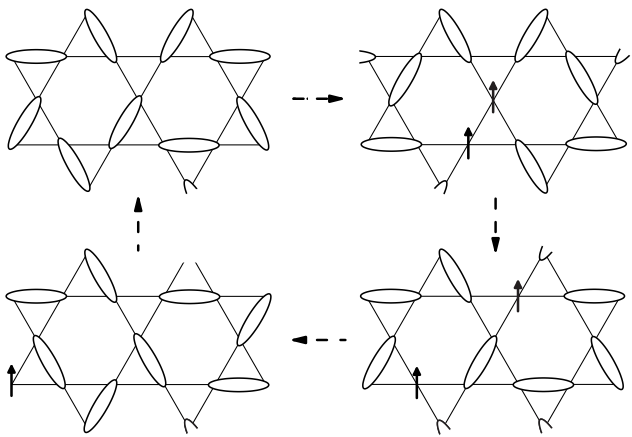}
      \caption{ \label{fig_spinon} Schematic representation of the
      creation and deconfinement of a spinon in a RVB ground
      state on a Kagom\'e lattice.
       }
      \end{figure}

Finding a phenomenological model reproducing the polarization
$P_{z}(t)$ for \emph{all} fields, \emph{all} temperatures and
\emph{all} dilutions has been for long a pending challenge in the
analysis of these $\mu$SR experiments in Kagom\'e frustrated
antiferromagnets. In the quantum framework of a singlet ground
state, one needs unpaired spins, i.e. isolated magnetic moments,
in order to generate magnetic excitations responsible for the muon
spin relaxation. Such excitations can be ascribed to unconfined
spinons \cite{Lhuillier01}, for which the location of spin
$\frac{1}{2}$ varies in time with no loss of coherence of the
excited state (Fig.~\ref{fig_spinon}). Hence, a given muon spin
couples to a spin only a short fraction $ft$ of the time $t$ after
implantation and one can use a model of ``sporadic'' field
fluctuations to describe $P_{z}(t)$.

This was the initial guess by Uemura \emph{et al.} to account for
the weakness of the field dependence of $P_{z}(t)$ at low
temperature in SCGO(0.89) \cite{Uemura94}, although the spins are
$S=\frac{3}{2}$. In the absence of any theoretical predictions, we
will just assume that the mechanism is comparable. This
polarization is given by a ``sporadic'' dynamical Kubo-Toyabe
function, $P_{z}^{K}(ft,\Delta,H_{LF},\nu)$, which rewrites simply
$P_{z}^{K}(t,f\Delta,fH_{LF},f\nu)$. $\Delta/\gamma_{\mu}$ is the
local field created on the muon site by a paramagnetic neighboring
released spin with a fluctuation frequency $\nu$ and a
corresponding exponential time correlation function $\exp(-\nu
t)$. $\Delta$ is therefore related to the muon location in the
unit cell and its average is assumed to be $p$-independent for a
given system but is expected to vary from BSZCGO to SCGO. The
Gaussian at early times, the weakness of the field dependence and
a dynamical relaxation down to 0 are altogether related to the $f$
factor and $\nu\sim \Delta$. In addition to this sporadic
relaxation, we found that a more conventional Markovian relaxation
needs to be introduced to fit the long times tail
($t\gtrsim3~\mu$s) for all fields and temperatures. We therefore
write
\begin{equation}
\label{eqmodel}
    P_{z}(t)=xP_{z}^{K}(t,f\Delta,fH_{LF},f\nu)+(1-x)e^{-\lambda^{\prime} t} \ ,
\end{equation}
where $x$ represents the weight of the short time sporadic
dynamical function, associated with the spinons dynamics. In the
following, we first detail how all these parameters can be
reliably deduced from the data and sketch a physical picture
consistent with our results.

We first present our analysis for $T\ll T_{g}$ in BSZCGO$(p)$. In
order to limit the number of free parameters, we make the minimal
assumption that the external field does not influence the dynamics
of the coherent spinon term. The dynamics ($\nu$) and the average
dipolar field created by a spinon on a muon site ($\Delta$) are
shared for all $p$ and $H_{LF}$. For low fields, $x$ is found
close to 1 for the purest samples [Fig.~\ref{fig_Asym_T}(a,b)],
making $\lambda^{\prime}$ a non-relevant fitting parameter. It
also enables us to determine $\nu$ and $\Delta$ when
$H_{LF}<500$~G, for various $p$. The parameter $f$ is adjusted for
each $p$ and its variation accounts for the evolution of
$\lambda_{T\rightarrow0}(p)$. On the contrary, the high field data
enable us to monitor the evolution of $x$ with $H_{LF}$ and to
determine a value for $\lambda^{\prime}$. We could not extend the
fits below $p=0.71$ since the weak field dependence prevents an
unambiguous determination of the parameters.

\begin{figure}[tbp!] \center
\includegraphics[width=1\linewidth]{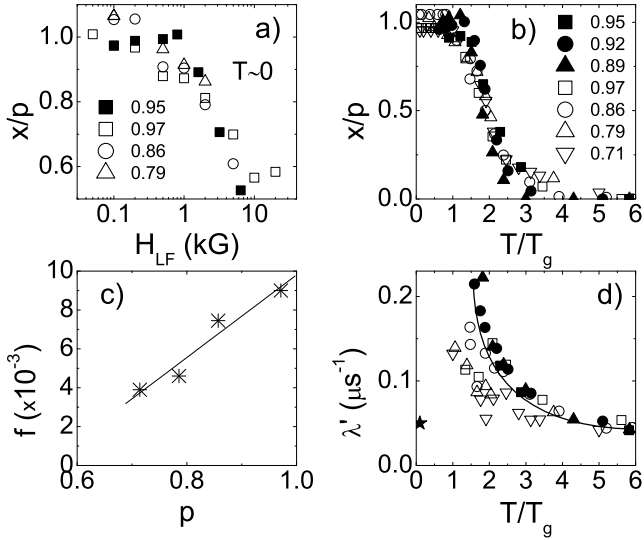}
      \caption{ \label{fig_fitx} Fitting parameters of Eq.~\ref{eqmodel}
      for BSZCGO and SCGO (open and full symbols). $f$ is common for both systems.
       }
      \end{figure}

We find perfect fits of our data [Fig.~\ref{fig_Asym_T}(d)] with
$\nu\sim 1000~\mu$s$^{-1}$,
$\Delta\sim350~\mu$s$^{-1}\sim\gamma_{\mu}\times4$~kG and an
average value of $f\sim0.006$. We find a nearly flat
$p$-independent long time relaxation rate for $T\ll T_{g}$
[$\lambda^{\prime}\sim0.05~\mu$s$^{-1}$, $\bigstar$ in
Fig.~\ref{fig_fitx}(d)]. An important finding is that $x$ is of
the order of $p$ at low fields [Fig.~\ref{fig_fitx}(a,b)]. This
may be related to the theoretical computations showing that the
correlations are enhanced around spin vacancies in the Kagom\'e
lattice \cite{Dommange03}, which would destroy locally the spin
liquid state. Besides, $x$ decreases appreciably for
$H_{LF}\sim10$~kG whatever the value of $p$. We can associate this
decrease to the existence of an energy scale $\sim$ 1~K, which is
of the order of $T_{g}$. Finally, we observe a linear variation of
$f$ with $p$ [Fig.~\ref{fig_fitx}(c)] and $f$ tends to vanish
around $p\sim0.5$, a limit consistent with our classical approach
(Fig.~\ref{fig_lambda_p}). Indeed, since we find $\nu\sim\Delta$
when $T\rightarrow0$, the relationship
$f(p)\Delta\sim\lambda_{T\rightarrow0}(p)$ is expected. This
indicates that even far from the substituted sites, the coherent
state is somehow affected, e.g the density of spinons could be
smaller.

As suggested by the similar $p$-variation of
$\lambda_{T\rightarrow0}$ in both BSZCGO$(p)$ and SCGO$(p)$, we
assume that the excitation modes are identical in both systems,
i.e., $f$ and $\nu$ are kept the same for comparable $p$. We find
$\Delta\sim1200~\mu$s$^{-1}$ for our SCGO($p\geq0.89$) samples, in
agreement with previous work on SCGO(0.89) \cite{Uemura94}. It is
quite rewarding to find that a \emph{common} physical picture
underlies all the sets of data at low temperature for \emph{both}
Kagom\'e bilayers.

We now extend the fits of the muon spin polarization $P_{z}(t)$ to
the whole temperature-range [Fig.~\ref{fig_Asym_T}(b)], fixing $f$
to its low-temperature value in order to limit the number of free
parameters. It would also, phenomenologically, mean that once a
spinon is created, its deconfinement process remains temperature
independent, which is a quite reasonable assumption from a quantum
point of view. As expected from the change of shape around $T_{g}$
\cite{Uemura94,Keren00}, the weight $x$ of the sporadic term
decreases to finally enter a high-temperature regime
[Fig.~\ref{fig_fitx}(b)] with an exponential muon relaxation
($x\rightarrow0$). The similarity between the field- and
temperature-dependence of $x$ indicates that the sporadic regime
is destroyed with an energy of the order of the freezing
temperature $T_\text{g}$. Figure~\ref{fig_lambda_p}(b) displays a
sharp crossover from one state to the other, between $T_{g}$ and $
3T_{g}$, corresponding to the $\lambda$ steep decrease
(Fig.~\ref{fig_lambda_scale}), for both BSZCGO$(p)$ and SCGO$(p)$.
For $T>T_{g}$, $\nu$ and $\lambda^{\prime}$
[Fig.~\ref{fig_fitx}(d)] decrease by one order of magnitude up to
$10T_{g}$.

At high temperature, it is natural to think in terms of
paramagnetic fluctuating spins. At lower temperatures,
$\lambda^{\prime}$ seems to diverge at $T_{g}$ and below, the
weight of the exponential term at low fields ($1-x\sim 1-p$) could
correspond to localized frozen spins, reflecting the glassy
component measured by SQUID. It is noteworthy that $T_{g}$ does
not increase with $1-p$ but rather decreases, at the opposite of
the case of canonical spin glasses. We could further confirm the
existence of such a frozen component in SCGO$(p=0.95$-0.89), since
a clear recovery of a small part ($\sim 4$\%) of the asymmetry is
found at long times [e.g. Fig.~\ref{fig_lambda_scale}(inset)]. It
is observed for $t\sim100\,\lambda^{-1}$, which is out of the
experimental range for BSZCGO$(p)$.

To summarize, the picture based on a coherent RVB state, which
magnetic excitations are mobile fluctuating spins $\frac{1}{2}$ on
the Kagom\'e lattice, explains well the data, provided that
(i)~these excitations can be generated even for $T\rightarrow0$.
This underlines the smallness of the ``magnetic'' gap, if any,
typically $\Delta < J/1000$~; (ii)~an energy scale related to
$T_{g}$ (fields of the order of 10~kG or, equivalently,
temperatures of the order of $T_{g}$, which vary very little with
$p$) are high enough to destroy the coherent RVB type state~;
(iii)~the substitution defects are accounted for by an additional
classical relaxation process.

\subsection{Spin-Glass-like transition}

The ``intrinsic'' spin-glass-like transition at $T_{g}$ is one of
the most puzzling observation in these frustrated magnets. Indeed,
the origin of such a spin glass state in a disorder-free system
still awaits for a complete understanding, although recent
theoretical approaches catch some of the facets of this original
ground state \cite{Mila02}. The most realistic model, presented by
Ferrero \emph{et al.} in Ref.~\onlinecite{Ferrero03}, uses the so
called ``dimerized'' approach \cite{Mila98,Mambrini00}, i.e.,
considers a geometry of the Kagom\'e lattice with two different
kinds of equilateral triangles, corresponding to the Cr(b) layers
in SCGO$(p)$ and BSZCGO$(p)$ (Fig.~\ref{cell}). It predicts a spin
glass-like transition in the $S=\frac{1}{2}$ Kagom\'e lattice,
related to the freezing of the chirality, with
$T_{g}\sim0.5J^{\prime}$. Here, $J^{\prime}$ is the largest
coupling in the Kagom\'e planes (corresponding to the thick lines
in Fig.~\ref{cell}). Although $0.05 J^{\prime}$ is of the order of
magnitude of the experimental $T_{g}$, the very similar values of
the Cr-Cr bonds in both systems and Eq.~\ref{eqcoupling} do not
allow to predict a factor 2 in $T_{g}$.

From our $\mu$SR results, we propose an other interpretation which
still awaits for theoretical confirmation. We find no energy gap
for the spin $\frac{1}{2}$ excitations. This could explain quite
well why the transition to the spin glass state is fairly
independent of $p$. Indeed, in this framework, spinons could
mediate the interactions between magnetic defects localized around
spin vacancies. This is corroborated by all the scaling properties
in $T/T_{g}$. The transition to the spin glass state would
correspond then to the formation of the coherent singlet state
rather than any interaction strength between defects. The bond
defects inherent to BSZCGO$(p)$ probably decrease the value of
this coherent state energy scale since it is more favorable to
create localized singlets in this geometry. This is qualitatively
consistent with the twice lower spin glass transition temperature
in this family.

\subsection{Extension to other compounds?}

We already noticed that the $\mu$SR undecouplable Gaussian line
shape, altogether with a plateau of the relaxation rate, have been
reported in other compounds, and now elaborate about the possible
links between these systems.

The other Kagom\'e compounds, volborthite
\cite{Hiroi01,Fukaya03,BertHFM} and Cr jarosite \cite{Keren96}
display the same properties as the Kagom\'e bilayers and our
phenomenological model seems correct.

Kojima \emph{et al.} already noticed the similarity between SCGO
and the doped Haldane chain Y$_{2-x}$Ca$_{x}$BaNiO$_{5}$ and
SCGO(0.89) \cite{Kojima95,KojimaPhD}. However, contrary to the
Kagom\'e bilayers, the relaxation function becomes more
``conventional'' when the system becomes purer, i.e., the muon
spin relaxation function is a square root exponential in the pure
system at low temperature. It is not surprising since in the pure
Haldane chain one expects a Valence Bond Crystal
\cite{Lhuillier01}. Therefore no magnetic excitation are allowed
at low temperature and no depolarization is expected in $\mu$SR.
However, mobile excitations are added when the chains are doped
\cite{Ammon00}, i.e. the muon spin relaxation becomes
unconventional. This corresponds to the data and is consistent
with our unconfined spinon phenomenological model.

Finally, Fukaya \emph{et al.} reported the same behavior in
Sr(Cu$_{1-x}$Zn$_{x}$)$_{2}$(BO$_{3}$)$_{2}$ for $x=0$ and 0.02
\cite{Fukaya03B}. This system displays, theoretically and
experimentally, an exact singlet dimer state and its intrinsic
susceptibility vanishes for $T\lesssim3$~K
\cite{Kageyama99,Kodama02}. Nonetheless, a CW term, corresponding
to 0.72\% of $S=\frac{1}{2}$ impurities with respect to the Cu
sites, is observed below 4~K \cite{Kageyama99}, which is finally
comparable, from the macroscopic susceptibility aspect, to the
former cases of pure volborthite and pure Y$_{2}$CaBaNiO$_{5}$.
However, according to us, there is today no clue about the
possibility of mobile excitations in this magnetic network with a
few percent of defects.

We conclude that \emph{all} these systems display (i)~a
theoretical and/or experimental singlet ground state; (ii)~a
defect term observed through a low-temperature CW like upturn in
the macroscopic susceptibility; (iii)~a spin glass-like transition
of this defect term at a temperature $T_{g}$ (but
Sr(Cu$_{1-x}$Zn$_{x}$)$_{2}$(BO$_{3}$)$_{2}$) \footnote{According
to us, no FC-ZFC magnetization data is published about
Sr(Cu$_{1-x}$Zn$_{x}$)$_{2}$(BO$_{3}$)$_{2}$.}; (iv)~a plateau of
the muon spin relaxation rate below $T_{g}$.

It is important to notice that the clear experimental correlation
between the muon spin relaxation rate plateau and the spin
glass-like transition is strongly against a pure muon-induced
effect argument for this plateau. Even if the muon has indeed an
effect on the physics of these compounds, the $\lambda$ plateau
below $T_{g}$ shows that a bulk transition occurs at this
temperature, i.e. that this freezing cannot be due to isolated
impurities.

Theoretical approaches are now required to link a possible
unconfined spinons state to a spin-glass-like behavior and a
relaxation plateau of the implanted muon spins. They could
therefore corroborate our phenomenological approach.


\section{Conclusion}
\label{conclusion}

With Heisenberg spins and nearest neighbor couplings of the order
of 40~K, the Kagom\'e bilayers of SCGO and BSZCCGO are today the
archetypes of Highly Frustrated Magnets. The comparison between
SQUID and NMR experiments on the two slightly different compounds
SCGO and BSZCGO allows us to isolate the intrinsic properties of
the frustrated Kagom\'e bilayer geometry and to understand better
the role of the (non)magnetic defects. We find that \emph{defects}
localized in the frustrated network induce a spatially extended
response of the magnetic system, consistent with theoretical
predictions. This response generates a Curie-like upturn in the
macroscopic susceptibility at low temperature. However, local
measurements of the susceptibility, using Ga NMR, show that the
\emph{intrinsic} susceptibility of the Kagom\'e bilayers decreases
below 45~K, a temperature of the order of $J$. This behavior was
also predicted theoretically and is consistent with short-ranged
spin-spin correlations, on the order of the lattice parameter. We
cannot give a definitive conclusion about the existence of a spin
gap with NMR experiments because the gallium nuclei cannot be
probed by NMR at low temperature. However a maximum value of $\sim
J/10$ is found for such a gap.

We measured the spin dynamics using Muon Spin Relaxation
experiments, down to 30~mK. We find that both systems SCGO and
BSZCGO display a dynamical magnetic state down to this
temperature, unconventional for the purest samples. Very simple
considerations show that the spin-glass-like state, measured at
$T_{g}\ll \theta_{\text{CW}}$ with macroscopic susceptibility
techniques, is correlated to the slowing down of the Cr spin
dynamics in the whole sample. The data are well accounted by a
description based on unconfined spinons as the magnetic
excitations and we suggest that this phenomenological approach
remains valid for several spin singlet compounds at low
temperature. Indeed, they all display striking similar aspects
such as a spin glass transition of the magnetic defect channel in
the macroscopic susceptibility and an undecouplable Gaussian line
shape with a relaxation rate plateau of the muon polarization
below this spin glass transition.

In all these systems the fast depolarization of the muon down to
the experimental limit of $T\sim30$~mK underlines the existence of
magnetic excitations at low temperature and hence the weakness of
an hypothetical unconfined spinon gap. An other spin dynamics
study with Neutron Spin Echo technique is in progress and could
give more details about this $T\rightarrow0$ spin liquid state
\cite{Mutkaprep}.

\begin{acknowledgments}
We thank H.~Alloul, F.~Bert, J.~Bobroff, R.J.~Cava, D.~Huber,
A.~Keren, C.~Lhuillier, M.~Mekata, G.~Misguich, R.~Moessner,
C.~Mondelli, H.~Mutka, B.~Ouladdiaf, C.~Payen, P.~Schiffer and
P.~Sindzingre for fruitful discussions. We also wish to thank
A.~Amato, C.~Baines and A.D.~Hillier, our $\mu$SR local contacts,
whose outstanding efforts have made these experiments possible.

The $\mu$SR experiments were performed at the Swiss Muon Source,
Paul Scherer Institute, Villigen, Switzerland, and financially
supported by the Federal Office for Education and Science, Berne,
Switzerland, as well as at the ISIS facility, Rutherford Appleton
Laboratory, Didcot, United Kingdom.

Part of this work was financially supported by the EC IHP
Programme for large scale facilities and by the EC Framework
Programmes V and VI.

We are grateful to the muon beamline groups and to the technical
staff of Laboratoire de Physique des Solides, Orsay.

\end{acknowledgments}


\begin{thebibliography}{92}
\expandafter\ifx\csname
natexlab\endcsname\relax\def\natexlab#1{#1}\fi
\expandafter\ifx\csname bibnamefont\endcsname\relax
  \def\bibnamefont#1{#1}\fi
\expandafter\ifx\csname bibfnamefont\endcsname\relax
  \def\bibfnamefont#1{#1}\fi
\expandafter\ifx\csname citenamefont\endcsname\relax
  \def\citenamefont#1{#1}\fi
\expandafter\ifx\csname url\endcsname\relax
  \def\url#1{\texttt{#1}}\fi
\expandafter\ifx\csname
urlprefix\endcsname\relax\def\urlprefix{URL }\fi
\providecommand{\bibinfo}[2]{#2}
\providecommand{\eprint}[2][]{\url{#2}}

\bibitem[{\citenamefont{{P. W. }Anderson}(1973)}]{Anderson73}
\bibinfo{author}{\bibnamefont{{P. W. }Anderson}}, \bibinfo{journal}{Mater. Res.
  Bull.} \textbf{\bibinfo{volume}{8}}, \bibinfo{pages}{153}
  (\bibinfo{year}{1973}).

\bibitem[{\citenamefont{{P. W. }Anderson}(1987)}]{Anderson87}
\bibinfo{author}{\bibnamefont{{P. W. }Anderson}}, \bibinfo{journal}{Science}
  \textbf{\bibinfo{volume}{235}}, \bibinfo{pages}{1196} (\bibinfo{year}{1987}).

\bibitem[{\citenamefont{Baskaran}(2003)}]{Baskaran03}
\bibinfo{author}{\bibfnamefont{G.}~\bibnamefont{Baskaran}},
  \bibinfo{journal}{Phys. Rev. Lett.} \textbf{\bibinfo{volume}{91}},
  \bibinfo{pages}{097003} (\bibinfo{year}{2003}).

\bibitem[{\citenamefont{Bernu et~al.}(1994)\citenamefont{Bernu, Lecheminant,
  Lhuillier, and Pierre}}]{Bernu94}
\bibinfo{author}{\bibfnamefont{B.}~\bibnamefont{Bernu}},
  \bibinfo{author}{\bibfnamefont{P.}~\bibnamefont{Lecheminant}},
  \bibinfo{author}{\bibfnamefont{C.}~\bibnamefont{Lhuillier}},
  \bibnamefont{and} \bibinfo{author}{\bibfnamefont{L.}~\bibnamefont{Pierre}},
  \bibinfo{journal}{Phys. Rev. B} \textbf{\bibinfo{volume}{50}},
  \bibinfo{pages}{10048} (\bibinfo{year}{1994}).

\bibitem[{\citenamefont{{A. P. }Ramirez}(2001)}]{Ramirez01}
\bibinfo{author}{\bibnamefont{{A. P. }Ramirez}}, in
  \emph{\bibinfo{booktitle}{Handbook on Magnetic Materials}}, edited by
  \bibinfo{editor}{\bibfnamefont{K.~J.~H.} \bibnamefont{Busch}}
  (\bibinfo{publisher}{Elsevier Science, Amsterdam}, \bibinfo{year}{2001}),
  vol.~\bibinfo{volume}{13}, p. \bibinfo{pages}{423}.

\bibitem[{\citenamefont{Stewart}(2004)}]{HFM04}
\bibinfo{editor}{\bibfnamefont{R.}~\bibnamefont{Stewart}}, ed.,
  \emph{\bibinfo{title}{Proceedings of the Highly Frustrated Magnetism 2003
  Conference, Grenoble, France}} (\bibinfo{year}{2004}), \bibinfo{note}{{J}.
  {P}hys.: {C}ondens. {M}atter, \textbf{16}, S553-922}.

\bibitem[{\citenamefont{Binder and Young}(1986)}]{Binder86}
\bibinfo{author}{\bibfnamefont{K.}~\bibnamefont{Binder}} \bibnamefont{and}
  \bibinfo{author}{\bibfnamefont{A.~P.} \bibnamefont{Young}},
  \bibinfo{journal}{Rev. Mod. Phys.} \textbf{\bibinfo{volume}{58}},
  \bibinfo{pages}{801} (\bibinfo{year}{1986}).

\bibitem[{\citenamefont{Huse and Rutenberg}(1992)}]{Huse92}
\bibinfo{author}{\bibfnamefont{D.~A.} \bibnamefont{Huse}} \bibnamefont{and}
  \bibinfo{author}{\bibfnamefont{A.~D.} \bibnamefont{Rutenberg}},
  \bibinfo{journal}{Phys. Rev. B} \textbf{\bibinfo{volume}{45}},
  \bibinfo{pages}{7536} (\bibinfo{year}{1992}).

\bibitem[{\citenamefont{Chalker et~al.}(1992)\citenamefont{Chalker, Holdsworth,
  and Shender}}]{Chalker92}
\bibinfo{author}{\bibfnamefont{J.~T.} \bibnamefont{Chalker}},
  \bibinfo{author}{\bibfnamefont{P.~C.~W.} \bibnamefont{Holdsworth}},
  \bibnamefont{and} \bibinfo{author}{\bibfnamefont{E.~F.}
  \bibnamefont{Shender}}, \bibinfo{journal}{Phys. Rev. Lett.}
  \textbf{\bibinfo{volume}{68}}, \bibinfo{pages}{855} (\bibinfo{year}{1992}).

\bibitem[{\citenamefont{Ritchey et~al.}(1993)\citenamefont{Ritchey, Chandra,
  and Coleman}}]{Ritchey93}
\bibinfo{author}{\bibfnamefont{I.}~\bibnamefont{Ritchey}},
  \bibinfo{author}{\bibfnamefont{P.}~\bibnamefont{Chandra}}, \bibnamefont{and}
  \bibinfo{author}{\bibfnamefont{P.}~\bibnamefont{Coleman}},
  \bibinfo{journal}{Phys. Rev. B} \textbf{\bibinfo{volume}{47}},
  \bibinfo{pages}{15342} (\bibinfo{year}{1993}).

\bibitem[{\citenamefont{Misguich and Lhuillier}(2003)}]{MisguichDiep}
\bibinfo{author}{\bibfnamefont{G.}~\bibnamefont{Misguich}} \bibnamefont{and}
  \bibinfo{author}{\bibfnamefont{C.}~\bibnamefont{Lhuillier}}, in
  \emph{\bibinfo{booktitle}{Frustrated Spin systems}}, edited by
  \bibinfo{editor}{\bibfnamefont{H.~T.} \bibnamefont{Diep}}
  (\bibinfo{publisher}{World Scientific, Singapore}, \bibinfo{year}{2003}),
  \bibinfo{note}{(cond-mat/0310405), and references therein}.

\bibitem[{\citenamefont{Moessner}(2001)}]{MoessnerHFM01}
\bibinfo{author}{\bibfnamefont{R.}~\bibnamefont{Moessner}},
  \bibinfo{journal}{Can. J. Phys.} \textbf{\bibinfo{volume}{79}},
  \bibinfo{pages}{1283} (\bibinfo{year}{2001}).

\bibitem[{\citenamefont{Zeng and Elser}(1995)}]{Zeng95}
\bibinfo{author}{\bibfnamefont{C.}~\bibnamefont{Zeng}} \bibnamefont{and}
  \bibinfo{author}{\bibfnamefont{V.}~\bibnamefont{Elser}},
  \bibinfo{journal}{Phys. Rev. B} \textbf{\bibinfo{volume}{51}},
  \bibinfo{pages}{8318} (\bibinfo{year}{1995}).

\bibitem[{\citenamefont{Mambrini and Mila}(2000)}]{Mambrini00}
\bibinfo{author}{\bibfnamefont{M.}~\bibnamefont{Mambrini}} \bibnamefont{and}
  \bibinfo{author}{\bibfnamefont{F.}~\bibnamefont{Mila}},
  \bibinfo{journal}{Eur. Phys. J. B} \textbf{\bibinfo{volume}{17}},
  \bibinfo{pages}{651} (\bibinfo{year}{2000}).

\bibitem[{\citenamefont{Waldtmann et~al.}(1998)\citenamefont{Waldtmann, Everts,
  Bernu, Lhuillier, Sindzingre, Lecheminant, and Pierre}}]{Waldtmann98}
\bibinfo{author}{\bibfnamefont{C.}~\bibnamefont{Waldtmann}},
  \bibinfo{author}{\bibfnamefont{H.~U.} \bibnamefont{Everts}},
  \bibinfo{author}{\bibfnamefont{B.}~\bibnamefont{Bernu}},
  \bibinfo{author}{\bibfnamefont{C.}~\bibnamefont{Lhuillier}},
  \bibinfo{author}{\bibfnamefont{P.}~\bibnamefont{Sindzingre}},
  \bibinfo{author}{\bibfnamefont{P.}~\bibnamefont{Lecheminant}},
  \bibnamefont{and} \bibinfo{author}{\bibfnamefont{L.}~\bibnamefont{Pierre}},
  \bibinfo{journal}{Eur. Phys. J. B} \textbf{\bibinfo{volume}{2}},
  \bibinfo{pages}{501} (\bibinfo{year}{1998}).

\bibitem[{\citenamefont{Mila}(1998)}]{Mila98}
\bibinfo{author}{\bibfnamefont{F.}~\bibnamefont{Mila}}, \bibinfo{journal}{Phys.
  Rev. Lett.} \textbf{\bibinfo{volume}{81}}, \bibinfo{pages}{2356}
  (\bibinfo{year}{1998}).

\bibitem[{\citenamefont{Misguich et~al.}(2003)\citenamefont{Misguich, Serban,
  and Pasquier}}]{Misguich03}
\bibinfo{author}{\bibfnamefont{G.}~\bibnamefont{Misguich}},
  \bibinfo{author}{\bibfnamefont{D.}~\bibnamefont{Serban}}, \bibnamefont{and}
  \bibinfo{author}{\bibfnamefont{V.}~\bibnamefont{Pasquier}},
  \bibinfo{journal}{Phys. Rev. B} \textbf{\bibinfo{volume}{67}},
  \bibinfo{pages}{214413} (\bibinfo{year}{2003}).

\bibitem[{\citenamefont{Canals and Lacroix}(1998)}]{Canals98}
\bibinfo{author}{\bibfnamefont{B.}~\bibnamefont{Canals}} \bibnamefont{and}
  \bibinfo{author}{\bibfnamefont{C.}~\bibnamefont{Lacroix}},
  \bibinfo{journal}{Phys. Rev. Lett.} \textbf{\bibinfo{volume}{80}},
  \bibinfo{pages}{2933} (\bibinfo{year}{1998}).

\bibitem[{\citenamefont{Lhuillier and Sindzingre}(2001)}]{Lhuillier01}
\bibinfo{author}{\bibfnamefont{C.}~\bibnamefont{Lhuillier}} \bibnamefont{and}
  \bibinfo{author}{\bibfnamefont{P.}~\bibnamefont{Sindzingre}}, in
  \emph{\bibinfo{booktitle}{Quantum Properties of Low-Dimensional
  Antiferromagnets}}, edited by
  \bibinfo{editor}{\bibfnamefont{Y.}~\bibnamefont{Ajiro}} \bibnamefont{and}
  \bibinfo{editor}{\bibfnamefont{J.~P.} \bibnamefont{Boucher}}
  (\bibinfo{publisher}{Kyushu University Press, Fukuoka},
  \bibinfo{year}{2001}), p. \bibinfo{pages}{111},
  \bibinfo{note}{cond-mat/0212351}.

\bibitem[{\citenamefont{Palmer and Chalker}(2000)}]{Palmer00}
\bibinfo{author}{\bibfnamefont{S.~E.} \bibnamefont{Palmer}} \bibnamefont{and}
  \bibinfo{author}{\bibfnamefont{J.~T.} \bibnamefont{Chalker}},
  \bibinfo{journal}{Phys. Rev. B} \textbf{\bibinfo{volume}{62}},
  \bibinfo{pages}{488} (\bibinfo{year}{2000}).

\bibitem[{\citenamefont{Tchernyshyov et~al.}(2002)\citenamefont{Tchernyshyov,
  Moessner, and Sondhi}}]{Tchernyshyov02}
\bibinfo{author}{\bibfnamefont{O.}~\bibnamefont{Tchernyshyov}},
  \bibinfo{author}{\bibfnamefont{R.}~\bibnamefont{Moessner}}, \bibnamefont{and}
  \bibinfo{author}{\bibfnamefont{S.~L.} \bibnamefont{Sondhi}},
  \bibinfo{journal}{Phys. Rev. Lett.} \textbf{\bibinfo{volume}{88}},
  \bibinfo{pages}{067203} (\bibinfo{year}{2002}).

\bibitem[{\citenamefont{Elhajal et~al.}(2002)\citenamefont{Elhajal, Canals, and
  Lacroix}}]{Elhajal02}
\bibinfo{author}{\bibfnamefont{M.}~\bibnamefont{Elhajal}},
  \bibinfo{author}{\bibfnamefont{B.}~\bibnamefont{Canals}}, \bibnamefont{and}
  \bibinfo{author}{\bibfnamefont{C.}~\bibnamefont{Lacroix}},
  \bibinfo{journal}{Phys. Rev. B} \textbf{\bibinfo{volume}{66}},
  \bibinfo{pages}{014422} (\bibinfo{year}{2002}).

\bibitem[{\citenamefont{Dommange et~al.}(2003)\citenamefont{Dommange, Mambrini,
  Normand, and Mila}}]{Dommange03}
\bibinfo{author}{\bibfnamefont{S.}~\bibnamefont{Dommange}},
  \bibinfo{author}{\bibfnamefont{M.}~\bibnamefont{Mambrini}},
  \bibinfo{author}{\bibfnamefont{B.}~\bibnamefont{Normand}}, \bibnamefont{and}
  \bibinfo{author}{\bibfnamefont{F.}~\bibnamefont{Mila}},
  \bibinfo{journal}{Phys. Rev. B} \textbf{\bibinfo{volume}{68}},
  \bibinfo{pages}{224416} (\bibinfo{year}{2003}).

\bibitem[{\citenamefont{Zhitomirsky et~al.}(2000)\citenamefont{Zhitomirsky,
  Honecker, and Petrenko}}]{Zhitomirsky00}
\bibinfo{author}{\bibfnamefont{M.~E.} \bibnamefont{Zhitomirsky}},
  \bibinfo{author}{\bibfnamefont{A.}~\bibnamefont{Honecker}}, \bibnamefont{and}
  \bibinfo{author}{\bibfnamefont{O.~A.} \bibnamefont{Petrenko}},
  \bibinfo{journal}{Phys. Rev. Lett.} \textbf{\bibinfo{volume}{85}},
  \bibinfo{pages}{3269} (\bibinfo{year}{2000}).

\bibitem[{\citenamefont{Domenge et~al.}()\citenamefont{Domenge, Sindzingre, and
  Lhuillier}}]{Domenge05}
\bibinfo{author}{\bibfnamefont{J.-C.} \bibnamefont{Domenge}},
  \bibinfo{author}{\bibfnamefont{P.}~\bibnamefont{Sindzingre}},
  \bibnamefont{and}
  \bibinfo{author}{\bibfnamefont{C.}~\bibnamefont{Lhuillier}},
  \bibinfo{note}{cond-mat/0502414}.

\bibitem[{\citenamefont{Wills}(2001)}]{WillsHFM01}
\bibinfo{author}{\bibfnamefont{A.~S.} \bibnamefont{Wills}},
  \bibinfo{journal}{Can. J. Phys.} \textbf{\bibinfo{volume}{79}},
  \bibinfo{pages}{1501} (\bibinfo{year}{2001}).

\bibitem[{\citenamefont{Narumi et~al.}(2004)\citenamefont{Narumi, Katsumata,
  Honda, Domenge, Sindzingre, Lhuillier, Shimaoka, Kobayashi, and
  Kindo}}]{Narumi04}
\bibinfo{author}{\bibfnamefont{Y.}~\bibnamefont{Narumi}},
  \bibinfo{author}{\bibfnamefont{K.}~\bibnamefont{Katsumata}},
  \bibinfo{author}{\bibfnamefont{Z.}~\bibnamefont{Honda}},
  \bibinfo{author}{\bibfnamefont{J.~C.} \bibnamefont{Domenge}},
  \bibinfo{author}{\bibfnamefont{P.}~\bibnamefont{Sindzingre}},
  \bibinfo{author}{\bibfnamefont{C.}~\bibnamefont{Lhuillier}},
  \bibinfo{author}{\bibfnamefont{Y.}~\bibnamefont{Shimaoka}},
  \bibinfo{author}{\bibfnamefont{T.~C.} \bibnamefont{Kobayashi}},
  \bibnamefont{and} \bibinfo{author}{\bibfnamefont{K.}~\bibnamefont{Kindo}},
  \bibinfo{journal}{Europhys. Lett.} \textbf{\bibinfo{volume}{65}},
  \bibinfo{pages}{705} (\bibinfo{year}{2004}).

\bibitem[{\citenamefont{Lawes et~al.}(2004)\citenamefont{Lawes, Kenzelmann,
  Rogado, Kim, Jorge, Cava, Aharony, Entin-Wohlman, Harris, Yildirim
  et~al.}}]{Lawes04}
\bibinfo{author}{\bibfnamefont{G.}~\bibnamefont{Lawes}},
  \bibinfo{author}{\bibfnamefont{M.}~\bibnamefont{Kenzelmann}},
  \bibinfo{author}{\bibfnamefont{N.}~\bibnamefont{Rogado}},
  \bibinfo{author}{\bibfnamefont{K.~H.} \bibnamefont{Kim}},
  \bibinfo{author}{\bibfnamefont{G.~A.} \bibnamefont{Jorge}},
  \bibinfo{author}{\bibfnamefont{R.~J.} \bibnamefont{Cava}},
  \bibinfo{author}{\bibfnamefont{A.}~\bibnamefont{Aharony}},
  \bibinfo{author}{\bibfnamefont{O.}~\bibnamefont{Entin-Wohlman}},
  \bibinfo{author}{\bibfnamefont{A.~B.} \bibnamefont{Harris}},
  \bibinfo{author}{\bibfnamefont{T.}~\bibnamefont{Yildirim}},
  \bibnamefont{et~al.}, \bibinfo{journal}{Phys. Rev. Lett.}
  \textbf{\bibinfo{volume}{93}}, \bibinfo{pages}{247201}
  (\bibinfo{year}{2004}).

\bibitem[{\citenamefont{Raju et~al.}(1999)\citenamefont{Raju, Dion, Gingras,
  Mason, and Greedan}}]{Raju99}
\bibinfo{author}{\bibfnamefont{N.~P.} \bibnamefont{Raju}},
  \bibinfo{author}{\bibfnamefont{M.}~\bibnamefont{Dion}},
  \bibinfo{author}{\bibfnamefont{M.~J.~P.} \bibnamefont{Gingras}},
  \bibinfo{author}{\bibfnamefont{T.~E.} \bibnamefont{Mason}}, \bibnamefont{and}
  \bibinfo{author}{\bibfnamefont{J.~E.} \bibnamefont{Greedan}},
  \bibinfo{journal}{Phys. Rev. B} \textbf{\bibinfo{volume}{59}},
  \bibinfo{pages}{14489} (\bibinfo{year}{1999}).

\bibitem[{\citenamefont{Keren and Gardner}(2001)}]{Keren01}
\bibinfo{author}{\bibfnamefont{A.}~\bibnamefont{Keren}} \bibnamefont{and}
  \bibinfo{author}{\bibfnamefont{J.}~\bibnamefont{Gardner}},
  \bibinfo{journal}{Phys. Rev. Lett.} \textbf{\bibinfo{volume}{17}},
  \bibinfo{pages}{177201} (\bibinfo{year}{2001}).

\bibitem[{\citenamefont{Hodges et~al.}(2002)\citenamefont{Hodges, Bonville,
  Forget, Yaouanc, {Dalmas de R\'eotier}, Andr\'e, Rams, Kr\'olas, Ritter,
  Gubbens et~al.}}]{Hodges02}
\bibinfo{author}{\bibfnamefont{J.~A.} \bibnamefont{Hodges}},
  \bibinfo{author}{\bibfnamefont{P.}~\bibnamefont{Bonville}},
  \bibinfo{author}{\bibfnamefont{A.}~\bibnamefont{Forget}},
  \bibinfo{author}{\bibfnamefont{A.}~\bibnamefont{Yaouanc}},
  \bibinfo{author}{\bibfnamefont{P.}~\bibnamefont{{Dalmas de R\'eotier}}},
  \bibinfo{author}{\bibfnamefont{G.}~\bibnamefont{Andr\'e}},
  \bibinfo{author}{\bibfnamefont{M.}~\bibnamefont{Rams}},
  \bibinfo{author}{\bibfnamefont{K.}~\bibnamefont{Kr\'olas}},
  \bibinfo{author}{\bibfnamefont{C.}~\bibnamefont{Ritter}},
  \bibinfo{author}{\bibfnamefont{P.~C.~M.} \bibnamefont{Gubbens}},
  \bibnamefont{et~al.}, \bibinfo{journal}{Phys. Rev. Lett.}
  \textbf{\bibinfo{volume}{88}}, \bibinfo{pages}{077204}
  (\bibinfo{year}{2002}).

\bibitem[{\citenamefont{Keren et~al.}(2004)\citenamefont{Keren, Gardner,
  Ehlers, Fukaya, Segal, and Uemura}}]{Keren04}
\bibinfo{author}{\bibfnamefont{A.}~\bibnamefont{Keren}},
  \bibinfo{author}{\bibfnamefont{J.~S.} \bibnamefont{Gardner}},
  \bibinfo{author}{\bibfnamefont{G.}~\bibnamefont{Ehlers}},
  \bibinfo{author}{\bibfnamefont{A.}~\bibnamefont{Fukaya}},
  \bibinfo{author}{\bibfnamefont{E.}~\bibnamefont{Segal}}, \bibnamefont{and}
  \bibinfo{author}{\bibfnamefont{Y.~J.} \bibnamefont{Uemura}},
  \bibinfo{journal}{Phys. Rev. Lett.} \textbf{\bibinfo{volume}{92}},
  \bibinfo{pages}{107204} (\bibinfo{year}{2004}).

\bibitem[{\citenamefont{Lee et~al.}(2002)\citenamefont{Lee, Broholm, Ratcliff,
  Gasparovic, Huang, Kim, and Cheong}}]{Lee02}
\bibinfo{author}{\bibfnamefont{S.~H.} \bibnamefont{Lee}},
  \bibinfo{author}{\bibfnamefont{C.}~\bibnamefont{Broholm}},
  \bibinfo{author}{\bibfnamefont{W.}~\bibnamefont{Ratcliff}},
  \bibinfo{author}{\bibfnamefont{G.}~\bibnamefont{Gasparovic}},
  \bibinfo{author}{\bibfnamefont{Q.}~\bibnamefont{Huang}},
  \bibinfo{author}{\bibfnamefont{T.~H.} \bibnamefont{Kim}}, \bibnamefont{and}
  \bibinfo{author}{\bibfnamefont{S.~W.} \bibnamefont{Cheong}},
  \bibinfo{journal}{{N}ature} \textbf{\bibinfo{volume}{418}},
  \bibinfo{pages}{856} (\bibinfo{year}{2002}).

\bibitem[{\citenamefont{Harris}(1999)}]{Harris99}
\bibinfo{author}{\bibfnamefont{M.}~\bibnamefont{Harris}},
  \bibinfo{journal}{{N}ature} \textbf{\bibinfo{volume}{399}},
  \bibinfo{pages}{311} (\bibinfo{year}{1999}).

\bibitem[{\citenamefont{{A. P. }Ramirez et~al.}(1999)\citenamefont{{A. P.
  }Ramirez, Hayashi, {R. J. }Cava, Siddhartan, and {B. S.
  }Shastry}}]{Ramirez99}
\bibinfo{author}{\bibnamefont{{A. P. }Ramirez}},
  \bibinfo{author}{\bibfnamefont{A.}~\bibnamefont{Hayashi}},
  \bibinfo{author}{\bibnamefont{{R. J. }Cava}},
  \bibinfo{author}{\bibfnamefont{R.}~\bibnamefont{Siddhartan}},
  \bibnamefont{and} \bibinfo{author}{\bibnamefont{{B. S. }Shastry}},
  \bibinfo{journal}{{N}ature} \textbf{\bibinfo{volume}{399}},
  \bibinfo{pages}{333} (\bibinfo{year}{1999}).

\bibitem[{\citenamefont{Obradors et~al.}(1988)\citenamefont{Obradors, Labarta,
  Isalgu\'e, Tejada, Rodriguez, and Pernet}}]{Obradors88}
\bibinfo{author}{\bibfnamefont{X.}~\bibnamefont{Obradors}},
  \bibinfo{author}{\bibfnamefont{A.}~\bibnamefont{Labarta}},
  \bibinfo{author}{\bibfnamefont{A.}~\bibnamefont{Isalgu\'e}},
  \bibinfo{author}{\bibfnamefont{J.}~\bibnamefont{Tejada}},
  \bibinfo{author}{\bibfnamefont{J.}~\bibnamefont{Rodriguez}},
  \bibnamefont{and} \bibinfo{author}{\bibfnamefont{M.}~\bibnamefont{Pernet}},
  \bibinfo{journal}{Solid State Commun.} \textbf{\bibinfo{volume}{65}},
  \bibinfo{pages}{189} (\bibinfo{year}{1988}).

\bibitem[{\citenamefont{{I. S. }Hagemann et~al.}(2001)\citenamefont{{I. S.
  }Hagemann, Huang, {X. P. A. }Gao, {A. P. }Ramirez, and {R. J.
  }Cava}}]{Hagemann01}
\bibinfo{author}{\bibnamefont{{I. S. }Hagemann}},
  \bibinfo{author}{\bibfnamefont{Q.}~\bibnamefont{Huang}},
  \bibinfo{author}{\bibnamefont{{X. P. A. }Gao}},
  \bibinfo{author}{\bibnamefont{{A. P. }Ramirez}}, \bibnamefont{and}
  \bibinfo{author}{\bibnamefont{{R. J. }Cava}}, \bibinfo{journal}{Phys. Rev.
  Lett.} \textbf{\bibinfo{volume}{86}}, \bibinfo{pages}{894}
  (\bibinfo{year}{2001}).

\bibitem[{\citenamefont{{A. P. }Ramirez et~al.}(1992)\citenamefont{{A. P.
  }Ramirez, {G. P. }Espinosa, and {A. S. }Cooper}}]{Ramirez92}
\bibinfo{author}{\bibnamefont{{A. P. }Ramirez}},
  \bibinfo{author}{\bibnamefont{{G. P. }Espinosa}}, \bibnamefont{and}
  \bibinfo{author}{\bibnamefont{{A. S. }Cooper}}, \bibinfo{journal}{Phys. Rev.
  B} \textbf{\bibinfo{volume}{45}}, \bibinfo{pages}{2505}
  (\bibinfo{year}{1992}).

\bibitem[{\citenamefont{Ohta et~al.}(1996)\citenamefont{Ohta, Sumikawa,
  Motokawa, Kikuchi, and Nagasawa}}]{Ohta96}
\bibinfo{author}{\bibfnamefont{H.}~\bibnamefont{Ohta}},
  \bibinfo{author}{\bibfnamefont{M.}~\bibnamefont{Sumikawa}},
  \bibinfo{author}{\bibfnamefont{M.}~\bibnamefont{Motokawa}},
  \bibinfo{author}{\bibfnamefont{H.}~\bibnamefont{Kikuchi}}, \bibnamefont{and}
  \bibinfo{author}{\bibfnamefont{H.}~\bibnamefont{Nagasawa}},
  \bibinfo{journal}{J. Phys. Soc. Jpn.} \textbf{\bibinfo{volume}{65}},
  \bibinfo{pages}{848} (\bibinfo{year}{1996}).

\bibitem[{\citenamefont{Limot et~al.}(2002)\citenamefont{Limot, Mendels,
  Collin, Mondelli, Ouladdiaf, Mutka, Blanchard, and Mekata}}]{Limot02}
\bibinfo{author}{\bibfnamefont{L.}~\bibnamefont{Limot}},
  \bibinfo{author}{\bibfnamefont{P.}~\bibnamefont{Mendels}},
  \bibinfo{author}{\bibfnamefont{G.}~\bibnamefont{Collin}},
  \bibinfo{author}{\bibfnamefont{C.}~\bibnamefont{Mondelli}},
  \bibinfo{author}{\bibfnamefont{B.}~\bibnamefont{Ouladdiaf}},
  \bibinfo{author}{\bibfnamefont{H.}~\bibnamefont{Mutka}},
  \bibinfo{author}{\bibfnamefont{N.}~\bibnamefont{Blanchard}},
  \bibnamefont{and} \bibinfo{author}{\bibfnamefont{M.}~\bibnamefont{Mekata}},
  \bibinfo{journal}{Phys. Rev. B} \textbf{\bibinfo{volume}{65}},
  \bibinfo{pages}{144447} (\bibinfo{year}{2002}).

\bibitem[{\citenamefont{Bono et~al.}(2004{\natexlab{a}})\citenamefont{Bono,
  Mendels, Collin, and Blanchard}}]{BonoRMN}
\bibinfo{author}{\bibfnamefont{D.}~\bibnamefont{Bono}},
  \bibinfo{author}{\bibfnamefont{P.}~\bibnamefont{Mendels}},
  \bibinfo{author}{\bibfnamefont{G.}~\bibnamefont{Collin}}, \bibnamefont{and}
  \bibinfo{author}{\bibfnamefont{N.}~\bibnamefont{Blanchard}},
  \bibinfo{journal}{Phys. Rev. Lett.} \textbf{\bibinfo{volume}{92}},
  \bibinfo{pages}{217202} (\bibinfo{year}{2004}{\natexlab{a}}).

\bibitem[{\citenamefont{Lee et~al.}(1996)\citenamefont{Lee, Broholm, Aeppli,
  Perring, Hessen, and Taylor}}]{Lee96}
\bibinfo{author}{\bibfnamefont{S.~H.} \bibnamefont{Lee}},
  \bibinfo{author}{\bibfnamefont{C.}~\bibnamefont{Broholm}},
  \bibinfo{author}{\bibfnamefont{G.}~\bibnamefont{Aeppli}},
  \bibinfo{author}{\bibfnamefont{T.~G.} \bibnamefont{Perring}},
  \bibinfo{author}{\bibfnamefont{B.}~\bibnamefont{Hessen}}, \bibnamefont{and}
  \bibinfo{author}{\bibfnamefont{A.}~\bibnamefont{Taylor}},
  \bibinfo{journal}{Phys. Rev. Lett.} \textbf{\bibinfo{volume}{76}},
  \bibinfo{pages}{4424} (\bibinfo{year}{1996}).

\bibitem[{\citenamefont{{L. J. }{de Jongh} and {A. R.
  }Miedema}(2001)}]{deJongh01}
\bibinfo{author}{\bibnamefont{{L. J. }{de Jongh}}} \bibnamefont{and}
  \bibinfo{author}{\bibnamefont{{A. R. }Miedema}}, \bibinfo{journal}{Adv.
  Phys.} \textbf{\bibinfo{volume}{50}}, \bibinfo{pages}{947}
  (\bibinfo{year}{2001}).

\bibitem[{\citenamefont{{A. P. }Ramirez et~al.}(1990)\citenamefont{{A. P.
  }Ramirez, {G. P. }Espinosa, and {A. S. }Cooper}}]{Ramirez90}
\bibinfo{author}{\bibnamefont{{A. P. }Ramirez}},
  \bibinfo{author}{\bibnamefont{{G. P. }Espinosa}}, \bibnamefont{and}
  \bibinfo{author}{\bibnamefont{{A. S. }Cooper}}, \bibinfo{journal}{Phys. Rev.
  Lett.} \textbf{\bibinfo{volume}{64}}, \bibinfo{pages}{2070}
  (\bibinfo{year}{1990}).

\bibitem[{\citenamefont{Lecheminant et~al.}(1997)\citenamefont{Lecheminant,
  Bernu, Lhuillier, Pierre, and Sindzingre}}]{Lecheminant97}
\bibinfo{author}{\bibfnamefont{P.}~\bibnamefont{Lecheminant}},
  \bibinfo{author}{\bibfnamefont{B.}~\bibnamefont{Bernu}},
  \bibinfo{author}{\bibfnamefont{C.}~\bibnamefont{Lhuillier}},
  \bibinfo{author}{\bibfnamefont{L.}~\bibnamefont{Pierre}}, \bibnamefont{and}
  \bibinfo{author}{\bibfnamefont{P.}~\bibnamefont{Sindzingre}},
  \bibinfo{journal}{Phys. Rev. B} \textbf{\bibinfo{volume}{56}},
  \bibinfo{pages}{2521} (\bibinfo{year}{1997}).

\bibitem[{\citenamefont{{A. P. }Ramirez et~al.}(2000)\citenamefont{{A. P.
  }Ramirez, Hessen, and Winklemann}}]{Ramirez00}
\bibinfo{author}{\bibnamefont{{A. P. }Ramirez}},
  \bibinfo{author}{\bibfnamefont{B.}~\bibnamefont{Hessen}}, \bibnamefont{and}
  \bibinfo{author}{\bibfnamefont{M.}~\bibnamefont{Winklemann}},
  \bibinfo{journal}{Phys. Rev. Lett.} \textbf{\bibinfo{volume}{84}},
  \bibinfo{pages}{2957} (\bibinfo{year}{2000}).

\bibitem[{\citenamefont{Sindzingre et~al.}(2000)\citenamefont{Sindzingre,
  Misguich, Lhuillier, Bernu, Pierre, Waldtmann, and Everts}}]{Sindzingre00}
\bibinfo{author}{\bibfnamefont{P.}~\bibnamefont{Sindzingre}},
  \bibinfo{author}{\bibfnamefont{G.}~\bibnamefont{Misguich}},
  \bibinfo{author}{\bibfnamefont{C.}~\bibnamefont{Lhuillier}},
  \bibinfo{author}{\bibfnamefont{B.}~\bibnamefont{Bernu}},
  \bibinfo{author}{\bibfnamefont{L.}~\bibnamefont{Pierre}},
  \bibinfo{author}{\bibfnamefont{C.}~\bibnamefont{Waldtmann}},
  \bibnamefont{and} \bibinfo{author}{\bibfnamefont{H.~U.}
  \bibnamefont{Everts}}, \bibinfo{journal}{Phys. Rev. Lett.}
  \textbf{\bibinfo{volume}{84}}, \bibinfo{pages}{2953} (\bibinfo{year}{2000}).

\bibitem[{\citenamefont{Broholm et~al.}(1990)\citenamefont{Broholm, Aeppli,
  Espinosa, and Cooper}}]{Broholm90}
\bibinfo{author}{\bibfnamefont{C.}~\bibnamefont{Broholm}},
  \bibinfo{author}{\bibfnamefont{G.}~\bibnamefont{Aeppli}},
  \bibinfo{author}{\bibfnamefont{G.~P.} \bibnamefont{Espinosa}},
  \bibnamefont{and} \bibinfo{author}{\bibfnamefont{A.~S.}
  \bibnamefont{Cooper}}, \bibinfo{journal}{Phys. Rev. Lett.}
  \textbf{\bibinfo{volume}{65}}, \bibinfo{pages}{3173} (\bibinfo{year}{1990}).

\bibitem[{\citenamefont{Uemura et~al.}(1994)\citenamefont{Uemura, Keren,
  Kojima, Le, Luke, Wu, Ajiro, Asano, Kuriyama, Mekata et~al.}}]{Uemura94}
\bibinfo{author}{\bibfnamefont{Y.~J.} \bibnamefont{Uemura}},
  \bibinfo{author}{\bibfnamefont{A.}~\bibnamefont{Keren}},
  \bibinfo{author}{\bibfnamefont{K.}~\bibnamefont{Kojima}},
  \bibinfo{author}{\bibfnamefont{L.~P.} \bibnamefont{Le}},
  \bibinfo{author}{\bibfnamefont{G.~M.} \bibnamefont{Luke}},
  \bibinfo{author}{\bibfnamefont{W.~D.} \bibnamefont{Wu}},
  \bibinfo{author}{\bibfnamefont{Y.}~\bibnamefont{Ajiro}},
  \bibinfo{author}{\bibfnamefont{T.}~\bibnamefont{Asano}},
  \bibinfo{author}{\bibfnamefont{Y.}~\bibnamefont{Kuriyama}},
  \bibinfo{author}{\bibfnamefont{M.}~\bibnamefont{Mekata}},
  \bibnamefont{et~al.}, \bibinfo{journal}{Phys. Rev. Lett.}
  \textbf{\bibinfo{volume}{73}}, \bibinfo{pages}{3306} (\bibinfo{year}{1994}).

\bibitem[{\citenamefont{Mart\'inez et~al.}(1992)\citenamefont{Mart\'inez,
  Sandiumenge, Rouco, Labarta, , Rodr\'iguez-Carvajal, Tovar, Causa, Gal\'i,
  and Obradors}}]{Martinez92}
\bibinfo{author}{\bibfnamefont{B.}~\bibnamefont{Mart\'\i nez}},
  \bibinfo{author}{\bibfnamefont{F.}~\bibnamefont{Sandiumenge}},
  \bibinfo{author}{\bibfnamefont{A.}~\bibnamefont{Rouco}},
  \bibinfo{author}{\bibfnamefont{A.}~\bibnamefont{Labarta}},
  \bibinfo{author}{\bibfnamefont{J.}~\bibnamefont{Rodr\'\i guez-Carvajal}},
  \bibinfo{author}{\bibfnamefont{M.}~\bibnamefont{Tovar}},
  \bibinfo{author}{\bibfnamefont{M.~T.} \bibnamefont{Causa}},
  \bibinfo{author}{\bibfnamefont{S.}~\bibnamefont{Gal\'\i}}, \bibnamefont{and}
  \bibinfo{author}{\bibfnamefont{X.}~\bibnamefont{Obradors}},
  \bibinfo{journal}{Phys. Rev. B} \textbf{\bibinfo{volume}{46}},
  \bibinfo{pages}{10786} (\bibinfo{year}{1992}).

\bibitem[{Mon()}]{Mondellipriv}
\bibinfo{note}{C. Mondelli et al., \emph{private communication}}.

\bibitem[{\citenamefont{{A. P. }Ramirez}(1994)}]{Ramirez94}
\bibinfo{author}{\bibnamefont{{A. P. }Ramirez}}, \bibinfo{journal}{Annu. Rev.
  Mater. Sci.} \textbf{\bibinfo{volume}{24}}, \bibinfo{pages}{453}
  (\bibinfo{year}{1994}).

\bibitem[{\citenamefont{Keren et~al.}(1998)\citenamefont{Keren, Mendels,
  Horvati\'c, Ferrer, Uemura, Mekata, and Asano}}]{Keren98}
\bibinfo{author}{\bibfnamefont{A.}~\bibnamefont{Keren}},
  \bibinfo{author}{\bibfnamefont{P.}~\bibnamefont{Mendels}},
  \bibinfo{author}{\bibfnamefont{M.}~\bibnamefont{Horvati\'c}},
  \bibinfo{author}{\bibfnamefont{F.}~\bibnamefont{Ferrer}},
  \bibinfo{author}{\bibfnamefont{Y.~J.} \bibnamefont{Uemura}},
  \bibinfo{author}{\bibfnamefont{M.}~\bibnamefont{Mekata}}, \bibnamefont{and}
  \bibinfo{author}{\bibfnamefont{T.}~\bibnamefont{Asano}},
  \bibinfo{journal}{Phys. Rev. B} \textbf{\bibinfo{volume}{57}},
  \bibinfo{pages}{10745} (\bibinfo{year}{1998}).

\bibitem[{\citenamefont{Mendels et~al.}(2000)\citenamefont{Mendels, Keren,
  Limot, Mekata, Collin, and Horvati\'c}}]{Mendels00}
\bibinfo{author}{\bibfnamefont{P.}~\bibnamefont{Mendels}},
  \bibinfo{author}{\bibfnamefont{A.}~\bibnamefont{Keren}},
  \bibinfo{author}{\bibfnamefont{L.}~\bibnamefont{Limot}},
  \bibinfo{author}{\bibfnamefont{M.}~\bibnamefont{Mekata}},
  \bibinfo{author}{\bibfnamefont{G.}~\bibnamefont{Collin}}, \bibnamefont{and}
  \bibinfo{author}{\bibfnamefont{M.}~\bibnamefont{Horvati\'c}},
  \bibinfo{journal}{Phys. Rev. Lett.} \textbf{\bibinfo{volume}{85}},
  \bibinfo{pages}{3496} (\bibinfo{year}{2000}).

\bibitem[{\citenamefont{Limot et~al.}(2001)\citenamefont{Limot, Mendels,
  Collin, Mondelli, Mutka, and Blanchard}}]{Limot01}
\bibinfo{author}{\bibfnamefont{L.}~\bibnamefont{Limot}},
  \bibinfo{author}{\bibfnamefont{P.}~\bibnamefont{Mendels}},
  \bibinfo{author}{\bibfnamefont{G.}~\bibnamefont{Collin}},
  \bibinfo{author}{\bibfnamefont{C.}~\bibnamefont{Mondelli}},
  \bibinfo{author}{\bibfnamefont{H.}~\bibnamefont{Mutka}}, \bibnamefont{and}
  \bibinfo{author}{\bibfnamefont{N.}~\bibnamefont{Blanchard}},
  \bibinfo{journal}{Can. J. Phys.} \textbf{\bibinfo{volume}{79}},
  \bibinfo{pages}{1393} (\bibinfo{year}{2001}).

\bibitem[{\citenamefont{Bono et~al.}(2004{\natexlab{b}})\citenamefont{Bono,
  Mendels, Collin, and Blanchard}}]{BonoHFM}
\bibinfo{author}{\bibfnamefont{D.}~\bibnamefont{Bono}},
  \bibinfo{author}{\bibfnamefont{P.}~\bibnamefont{Mendels}},
  \bibinfo{author}{\bibfnamefont{G.}~\bibnamefont{Collin}}, \bibnamefont{and}
  \bibinfo{author}{\bibfnamefont{N.}~\bibnamefont{Blanchard}},
  \bibinfo{journal}{J. Phys.: Condens. Matter} \textbf{\bibinfo{volume}{16}},
  \bibinfo{pages}{S817} (\bibinfo{year}{2004}{\natexlab{b}}).

\bibitem[{\citenamefont{Keren et~al.}(2000)\citenamefont{Keren, Uemura, Luke,
  Mendels, Mekata, and Asano}}]{Keren00}
\bibinfo{author}{\bibfnamefont{A.}~\bibnamefont{Keren}},
  \bibinfo{author}{\bibfnamefont{Y.~J.} \bibnamefont{Uemura}},
  \bibinfo{author}{\bibfnamefont{G.}~\bibnamefont{Luke}},
  \bibinfo{author}{\bibfnamefont{P.}~\bibnamefont{Mendels}},
  \bibinfo{author}{\bibfnamefont{M.}~\bibnamefont{Mekata}}, \bibnamefont{and}
  \bibinfo{author}{\bibfnamefont{T.}~\bibnamefont{Asano}},
  \bibinfo{journal}{Phys. Rev. Lett.} \textbf{\bibinfo{volume}{84}},
  \bibinfo{pages}{3450} (\bibinfo{year}{2000}).

\bibitem[{\citenamefont{Bono et~al.}(2004{\natexlab{c}})\citenamefont{Bono,
  Mendels, Collin, Blanchard, Bert, Amato, Baines, and Hillier}}]{BonoMuSR}
\bibinfo{author}{\bibfnamefont{D.}~\bibnamefont{Bono}},
  \bibinfo{author}{\bibfnamefont{P.}~\bibnamefont{Mendels}},
  \bibinfo{author}{\bibfnamefont{G.}~\bibnamefont{Collin}},
  \bibinfo{author}{\bibfnamefont{N.}~\bibnamefont{Blanchard}},
  \bibinfo{author}{\bibfnamefont{F.}~\bibnamefont{Bert}},
  \bibinfo{author}{\bibfnamefont{A.}~\bibnamefont{Amato}},
  \bibinfo{author}{\bibfnamefont{C.}~\bibnamefont{Baines}}, \bibnamefont{and}
  \bibinfo{author}{\bibfnamefont{A.~D.} \bibnamefont{Hillier}},
  \bibinfo{journal}{Phys. Rev. Lett.} \textbf{\bibinfo{volume}{93}}
  (\bibinfo{year}{2004}{\natexlab{c}}).

\bibitem[{\citenamefont{{D. E. }MacLaughlin and Alloul}(1976)}]{Alloul76}
\bibinfo{author}{\bibnamefont{{D. E. }MacLaughlin}} \bibnamefont{and}
  \bibinfo{author}{\bibfnamefont{H.}~\bibnamefont{Alloul}},
  \bibinfo{journal}{Phys. Rev. Lett.} \textbf{\bibinfo{volume}{36}},
  \bibinfo{pages}{1158} (\bibinfo{year}{1976}).

\bibitem[{\citenamefont{Harris et~al.}(1992)\citenamefont{Harris, Kallin, and
  Berlinsky}}]{Harris92}
\bibinfo{author}{\bibfnamefont{A.~B.} \bibnamefont{Harris}},
  \bibinfo{author}{\bibfnamefont{C.}~\bibnamefont{Kallin}}, \bibnamefont{and}
  \bibinfo{author}{\bibfnamefont{A.~J.} \bibnamefont{Berlinsky}},
  \bibinfo{journal}{Phys. Rev. B} \textbf{\bibinfo{volume}{45}},
  \bibinfo{pages}{2899} (\bibinfo{year}{1992}).

\bibitem[{\citenamefont{Motida and Miyahara}(1970)}]{Motida70}
\bibinfo{author}{\bibfnamefont{K.}~\bibnamefont{Motida}} \bibnamefont{and}
  \bibinfo{author}{\bibfnamefont{S.}~\bibnamefont{Miyahara}},
  \bibinfo{journal}{J. Phys. Soc. Jpn.} \textbf{\bibinfo{volume}{28}},
  \bibinfo{pages}{1188} (\bibinfo{year}{1970}).

\bibitem[{\citenamefont{Samuelsen et~al.}(1970)\citenamefont{Samuelsen,
  Hutchings, and Shirane}}]{Samuelsen70}
\bibinfo{author}{\bibfnamefont{E.~J.} \bibnamefont{Samuelsen}},
  \bibinfo{author}{\bibfnamefont{M.~T.} \bibnamefont{Hutchings}},
  \bibnamefont{and} \bibinfo{author}{\bibfnamefont{G.}~\bibnamefont{Shirane}},
  \bibinfo{journal}{Physica} \textbf{\bibinfo{volume}{48}}, \bibinfo{pages}{13}
  (\bibinfo{year}{1970}).

\bibitem[{\citenamefont{Mondelli et~al.}(1999)\citenamefont{Mondelli, Andersen,
  Mutka, Payen, and Frick}}]{Mondelli99B}
\bibinfo{author}{\bibfnamefont{C.}~\bibnamefont{Mondelli}},
  \bibinfo{author}{\bibfnamefont{K.}~\bibnamefont{Andersen}},
  \bibinfo{author}{\bibfnamefont{H.}~\bibnamefont{Mutka}},
  \bibinfo{author}{\bibfnamefont{C.}~\bibnamefont{Payen}}, \bibnamefont{and}
  \bibinfo{author}{\bibfnamefont{B.}~\bibnamefont{Frick}},
  \bibinfo{journal}{Physica B} \textbf{\bibinfo{volume}{267-268}},
  \bibinfo{pages}{139} (\bibinfo{year}{1999}).

\bibitem[{\citenamefont{Garcia-Adeva and Huber}(2001)}]{Garcia01}
\bibinfo{author}{\bibfnamefont{A.~J.} \bibnamefont{Garcia-Adeva}}
  \bibnamefont{and} \bibinfo{author}{\bibfnamefont{D.~L.} \bibnamefont{Huber}},
  \bibinfo{journal}{Phys. Rev. B} \textbf{\bibinfo{volume}{63}},
  \bibinfo{pages}{174433} (\bibinfo{year}{2001}).

\bibitem[{\citenamefont{Schiffer and Daruka}(1997)}]{Schiffer97}
\bibinfo{author}{\bibfnamefont{P.}~\bibnamefont{Schiffer}} \bibnamefont{and}
  \bibinfo{author}{\bibfnamefont{I.}~\bibnamefont{Daruka}},
  \bibinfo{journal}{Phys. Rev. B} \textbf{\bibinfo{volume}{56}},
  \bibinfo{pages}{13712} (\bibinfo{year}{1997}).

\bibitem[{\citenamefont{Takigawa et~al.}(1999)\citenamefont{Takigawa, Motoyama,
  Eisaki, and Uchida}}]{Takigawa97}
\bibinfo{author}{\bibfnamefont{M.}~\bibnamefont{Takigawa}},
  \bibinfo{author}{\bibfnamefont{N.}~\bibnamefont{Motoyama}},
  \bibinfo{author}{\bibfnamefont{H.}~\bibnamefont{Eisaki}}, \bibnamefont{and}
  \bibinfo{author}{\bibfnamefont{S.}~\bibnamefont{Uchida}},
  \bibinfo{journal}{Phys. Rev. B} \textbf{\bibinfo{volume}{55}},
  \bibinfo{pages}{14129} (\bibinfo{year}{1999}).

\bibitem[{\citenamefont{Tedoldi et~al.}(1999)\citenamefont{Tedoldi,
  Santachiara, and Horvatic}}]{Tedoldi99}
\bibinfo{author}{\bibfnamefont{F.}~\bibnamefont{Tedoldi}},
  \bibinfo{author}{\bibfnamefont{R.}~\bibnamefont{Santachiara}},
  \bibnamefont{and} \bibinfo{author}{\bibfnamefont{M.}~\bibnamefont{Horvatic}},
  \bibinfo{journal}{Phys. Rev. Lett.} \textbf{\bibinfo{volume}{83}},
  \bibinfo{pages}{412} (\bibinfo{year}{1999}).

\bibitem[{\citenamefont{Azuma et~al.}(1994)\citenamefont{Azuma, Hiroi, Takano,
  Ishida, and Kitaoka}}]{Azuma94}
\bibinfo{author}{\bibfnamefont{M.}~\bibnamefont{Azuma}},
  \bibinfo{author}{\bibfnamefont{Z.}~\bibnamefont{Hiroi}},
  \bibinfo{author}{\bibfnamefont{M.}~\bibnamefont{Takano}},
  \bibinfo{author}{\bibfnamefont{K.}~\bibnamefont{Ishida}}, \bibnamefont{and}
  \bibinfo{author}{\bibfnamefont{Y.}~\bibnamefont{Kitaoka}},
  \bibinfo{journal}{Phys. Rev. Lett.} \textbf{\bibinfo{volume}{73}},
  \bibinfo{pages}{3463} (\bibinfo{year}{1994}).

\bibitem[{\citenamefont{Ouazi et~al.}(2004)\citenamefont{Ouazi, Bobroff,
  Alloul, and MacFarlane}}]{Ouazi04}
\bibinfo{author}{\bibfnamefont{S.}~\bibnamefont{Ouazi}},
  \bibinfo{author}{\bibfnamefont{J.}~\bibnamefont{Bobroff}},
  \bibinfo{author}{\bibfnamefont{H.}~\bibnamefont{Alloul}}, \bibnamefont{and}
  \bibinfo{author}{\bibfnamefont{W.~A.} \bibnamefont{MacFarlane}},
  \bibinfo{journal}{Phys. Rev. B} \textbf{\bibinfo{volume}{70}},
  \bibinfo{pages}{104515} (\bibinfo{year}{2004}).

\bibitem[{\citenamefont{Moessner and Berlinsky}(1999)}]{Moessner99}
\bibinfo{author}{\bibfnamefont{R.}~\bibnamefont{Moessner}} \bibnamefont{and}
  \bibinfo{author}{\bibfnamefont{A.~J.} \bibnamefont{Berlinsky}},
  \bibinfo{journal}{Phys. Rev. Lett.} \textbf{\bibinfo{volume}{83}},
  \bibinfo{pages}{3293} (\bibinfo{year}{1999}).

\bibitem[{\citenamefont{Abragam}(1961)}]{Abragam}
\bibinfo{author}{\bibfnamefont{A.}~\bibnamefont{Abragam}},
  \emph{\bibinfo{title}{Principles of Nuclear Magnetism}}
  (\bibinfo{publisher}{Clarendon Press-Oxford}, \bibinfo{year}{1961}).

\bibitem[{\citenamefont{Lee et~al.}(1999)\citenamefont{Lee, Kilcoyne, and
  Cywinski}}]{Muons}
\bibinfo{editor}{\bibfnamefont{S.~L.} \bibnamefont{Lee}},
  \bibinfo{editor}{\bibfnamefont{S.~H.} \bibnamefont{Kilcoyne}},
  \bibnamefont{and} \bibinfo{editor}{\bibfnamefont{R.}~\bibnamefont{Cywinski}},
  eds., \emph{\bibinfo{title}{Muon Science -- Muons in Physics, Chemistry and
  Materials}} (\bibinfo{publisher}{Scottish Universities Summer School in
  Physics \& Institute of Physics Publishing, Bristol and Philadelphia},
  \bibinfo{year}{1999}), \bibinfo{note}{proceedings of the Fifty First Scottish
  Universities Summer School in Physics, August 1998}.

\bibitem[{\citenamefont{Hayano et~al.}(1979)\citenamefont{Hayano, Uemura,
  Imazato, Nishida, Yamazaki, and Kubo}}]{Hayano79}
\bibinfo{author}{\bibfnamefont{R.~S.} \bibnamefont{Hayano}},
  \bibinfo{author}{\bibfnamefont{Y.~J.} \bibnamefont{Uemura}},
  \bibinfo{author}{\bibfnamefont{J.}~\bibnamefont{Imazato}},
  \bibinfo{author}{\bibfnamefont{N.}~\bibnamefont{Nishida}},
  \bibinfo{author}{\bibfnamefont{T.}~\bibnamefont{Yamazaki}}, \bibnamefont{and}
  \bibinfo{author}{\bibfnamefont{R.}~\bibnamefont{Kubo}},
  \bibinfo{journal}{Phys. Rev. B} \textbf{\bibinfo{volume}{20}},
  \bibinfo{pages}{850} (\bibinfo{year}{1979}).

\bibitem[{\citenamefont{Uemura et~al.}(1985)\citenamefont{Uemura, Yamazaki,
  Harshman, Senba, and Ansaldo}}]{Uemura85}
\bibinfo{author}{\bibfnamefont{Y.~J.} \bibnamefont{Uemura}},
  \bibinfo{author}{\bibfnamefont{T.}~\bibnamefont{Yamazaki}},
  \bibinfo{author}{\bibfnamefont{D.~R.} \bibnamefont{Harshman}},
  \bibinfo{author}{\bibfnamefont{M.}~\bibnamefont{Senba}}, \bibnamefont{and}
  \bibinfo{author}{\bibfnamefont{E.~J.} \bibnamefont{Ansaldo}},
  \bibinfo{journal}{Phys. Rev. B} \textbf{\bibinfo{volume}{31}},
  \bibinfo{pages}{546} (\bibinfo{year}{1985}).

\bibitem[{\citenamefont{Keren}(1994)}]{Keren94}
\bibinfo{author}{\bibfnamefont{A.}~\bibnamefont{Keren}},
  \bibinfo{journal}{Phys. Rev. B} \textbf{\bibinfo{volume}{50}},
  \bibinfo{pages}{10039} (\bibinfo{year}{1994}).

\bibitem[{\citenamefont{Keren et~al.}(1996)\citenamefont{Keren, Kojima, Le,
  Luke, Wu, Uemura, Takano, Dabkowska, and {M. J. P. }Gingras}}]{Keren96}
\bibinfo{author}{\bibfnamefont{A.}~\bibnamefont{Keren}},
  \bibinfo{author}{\bibfnamefont{K.}~\bibnamefont{Kojima}},
  \bibinfo{author}{\bibfnamefont{L.~P.} \bibnamefont{Le}},
  \bibinfo{author}{\bibfnamefont{G.~M.} \bibnamefont{Luke}},
  \bibinfo{author}{\bibfnamefont{W.~D.} \bibnamefont{Wu}},
  \bibinfo{author}{\bibfnamefont{Y.~J.} \bibnamefont{Uemura}},
  \bibinfo{author}{\bibfnamefont{M.}~\bibnamefont{Takano}},
  \bibinfo{author}{\bibfnamefont{H.}~\bibnamefont{Dabkowska}},
  \bibnamefont{and} \bibinfo{author}{\bibnamefont{{M. J. P. }Gingras}},
  \bibinfo{journal}{Phys. Rev. B} \textbf{\bibinfo{volume}{53}},
  \bibinfo{pages}{2451} (\bibinfo{year}{1996}).

\bibitem[{\citenamefont{Fukaya et~al.}(2003{\natexlab{a}})\citenamefont{Fukaya,
  Fudamoto, Gat, Ito, Larkin, Savici, Uemura, Kyriakou, Luke, {M. T. }Rovers
  et~al.}}]{Fukaya03}
\bibinfo{author}{\bibfnamefont{A.}~\bibnamefont{Fukaya}},
  \bibinfo{author}{\bibfnamefont{Y.}~\bibnamefont{Fudamoto}},
  \bibinfo{author}{\bibfnamefont{I.~M.} \bibnamefont{Gat}},
  \bibinfo{author}{\bibfnamefont{T.}~\bibnamefont{Ito}},
  \bibinfo{author}{\bibfnamefont{M.~I.} \bibnamefont{Larkin}},
  \bibinfo{author}{\bibfnamefont{A.~T.} \bibnamefont{Savici}},
  \bibinfo{author}{\bibfnamefont{Y.~J.} \bibnamefont{Uemura}},
  \bibinfo{author}{\bibfnamefont{P.~P.} \bibnamefont{Kyriakou}},
  \bibinfo{author}{\bibfnamefont{G.~M.} \bibnamefont{Luke}},
  \bibinfo{author}{\bibnamefont{{M. T. }Rovers}}, \bibnamefont{et~al.},
  \bibinfo{journal}{Phys. Rev. Lett.} \textbf{\bibinfo{volume}{91}},
  \bibinfo{pages}{207603} (\bibinfo{year}{2003}{\natexlab{a}}).

\bibitem[{\citenamefont{Kojima et~al.}(1995)\citenamefont{Kojima, Keren, Le,
  Luke, Nachumi, Wu, Uemura, Kiyono, Miyasaka, Takagi et~al.}}]{Kojima95}
\bibinfo{author}{\bibfnamefont{K.}~\bibnamefont{Kojima}},
  \bibinfo{author}{\bibfnamefont{A.}~\bibnamefont{Keren}},
  \bibinfo{author}{\bibfnamefont{L.~P.} \bibnamefont{Le}},
  \bibinfo{author}{\bibfnamefont{G.~M.} \bibnamefont{Luke}},
  \bibinfo{author}{\bibfnamefont{B.}~\bibnamefont{Nachumi}},
  \bibinfo{author}{\bibfnamefont{W.~D.} \bibnamefont{Wu}},
  \bibinfo{author}{\bibfnamefont{Y.~J.} \bibnamefont{Uemura}},
  \bibinfo{author}{\bibfnamefont{K.}~\bibnamefont{Kiyono}},
  \bibinfo{author}{\bibfnamefont{S.}~\bibnamefont{Miyasaka}},
  \bibinfo{author}{\bibfnamefont{H.}~\bibnamefont{Takagi}},
  \bibnamefont{et~al.}, \bibinfo{journal}{Phys. Rev. Lett.}
  \textbf{\bibinfo{volume}{74}}, \bibinfo{pages}{3471} (\bibinfo{year}{1995}).

\bibitem[{\citenamefont{Chubukov}(1992)}]{Chubukov92}
\bibinfo{author}{\bibfnamefont{A.}~\bibnamefont{Chubukov}},
  \bibinfo{journal}{Phys. Rev. Lett.} \textbf{\bibinfo{volume}{69}},
  \bibinfo{pages}{832} (\bibinfo{year}{1992}).

\bibitem[{\citenamefont{Mila and Dean}(2002)}]{Mila02}
\bibinfo{author}{\bibfnamefont{F.}~\bibnamefont{Mila}} \bibnamefont{and}
  \bibinfo{author}{\bibfnamefont{D.}~\bibnamefont{Dean}},
  \bibinfo{journal}{Eur. Phys. J. B} \textbf{\bibinfo{volume}{26}},
  \bibinfo{pages}{301} (\bibinfo{year}{2002}).

\bibitem[{\citenamefont{Ferrero et~al.}(2003)\citenamefont{Ferrero, Becca, and
  Mila}}]{Ferrero03}
\bibinfo{author}{\bibfnamefont{M.}~\bibnamefont{Ferrero}},
  \bibinfo{author}{\bibfnamefont{F.}~\bibnamefont{Becca}}, \bibnamefont{and}
  \bibinfo{author}{\bibfnamefont{F.}~\bibnamefont{Mila}},
  \bibinfo{journal}{Phys. Rev. B} \textbf{\bibinfo{volume}{68}},
  \bibinfo{pages}{214431} (\bibinfo{year}{2003}).

\bibitem[{\citenamefont{Hiroi et~al.}(2001)\citenamefont{Hiroi, Hanawa,
  Kobayashi, Nohara, Takagi, Kato, and Takigawa}}]{Hiroi01}
\bibinfo{author}{\bibfnamefont{Z.}~\bibnamefont{Hiroi}},
  \bibinfo{author}{\bibfnamefont{M.}~\bibnamefont{Hanawa}},
  \bibinfo{author}{\bibfnamefont{N.}~\bibnamefont{Kobayashi}},
  \bibinfo{author}{\bibfnamefont{M.}~\bibnamefont{Nohara}},
  \bibinfo{author}{\bibfnamefont{H.}~\bibnamefont{Takagi}},
  \bibinfo{author}{\bibfnamefont{Y.}~\bibnamefont{Kato}}, \bibnamefont{and}
  \bibinfo{author}{\bibfnamefont{M.}~\bibnamefont{Takigawa}},
  \bibinfo{journal}{J. Phys. Soc. Jpn.} \textbf{\bibinfo{volume}{70}},
  \bibinfo{pages}{3377} (\bibinfo{year}{2001}).

\bibitem[{\citenamefont{Bert et~al.}(2004)\citenamefont{Bert, Bono, Mendels,
  Trombe, Millet, Amato, Baines, and Hillier}}]{BertHFM}
\bibinfo{author}{\bibfnamefont{F.}~\bibnamefont{Bert}},
  \bibinfo{author}{\bibfnamefont{D.}~\bibnamefont{Bono}},
  \bibinfo{author}{\bibfnamefont{P.}~\bibnamefont{Mendels}},
  \bibinfo{author}{\bibfnamefont{J.~C.} \bibnamefont{Trombe}},
  \bibinfo{author}{\bibfnamefont{P.}~\bibnamefont{Millet}},
  \bibinfo{author}{\bibfnamefont{A.}~\bibnamefont{Amato}},
  \bibinfo{author}{\bibfnamefont{C.}~\bibnamefont{Baines}}, \bibnamefont{and}
  \bibinfo{author}{\bibfnamefont{A.}~\bibnamefont{Hillier}},
  \bibinfo{journal}{J. Phys.: Condens. Matter} \textbf{\bibinfo{volume}{16}},
  \bibinfo{pages}{S829} (\bibinfo{year}{2004}).

\bibitem[{\citenamefont{Kojima}(1995)}]{KojimaPhD}
\bibinfo{author}{\bibfnamefont{K.}~\bibnamefont{Kojima}}, Ph.D. thesis,
  \bibinfo{school}{The University of Tokyo} (\bibinfo{year}{1995}).

\bibitem[{\citenamefont{Ammon and Imada}(2000)}]{Ammon00}
\bibinfo{author}{\bibfnamefont{B.}~\bibnamefont{Ammon}} \bibnamefont{and}
  \bibinfo{author}{\bibfnamefont{M.}~\bibnamefont{Imada}},
  \bibinfo{journal}{Phys. Rev. Lett.} \textbf{\bibinfo{volume}{85}},
  \bibinfo{pages}{1056} (\bibinfo{year}{2000}).

\bibitem[{\citenamefont{Fukaya et~al.}(2003{\natexlab{b}})\citenamefont{Fukaya,
  Fudamoto, Gat, Ito, Larkin, Savici, Uemura, Kyriakou, Luke, Rovers
  et~al.}}]{Fukaya03B}
\bibinfo{author}{\bibfnamefont{A.}~\bibnamefont{Fukaya}},
  \bibinfo{author}{\bibfnamefont{Y.}~\bibnamefont{Fudamoto}},
  \bibinfo{author}{\bibfnamefont{I.~M.} \bibnamefont{Gat}},
  \bibinfo{author}{\bibfnamefont{T.}~\bibnamefont{Ito}},
  \bibinfo{author}{\bibfnamefont{M.~I.} \bibnamefont{Larkin}},
  \bibinfo{author}{\bibfnamefont{A.~T.} \bibnamefont{Savici}},
  \bibinfo{author}{\bibfnamefont{Y.~J.} \bibnamefont{Uemura}},
  \bibinfo{author}{\bibfnamefont{P.~P.} \bibnamefont{Kyriakou}},
  \bibinfo{author}{\bibfnamefont{G.~M.} \bibnamefont{Luke}},
  \bibinfo{author}{\bibfnamefont{M.~T.} \bibnamefont{Rovers}},
  \bibnamefont{et~al.}, \bibinfo{journal}{Physica B}
  \textbf{\bibinfo{volume}{326}}, \bibinfo{pages}{446}
  (\bibinfo{year}{2003}{\natexlab{b}}).

\bibitem[{\citenamefont{Kageyama et~al.}(1999)\citenamefont{Kageyama,
  Yoshimura, Stern, Mushnikov, Onizuka, Kato, Kosuge, Slichter, Goto, and
  Ueda}}]{Kageyama99}
\bibinfo{author}{\bibfnamefont{H.}~\bibnamefont{Kageyama}},
  \bibinfo{author}{\bibfnamefont{K.}~\bibnamefont{Yoshimura}},
  \bibinfo{author}{\bibfnamefont{R.}~\bibnamefont{Stern}},
  \bibinfo{author}{\bibfnamefont{N.~V.} \bibnamefont{Mushnikov}},
  \bibinfo{author}{\bibfnamefont{K.}~\bibnamefont{Onizuka}},
  \bibinfo{author}{\bibfnamefont{M.}~\bibnamefont{Kato}},
  \bibinfo{author}{\bibfnamefont{K.}~\bibnamefont{Kosuge}},
  \bibinfo{author}{\bibfnamefont{C.~P.} \bibnamefont{Slichter}},
  \bibinfo{author}{\bibfnamefont{T.}~\bibnamefont{Goto}}, \bibnamefont{and}
  \bibinfo{author}{\bibfnamefont{Y.}~\bibnamefont{Ueda}},
  \bibinfo{journal}{Phys. Rev. Lett.} \textbf{\bibinfo{volume}{82}},
  \bibinfo{pages}{3168} (\bibinfo{year}{1999}).

\bibitem[{\citenamefont{Kodama et~al.}(2002)\citenamefont{Kodama, Takigawa,
  Horvati\'c, Berthier, Kageyama, Ueda, Miyahara, Becca, and Mila}}]{Kodama02}
\bibinfo{author}{\bibfnamefont{K.}~\bibnamefont{Kodama}},
  \bibinfo{author}{\bibfnamefont{M.}~\bibnamefont{Takigawa}},
  \bibinfo{author}{\bibfnamefont{M.}~\bibnamefont{Horvati\'c}},
  \bibinfo{author}{\bibfnamefont{C.}~\bibnamefont{Berthier}},
  \bibinfo{author}{\bibfnamefont{H.}~\bibnamefont{Kageyama}},
  \bibinfo{author}{\bibfnamefont{Y.}~\bibnamefont{Ueda}},
  \bibinfo{author}{\bibfnamefont{S.}~\bibnamefont{Miyahara}},
  \bibinfo{author}{\bibfnamefont{F.}~\bibnamefont{Becca}}, \bibnamefont{and}
  \bibinfo{author}{\bibfnamefont{F.}~\bibnamefont{Mila}},
  \bibinfo{journal}{Science} \textbf{\bibinfo{volume}{298}},
  \bibinfo{pages}{395} (\bibinfo{year}{2002}).

\bibitem[{Mut()}]{Mutkaprep}
\bibinfo{note}{{H.~Mutka \emph{et al.}, in preparation.}}

\bibitem[{\citenamefont{Cohen and Reif}(1957)}]{Cohen57}
\bibinfo{author}{\bibfnamefont{M.~H.} \bibnamefont{Cohen}} \bibnamefont{and}
  \bibinfo{author}{\bibfnamefont{F.}~\bibnamefont{Reif}},
  \bibinfo{journal}{{S}olid {S}tate Physics} \textbf{\bibinfo{volume}{5}},
  \bibinfo{pages}{321} (\bibinfo{year}{1957}).

\bibitem[{\citenamefont{{R. E. }Walstedt and {L. R.
  }Walker}(1974)}]{Walstedt74}
\bibinfo{author}{\bibnamefont{{R. E. }Walstedt}} \bibnamefont{and}
  \bibinfo{author}{\bibnamefont{{L. R. }Walker}}, \bibinfo{journal}{Phys. Rev.
  B} \textbf{\bibinfo{volume}{9}}, \bibinfo{pages}{4857}
  (\bibinfo{year}{1974}).

\bibitem[{\citenamefont{Brewer et~al.}(1990)\citenamefont{Brewer, Kiefl,
  Carolan, Dosanjh, Hardy, Kreitzman, Li, Riseman, Schleger, Zhou
  et~al.}}]{Brewer90}
\bibinfo{author}{\bibfnamefont{J.~H.} \bibnamefont{Brewer}},
  \bibinfo{author}{\bibfnamefont{R.~F.} \bibnamefont{Kiefl}},
  \bibinfo{author}{\bibfnamefont{J.~F.} \bibnamefont{Carolan}},
  \bibinfo{author}{\bibfnamefont{P.}~\bibnamefont{Dosanjh}},
  \bibinfo{author}{\bibfnamefont{W.~N.} \bibnamefont{Hardy}},
  \bibinfo{author}{\bibfnamefont{S.~R.} \bibnamefont{Kreitzman}},
  \bibinfo{author}{\bibfnamefont{Q.}~\bibnamefont{Li}},
  \bibinfo{author}{\bibfnamefont{T.~M.} \bibnamefont{Riseman}},
  \bibinfo{author}{\bibfnamefont{P.}~\bibnamefont{Schleger}},
  \bibinfo{author}{\bibfnamefont{H.}~\bibnamefont{Zhou}}, \bibnamefont{et~al.},
  \bibinfo{journal}{{H}yperfine {I}nteract.} \textbf{\bibinfo{volume}{63}},
  \bibinfo{pages}{177} (\bibinfo{year}{1990}).

\end{thebibliography}
\end{document}